\documentstyle[12pt,fleqn,a4,twoside,fancyheadings]{report}
\def\ra{\rightarrow}
\def\be{\begin{equation}}
\def\ee{\end{equation}}
\def\bea{\begin{eqnarray}}
\def\eea{\end{eqnarray}}

\def\PP{ {\bf !!!P!!! \abc \end{document}} }
\title{\LARGE\bf The Isolated Photon Production\\
in the Deep Inelastic Compton Process \\
at the $ep$ collider DESY HERA}
\author{\\[1.5cm] {\Large Andrzej Zembrzuski} \\[2cm]
\\{Institute of Theoretical Physics,}\\ {Warsaw University, Poland} \\[2cm]
\\ {\small\sl A Dissertation submitted}\\
{\small\sl in partial fulfillment of the}\\
{\small\sl requirements for the degree of}\\
{\small\sl Doctor of Philosophy}\\[1cm]{}}
\date{July 2004\\[0cm]{}}

\begin{document}
\pagenumbering{roman}
\titlepage
\setcounter{page}{1}
\begin{center}
\Large Institute of Theoretical Physics\\
Faculty of Physics, Warsaw University\\
~\\~\\~\\~

%\LARGE\bf The NLO QCD predictions\\for the Compton process\\at DESY HERA
%\LARGE\bf Photoproduction of prompt photons\\at DESY HERA\\
%in next-to-leading order QCD
%\LARGE\bf The Isolated Photon Production\\
%in the Deep Inelastic Compton Process\\at the $ep$ collider DESY HERA \\
\LARGE\bf Isolated Photon Production\\
in Deep Inelastic Compton Process \\ at HERA \\
\Large
~\\~\\~\\~\\~

\Large Andrzej Zembrzuski\\
~\\~\\~\\~\\~

\small\sl A Dissertation submitted\\
in partial fulfillment of the\\
requirements for the degree of\\
Doctor of Philosophy\\
\Large
~\\~

\Large Supervisor: dr hab. Maria Krawczyk, prof. UW\\
~\\~

August 2004

\end{center}
%\maketitle
\newpage
\setcounter{page}{1}
\thispagestyle{empty}
~\\ \\ \\ \\ \\ \\ \\ \\ \\
\begin{center}
{\Large\bf Acknowledgments}

~\\~

\begin{minipage}[t]{13cm}

I would like to express my gratitude to my supervisor, Prof. Maria
Krawczyk, for her support, many helpful discussions and the suggestion
of the interesting topic. This dissertation could not have been
accomplished without the guidance of Prof. Maria Krawczyk.

~

I am grateful to Prof. Peter Bussey and Dr. Sergei Chekanov
from the ZEUS Collaboration, and Dr. J\"org Gayler and Dr. Rachid Lemrani
from the H1 Collaboration for discussions and explanations
concerning the experimental analyses.

~

I would like to thank Prof. Michel Fontannaz for providing me with the
fortran subroutines computing the AFG and AFG02 parton densities
in the photon and BFG fragmentation functions.

~

Many thanks to Dr. Pawe{\l} Jankowski for comments and discussions.

~

Finally, I am deeply grateful to my family, Zofia, Konrad and Marek
Zembrzuski, and to Gosia Rytelewska for their support and help.
\end{minipage}
\end{center}

\thispagestyle{empty}

\newpage
\setcounter{page}{1}
\tableofcontents

%%%%%%%%%%%%%%%%%%%%%%%%%%%%%%%%%%%%%%%%%%%%%%%%%%%%%%%%%%%%%%%%%%%%%%%%%%
\chapter{Introduction}\label{intro}
%%%%%%%%%%%%%%%%%%%%%%%%%%%%%%%%%%%%%%%%%%%%%%%%%%%%%%%%%%%%%%%%%%%%%%%%%%
\pagenumbering{arabic}

The Standard Model of particle physics is a successful
theory describing the elementary particles and their interactions.
It combines the SU(2)$_L\otimes$U(1)$_Y$
theory of electroweak interactions with the SU(3)
Quantum Chromodynamics (QCD) %\PP 
- the theory of strong interactions.
The strong interacting particles, {\sl hadrons} (e.g. protons),
are build from more elementary point-like objects, {\sl quarks} 
and {\sl gluons}. The quarks are charged and they interact 
also electromagnetically with other
charged particles, e.g. electrons, by exchange of the 
point-like {\sl photons}. The electromagnetic interactions
are described by the Quantum Electrodynamic (QED), being a part
of the unified SU(2)$_L\otimes$U(1)$_Y$ theory.

This work is devoted to the theoretical
study of the {\sl photoproduction} of the photon with a large
{\sl transverse momentum},
called the Deep Inelastic Compton  (DIC) process, 
in the electron-proton scatterings at the DESY HERA collider.
We %\PP  
calculate the cross sections for this process
with the beyond leading logarithmic accuracy within a framework
of the {\sl perturbative} QCD (see eg. \cite{Sterman:1994ce}).

The {\sl photoproduction} is a process in which a real or almost real
photon collides with another particle producing
some final state particles
(for recent reviews see~\cite{Klasen:2002xb,Krawczyk:2000mf,Nisius:2000cv}).
Currently, all high energy photon-proton collisions are realized
in the electron-proton scatterings.
They correspond predominantly to events with low 
virtuality of the exchanged photons, $Q^2\approx 0$.
%~~\footnote{The $Q^2$
%stands for the absolute value of the photon four-momentum squared.}. 
These quasi-real photons colliding with the proton
may lead for instance 
to a production of jets or particles with a large transverse 
momentum, $p_T$. If $p_T$ is much larger than the QCD scale,
$p_T\gg\Lambda_{QCD}$, then the perturbative QCD can be applied
to describe such a process.
%~\footnote{The $p_T$ and $\Lambda_{QCD}$
%stand for the transverse momentum and the QCD scale, respectively.}.

Processes in which the large transverse momentum 
of {\sl jets}, hadrons or photons is observed in a final state 
in the $e^+e^-$, hadron-hadron and electron-proton colliders
play an important role in testing QCD and in measuring 
the {\sl parton densities} or the {\sl parton fragmentation functions}
as well as the {\sl strong coupling constant} $\alpha_S$.
The process with the production of the photon is a special
one,  since the photon may couple 
directly to quarks involved in the {\sl hard} QCD process, so it 
may provide a relatively clear information about the QCD dynamics. 
%with only small uncertainties due to the {\sl long-distance non-perturbative}
%effects like the {\sl hadronization} or {\sl fragmentation}. 
Such photons arising
predominantly from the hard process are called {\sl prompt} photons.
Although the cross sections for the prompt photon production
are smaller than for the production of jets, these processes are considered
as an important source of complementary informations.

\begin{figure}[h] 
\vskip 5.2cm\relax\noindent\hskip 0cm
       \relax{\includegraphics{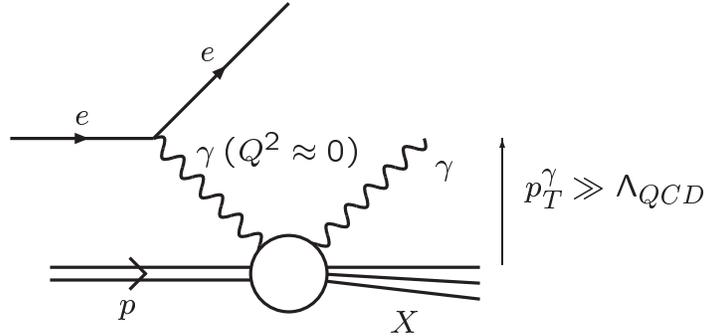}}
\vspace{-1cm}
\caption{\small\sl The $ep\ra e\gamma ~X$ photoproduction.}
\label{figcompton}
\end{figure}

\begin{figure}[h] 
\vskip 5.2cm\relax\noindent\hskip 0cm
       \relax{\includegraphics{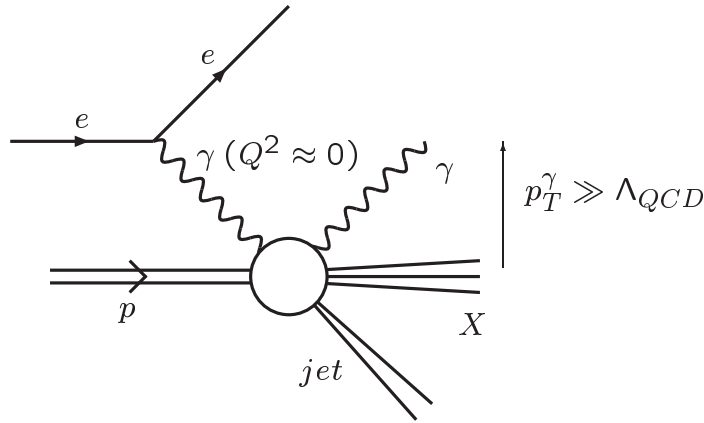}}
\vspace{0cm}
\caption{\small\sl The $ep\ra e\gamma ~jet ~X$ photoproduction.}
\label{figcomptonjet}
\end{figure}

The photoproduction of the prompt photon in the electron-proton scattering
is shown in Fig. \ref{figcompton}.
In such a reaction an almost real photon emitted from the electron
interacts with the proton leading to the production of large-$p_T$
photon and a hadronic system $X$: $ep\rightarrow e\gamma X$
\footnote{For a relatively small momentum transfer between the initial
and the final electron, $Q^2\approx 0$, 
the exchange of the photon dominates over 
the $Z$ boson exchange and we do not include the $Z$ boson in our
analysis. We also neglect the emission of the final
photon directly from the electron as it gives 
%\PP{\bf zawsze?}
a very small contribution to the cross section for the photoproduction
of large-$p_T$ photons \cite{ula}.}. 
Particles in the hadronic final state can
form a jet or jets which balance the photon transverse momentum.
If, beside the final photon, the jet is considered, 
we write: $ep\rightarrow e\gamma ~jet ~X$ 
(Fig.~\ref{figcomptonjet}), where $X$
stands for other hadronic final state.
In the work we study both types of processes: $ep\ra e\gamma ~X$
and $ep\ra e\gamma ~jet ~X$.

\begin{figure}[h] 
\vskip 2.5cm\relax\noindent\hskip -3.5cm
       \relax{\includegraphics{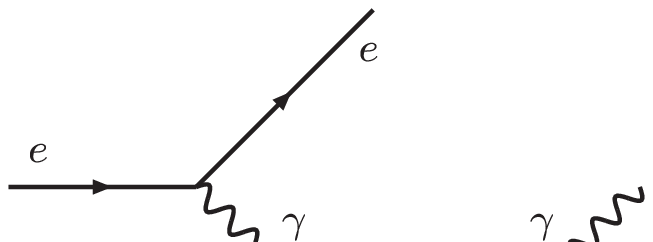}}
\vskip -0.55cm\relax\noindent\hskip 4cm
       \relax{\includegraphics{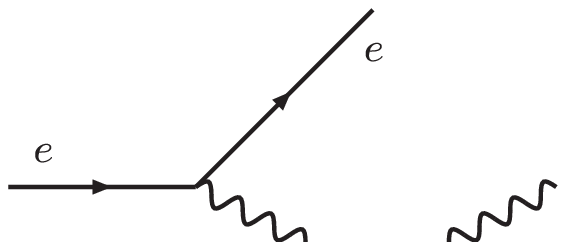}}
\vspace{3.5cm}
\caption{\small\sl The Born contributions to the $ep\ra e\gamma (jet)~X$
photoproduction.}
\label{figcomptonborn}
\end{figure}

In the lowest order of perturbative QCD the photon emitted by
the electron is scattered from the quark being a component of the proton
(the Born contribution), see  Fig.~\ref{figcomptonborn}.
The final state of the hard interaction between the photon
and the initial quark consists of the large-$p_T$
photon and the quark from which the jet arises.
In this lowest order process both the mediating and the final photon
interact directly with the quark. 

\begin{figure}[h]  
\vskip 5cm\relax\noindent\hskip -1cm
       \relax{\includegraphics{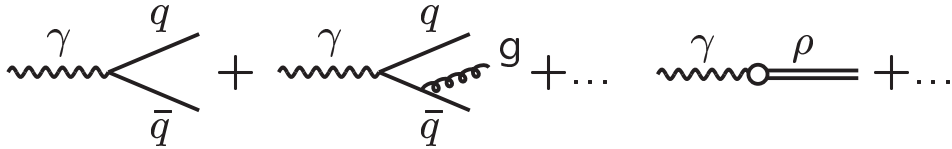}}
\vspace{-3cm}
\caption{\small\sl Various processes contributing to the 
hadron-like ``structure of the photon''.}
\label{phot-res}
\end{figure}

\begin{figure}[h] 
\vskip 4.8cm\relax\noindent\hskip 1.8cm
       \relax{\includegraphics{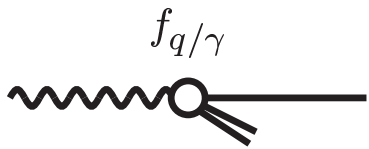}}
\vspace{-2.7cm}
\caption{\small\sl The resolved photon.}
\label{phot-str}
\end{figure}

However, there are also processes of a quite different nature.
At high energies the photon being a point-like {\sl gauge boson},
may interact like a hadron. 
This hadron-like properties of the photon
are due to fluctuations into a virtual pair quark-antiquark 
%\PP {\bf to nie sa dlugozasiegowe procesy dwa pierwsze z rys. 1.4!}  
or into a vector meson, see Fig. \ref{phot-res}.
Due to these fluctuations the photon
may behave as an object consisting of
quarks, antiquarks and gluons.
The interactions in which such a photon, called the {\sl resolved photon},  
exhibits its complex hadronic-like
``structure'' are effectively described using 
the formalism of the {\sl photon structure functions} 
\cite{Walsh:1973mz,DeWitt:1978wn} 
in full analogy to the corresponding formalism for the 
proton \cite{Bjorken:1969ja,Halzen:1984mc,Sterman:1994ce}.
The density of probability of ``finding''
the parton (quark or gluon) ``inside'' the photon is given in this
formalism by the parton density/distribution in the photon,
$f_{q(g)/\gamma}$. The process $\gamma\ra q + anything$
described by the function $f_{q/\gamma}$ is illustrated in Fig. \ref{phot-str}.
The parton densities are related to the {\sl structure functions},
as for the proton.
However, the photon structure functions, 
%\PP like 
e.g. $F_2^{\gamma}$,
in contrast to the corresponding proton structure functions,
are  calculable in the {\sl Parton Model}, based on the
$\gamma^*\gamma\ra q\bar{q}$ process. The $F_2^{\gamma}$
is proportional to the {\sl electromagnetic coupling constant}, 
$\alpha$, and it depends logarithmically on the energy scale of the process,
$\mu$, already in the {\sl Parton Model}
\footnote{In the case of the $\gamma^*\gamma\ra q\bar{q}$ process, 
the energy scale $\mu$ squared is equal to the virtuality, 
$Q^2$, of the virtual photon ($\gamma^*$) which probes the structure of the
real photon ($\gamma$).}:
\bea
F_2^{\gamma} \sim \alpha \ln {\mu^2\over \Lambda_{QCD}^2}+...
\eea
%\PP 
We stress that this logarithmic scaling violation in $F_2^{\gamma}$ 
in the Parton Model
arises from the purely electromagnetic coupling $\gamma\ra q\bar{q}$.
For more detailed discussion on the concept of the ``photon structure''
see e.g. \cite{Krawczyk:2000mf,Nisius:2000cv,Schienbein:2001cd}.

\begin{figure}[h] 
\vskip 4.8cm\relax\noindent\hskip 1.8cm
       \relax{\includegraphics{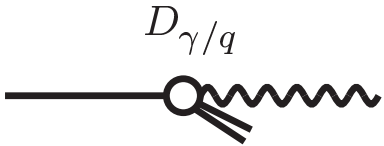}}
\vspace{-2.7cm}
\caption{\small\sl The quark-to-photon fragmentation.}
\label{phot-frag}
\end{figure}

So, taking into account the hadronic structure of the photon,
the photon mediating in the electron-proton scattering shown in Figs.
\ref{figcompton} and \ref{figcomptonjet} may interact with the 
constituents of the proton directly (as in Fig.~\ref{figcomptonborn})
or indirectly as the resolved photon. 
Similarly, the final photon can be produced directly in the 
hard interaction with the quark (as in Fig.~\ref{figcomptonborn})
or it may originate from {\sl fragmentation
processes}. The cross section of the latter processes involve 
relatively poorly known {\sl parton-to-photon fragmentation functions}
(see e.g. \cite{Gehrmann-DeRidder:1999yu} and references therein)
which describe the density of probability of the parton ``decay''
into the photon, see Fig. \ref{phot-frag}.
Fortunately, one can suppress this contribution introducing 
the {\sl isolation} of the final photon. 
The {\sl isolation} constraints, which separate the photon from
energetic hadrons, are introduced in both experimental and theoretical
analyses. We follow this approach %\PP 
in this work.

The standard experiments to extract the parton densities 
in the proton or photon are the Deep Inelastic Scatterings
in the $ep$ 
%(DIS$_{ep}$) 
or $e\gamma$ 
%(DIS$_{e\gamma}$)
collisions, where the structure functions $F_2^p$ and $F_2^{\gamma}$
are measured. From these structure functions the corresponding parton 
densities can be derived. The photoproduction processes, as the
prompt photon production considered herein, can provide
an additional information about the parton densities in both the proton
and the photon, see e.g. \cite{Klasen:2002xb,Krawczyk:2000mf}.
%\PP {\bf REFERENCJE?}

The production of the photons with large transverse momenta
in the $\gamma p$ scatterings,
and its importance for probing the proton structure,
was studied theoretically for the first time in 1969~\cite{Bjorken:1969ja}. 
The concept of the resolved photon was included
in the analyses of this process ten years latter~\cite{Tu:1979vg}.
Then, a few groups of authors studied the inclusive (i.e. non-isolated)
large-$p_T$ photon production in the $\gamma p$ or $ep$ scattering, 
and its importance
for testing QCD and constraining the 
parton densities in both the proton and photon
\cite{Iguchi:1980dm}-\cite{Gordon:1994sm}.
The next-to-leading order (NLO) QCD predictions for the 
photoproduction of isolated photons in the $ep$ process at the HERA collider
have been presented in the papers~\cite{Gordon:1995km,Krawczyk:1998it,
Fontannaz:2001ek,Krawczyk:2001tz}.
Finally, the photoproduction of the isolated photon and a jet 
at HERA has been calculated in NLO QCD in 
\cite{Gordon:1998yt}-\cite{Fontannaz:2004qv}.
%\cite{Gordon:1998yt,Fontannaz:2001nq,Zembrzuski:2003nu,Fontannaz:2003yn,
%Heinrich:2003vx,Fontannaz:2004qv}.
There is also a NLO calculation for the isolated photon
and the isolated photon plus a jet production in the deep inelastic
events ($Q^2\gg 1$ GeV$^2$)~\cite{Kramer:1998nb,Gehrmann-DeRidder:1999wy,
Gehrmann-DeRidder:1999yu,Gehrmann-DeRidder:2000ce},
which is in many theoretical aspects close to the calculations
for the corresponding photoproduction events.

The photoproduction of isolated photons without and with additional jets 
has been measured at the HERA collider by the 
ZEUS~\cite{Breitweg:1997pa}-\cite{Bussey2001} and 
H1~\cite{h11997}-\cite{unknown:2004uv}
Collaborations. 
All the NLO QCD predictions as well as the Monte Carlo
simulations for the isolated photon with \underline{no jet} requirement
tend to lie below the ZEUS data
for the photon rapidities $\eta^{\gamma}<0.1$ 
\cite{unknown:uj,Breitweg:1999su,SungWonLee:thesis,Bussey2001}, 
and below the final H1 data \cite{unknown:2004uv}
in the whole range of the photon rapidities~\footnote{The rapidity of
e.g. the photon is defined as $\eta^{\gamma}\equiv -\ln\tan (\theta/2)$,
where $\theta$ stands for the angle between the final photon momentum
and the momentum of the initial proton. 
The positive rapidity is pointed in the proton direction.}.
On the other hand the H1 data \cite{unknown:2004uv}
for the isolated photon \underline{plus jet} production
are somewhat better described by the QCD predictions.

In the paper~\cite{Chekanov:2001aq} the ZEUS Collaboration has 
implemented in Monte Carlo simulations for the prompt photon 
\underline{plus jet} production the intrinsic transverse 
momentum of partons in the proton. From 
%the comparison with 
a fit to the data 
it was found that the effective intrinsic transverse momentum
is very large 
%\PP from the point of view of standard QCD expectations
%\marginpar{inaczej}
with respect to the mass of the proton, namely
 $<k_T>=1.69\pm0.18^{+0.18}_{-0.20}$ GeV. 
In our work the intrinsic transverse momentum is
not included, since one can argue that the measured $<k_T>$ in the proton
describes effectively higher order emissions or multiple interactions
between particles involved in the process. 
Moreover, it was shown in \cite{Fontannaz:2001nq}
that the NLO QCD calculation is able to describe the ZEUS data
\cite{Chekanov:2001aq} with no need for the intrinsic transverse momentum. 

As it was mentioned above, there are some moderate discrepancies
between the QCD predictions and the data for the photoproduction
of the isolated photons in $ep$ scattering, especially when no jet 
in the final state is considered. Note, that some
differences between theory and data are also observed for 
other processes involving photons 
%\PP 
%in the initial or final state
%of the hard partonic processes 
in the hadron-hadron, $ep$ and $e^+e^-$ 
collisions, see e.g. 
\cite{Laenen:2000de,Acosta:2002ya,Abbiendi:2003kf,Chekanov:2004wr}.
It may indicate, that our understanding of such processes is not 
satisfactory yet, and further theoretical and experimental searches
are needed. This is one of reasons for our study.

%\PP 
This thesis contains a detailed theoretical analysis of the 
cross section for the photoproduction of the prompt photon and 
the prompt photon plus a jet including effects due to the 
structure of the photon and the fragmentation into photon.
In 
%\PP 
most of theoretical studies the parton densities in the photon, 
$f_{\gamma}$, and the parton-to-photon fragmentation functions, $D_{\gamma}$,
are treated as quantities of order 
$\mathcal{O}$$(\alpha/\alpha_S)$. In our calculations they are
 considered as quantities of order $\mathcal{O}$$(\alpha)$.
This leads to a different set of diagrams included in our
calculation \cite{Krawczyk:1998it,Krawczyk:2001tz,Zembrzuski:2003nu}
and in calculations of other authors
\cite{Gordon:1995km,Fontannaz:2001ek,Gordon:1998yt,Fontannaz:2001nq,
Fontannaz:2003yn,Heinrich:2003vx,Fontannaz:2004qv} for the photoproduction 
of the isolated photons at HERA.

We implement the isolation restrictions in the cross section using two
methods. First, we use the {\sl small cone approximation} method,
in which the isolation is implemented in an approximated way. This method
was previously used by other authors. However, not all 
expressions needed to calculate the higher order corrections 
in the small cone approximation exist in the literature, 
so we 
%\PP 
derive and present the missing formulae.
Next, we use the method, in which the phase space of produced particles is
divided into a few parts. This allows to implement the isolation, 
as well as other kinematic cuts, in
an exact way. We perform a comparison between both methods and show
that the small cone approximation is quite accurate and leads to 
reliable predictions for the prompt photon production at HERA.

The division (slicing) of the phase space is a standard
approach to calculate various cross sections, e.g. cross sections
for the isolated photon production. However, our method
differs in details from methods applied in other calculations
and allows to obtain relatively simple analytical formulae for the
cross sections including higher order corrections.

The formulae for the higher order corrections are different 
in each part of the divided phase space. Our method of the division
allows to apply in some parts of the phase space the 
%\PP 
known cross sections  for the non-isolated
photon production. In other parts of the phase space we derive
all needed analytical expressions for the higher order corrections
to the isolated photon production in the $ep$ collision. Some obtained
expressions are consistent with formulae presented previously
by other authors (for processes other than this considered herein).
We find it useful to present in this work 
in a compact analytical form all the 
expressions, which are necessary to obtain the cross section for 
the isolated photon production in the $ep$ scattering including
higher order corrections.

This thesis is based on the papers listed below
%\cite{Krawczyk:2000mf,Krawczyk:1998it,Krawczyk:2001tz,Zembrzuski:2003nu,
%Krawczyk:1997zv,Krawczyk:1999eq}, 
and it contains in addition some new results never published before:

\begin{itemize}
\item
M.~Krawczyk and A.~Zembrzuski,
``Probing the structure of virtual photon in the deep inelastic Compton
process at HERA,''
Phys.\ Rev.\ D {\bf 57} (1998) 10
[arXiv:hep-ph/9708274],
\item
M.~Krawczyk, A.~Zembrzuski and M.~Staszel,
``Survey of recent data on photon structure functions and resolved photon
processes,'' DESY 98-013,
arXiv:hep-ph/9806291,
\item
M.~Krawczyk and A.~Zembrzuski, in: A.~Astbury, D.~Axen, J.~Robinson (Eds.),
Proceedings of the 29th Int. Conference on High Energy Physics,
ICHEP'98, Vancouver, Canada, July 1998,
World Scientific, 1999, p.895,
``NLO prediction for the photoproduction of the isolated photon at HERA,''
arXiv:hep-ph/9810253,
\item
M.~Krawczyk and A.~Zembrzuski,
``The forward photon production and the gluonic content of the real and
virtual photon at the HERA collider,''
Nucl.\ Phys.\ Proc.\ Suppl.\  {\bf 82} (2000) 167
[arXiv:hep-ph/9912368],
\item
M.~Krawczyk, A.~Zembrzuski and M.~Staszel,
``Survey of present data on photon structure functions and resolved  photon processes,''
Phys.\ Rept.\  {\bf 345} (2001) 265
[arXiv:hep-ph/0011083].
\item
M.~Krawczyk and A.~Zembrzuski,
``Photoproduction of the isolated photon at DESY HERA in next-to-leading order QCD,''
Phys.\ Rev.\ D {\bf 64} (2001) 114017
[arXiv:hep-ph/0105166],
\item
A.~Zembrzuski and M.~Krawczyk,
``Photoproduction of isolated photon and jet at the DESY HERA,''
arXiv:hep-ph/0309308,
\end{itemize}

The
%\PP 
paper  
%work 
is organized as follows. First, in 
%\PP 
Chapter \ref{nlo} we discuss our choice of diagrams and
present some calculation details.
The formula for the cross section 
and our results for the inclusive (non-isolated) photon production
are briefly 
%\PP 
discussed in Chapter \ref{non}.
In Chapter~\ref{small} the isolation cuts are defined and the
calculation for the isolated photon in the small cone approximation
is presented;  we study the
influence of the isolation cut on the production rate of the photon and  
the role of other cuts applied in experiments.
Then, in Chapter~\ref{isol} 
the division (slicing) of the three body phase space
is discussed and we present our analytical results. 
%\PP 
Here, the numerical predictions 
for the isolated photon production with the exact implementation of 
the isolation cuts are also presented and compared with the 
previous approximated ones.
The results for the $ep\ra e\gamma ~jet ~X$ process
are presented in Chapter~\ref{jet}.
We discuss theoretical uncertainties and
compare our predictions with predictions of other 
NLO QCD calculations, as well as with existing data for the
photoproduction of the isolated photon or the isolated photon
plus a jet at HERA (Chapters \ref{small}, \ref{isol}, \ref{jet}).
In Chapter \ref{Sglu} we briefly discuss the sensitivity
of the leading order (LO) cross section to the gluon distribution
in the photon taking into account a non-zero virtuality of the
exchanged photon, $Q^2\ne 0$.
The summary is given in Chapter~\ref{Ssum}.
Finally, the Appendices contain the formulae for the cross sections
for all the processes included in our analysis. Some of these formulae
are taken from the literature and the other are briefly derived.
%\PP{\LARGE\bf gdzie o Appedixie??}

%%%%%%%%%%%%%%%%%%%%%%%%%%%%%%%%%%%%%%%%%%%%%%%%%%%%%%%%%%%%%%%%%%%%%%%%%%
\chapter{The Deep Inelastic Compton process in NLO QCD}\label{nlo}
%%%%%%%%%%%%%%%%%%%%%%%%%%%%%%%%%%%%%%%%%%%%%%%%%%%%%%%%%%%%%%%%%%%%%%%%%%

Our calculation bases on the standard leading twist perturbative QCD 
description of hard hadronic processes.
This means that we consider such diagrams where
one active parton from each initial particle
is involved in the hard process. However, in some Monte Carlo
simulations performed 
%\PP 
by experimental groups recently (2004) 
a large effect of multiple interactions was found 
for the prompt photon production at HERA. This effect is 
included in experimental analyses by introducing some models.
We present 
%\PP 
predictions based on our calculations
and ``corrected'' by an experimental group
for the multiple interactions as well as hadronization processes
in Chapters \ref{isol} and \ref{jet}.

There are several calculations
for the photoproduction of prompt (isolated) photons at the HERA collider,
namely: the calculation of Gordon and Vogelsang (GV)
\cite{Gordon:1995km}, Gordon (LG) \cite{Gordon:1998yt},
Krawczyk and Zembrzuski (K\&Z)
\cite{Krawczyk:1998it,Krawczyk:2001tz,Zembrzuski:2003nu}
and Fontannaz, Guillet and Heinrich (FGH)
\cite{Fontannaz:2001ek,Fontannaz:2001nq}.
%,Fontannaz:2003yn,Heinrich:2003vx,Fontannaz:2004qv}.
All these calculations differ from one another by set of diagrams 
included in next-to-leading order (NLO).
Below we discuss our choice of diagrams (Sec. \ref{Snlo:diagrams}), 
as well as some calculation details (Secs. \ref{det}, \ref{epa}).

%%%%%%%%%%%%%%%%%%%%%%%%%%%%%%%%%%%%%%%%%%%%%%%%%%%%%%%%%%%%%%%%%%%%%%%%%%
\section{Contributing processes}\label{Snlo:diagrams}
%%%%%%%%%%%%%%%%%%%%%%%%%%%%%%%%%%%%%%%%%%%%%%%%%%%%%%%%%%%%%%%%%%%%%%%%%%

%%%%%%%%%%%%%%%%%%%%%%%%%%%%%%%%%%%%%%%%%%%%%%%%%%%%%%%%%%%%%%%%%%%%%%%%%%
\subsection{Born process and $\mathcal{O}$$(\alpha_S)$ corrections,
resolved $\gamma$ {\underline{or}} fragmentation into $\gamma$}\label{CnloB}
%%%%%%%%%%%%%%%%%%%%%%%%%%%%%%%%%%%%%%%%%%%%%%%%%%%%%%%%%%%%%%%%%%%%%%%%%%

\begin{figure}[b]
\vskip 3cm\relax\noindent\hskip 0cm
       \relax{\includegraphics{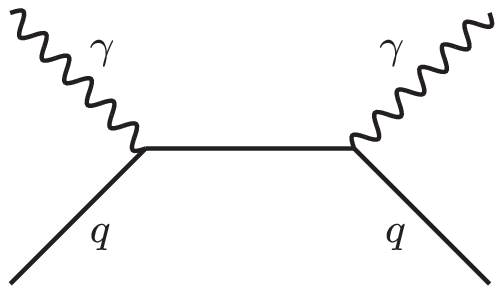}}
\begin{picture}(0,0)
\put(300,40){+ crossed diagram}
\end{picture}
\vspace{0cm}
\caption{\small\sl The Born process (the Compton scattering on the quark).}
\label{figborn}
%\end{figure}
%\begin{figure}[ht]
\vskip 3.3cm\relax\noindent\hskip -0.2cm
       \relax{\includegraphics{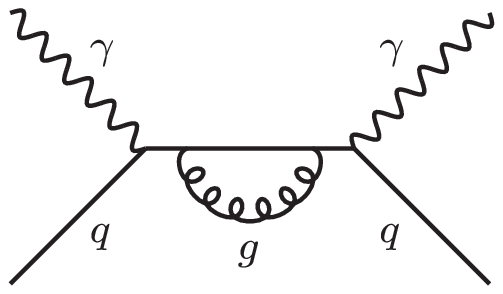}}
\vspace{0cm}
\caption{\small\sl An example of the virtual gluon
corrections to the Born process.}
\label{figvirt}
%\end{figure}
%\begin{figure}[ht]
\vskip 3.3cm\relax\noindent\hskip -0.2cm
       \relax{\includegraphics{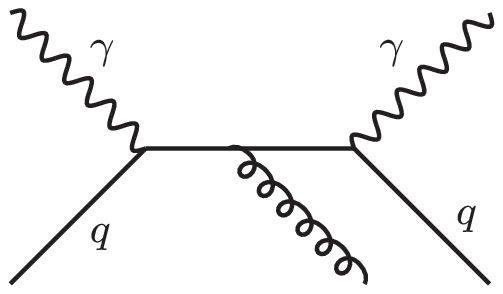}}
\vspace{0cm}
\caption{\small\sl An example of the real gluon 
corrections to the Born process,
$\gamma q\ra\gamma qg$.}
\label{figreal}
%\end{figure}
%\begin{figure}[ht]
\vskip 3.5cm\relax\noindent\hskip 0.7cm
       \relax{\includegraphics{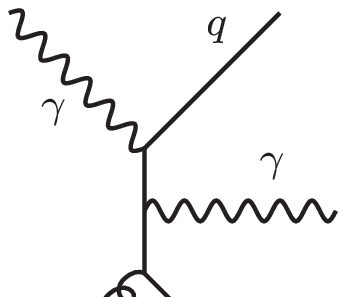}}
\vspace{1.cm}
\caption{\small\sl An example diagram for 
the process $\gamma g\ra\gamma q\bar{q}$.}
\label{fig23}
\end{figure}

We wish to study the hard electron-proton scattering 
leading to a production of a photon or a photon plus a jet
and anything else, $ep\ra e\gamma ~(jet) ~X$ 
(Figs. \ref{figcompton}-\ref{figcomptonborn}).
In this reaction the mediating photon arising from the electron
interacts with a partonic constituent (quark or gluon) inside the 
proton \cite{Bjorken:1969ja}. 
In the lowest order (Born) process, $\gamma q\ra\gamma q$, 
the photon is scattered from a quark yielding the final photon and the quark, 
see Fig.~\ref{figborn}.
This subprocess is of order $\mathcal{O}$$(\alpha^2)$ and has
a pure electromagnetic nature. To obtain the NLO QCD predictions,
the corrections of order $\mathcal{O}$$(\alpha_S)$ to the Born process
have to be taken into account. These corrections lead to partonic processes 
of order $\mathcal{O}$$(\alpha^2\alpha_S)$. They
include the virtual gluon exchange (Fig. \ref{figvirt}), the real gluon
emission (Fig. \ref{figreal}) and the process $\gamma g\ra\gamma q\bar{q}$
(Fig. \ref{fig23})~\cite{Duke:1982bj,mkcorr,Aurenche:1984hc,jan}. 

The corrections due to the virtual gluon
exchange and the real gluon emission contain {\sl infrared} singularities which
cancel when both contributions are added up properly in the scattering
amplitude squared (see Secs. \ref{Snon:x},
\ref{Sapprox}, \ref{pss}).
The contribution due to the processes $\gamma q\ra\gamma qg$ 
and $\gamma g\ra\gamma q\bar{q}$ contains another type of singularities,
so called {\sl mass} or {\sl collinear} singularities, which do not
cancel. In order to remove them from the cross section the
{\sl factorization} procedure is applied: the singularities are subtracted
and shifted into corresponding parton densities in the proton or
photon or into parton-to-photon fragmentation functions
\cite{Aurenche:1984hc,jan}. 
At this stage the {\sl factorization} scale appears. 
%As a consequence of the factorization procedure,
The {\sl bare} ({\sl scale invariant}) parton densities in the proton are 
replaced by renormalized {\sl scale dependent} densities. Moreover,
the {\sl scale dependent} parton densities in the photon and 
the {\sl scale dependent} fragmentation functions appear in the
calculation. They are necessary ingredients in the NLO calculation
to absorb the 
%\PP 
%unphysical 
mass singularities.

\begin{figure}[b]
\vskip 5cm\relax\noindent\hskip -2cm
       \relax{\includegraphics{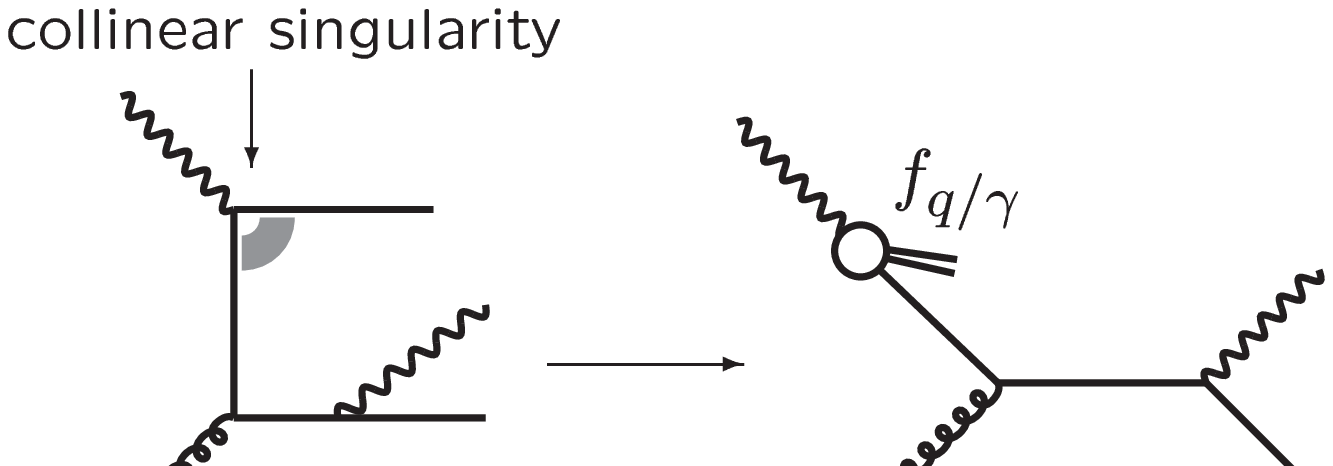}}
\vspace{1.5cm}
\caption{\small\sl A sample of diagrams illustrating the factorization
of the collinear singularities from the process $\gamma g\ra\gamma q\bar{q}$
into the parton density in the photon, $f_{q/\gamma}$.}
\label{fact-phot}
%\end{figure}
%\begin{figure}[ht]
\vskip 4.5cm\relax\noindent\hskip -2cm
       \relax{\includegraphics{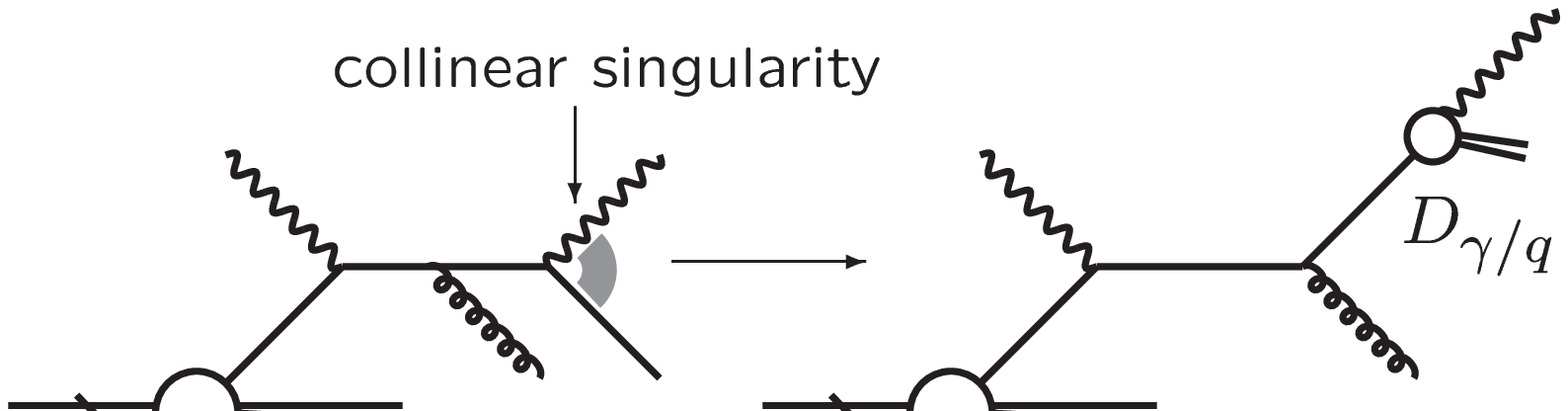}}
\vspace{0.7cm}
\caption{\small\sl A sample of diagrams illustrating the factorization
of the collinear singularities from the process $\gamma q\ra\gamma qg$
into the fragmentation function, $D_{\gamma/q}$.}
\label{fact-frag}
\end{figure}

The factorization procedure is illustrated in Figs. \ref{fact-phot} 
and \ref{fact-frag}: the singularities due to the collinear configurations
in the vertexes $\gamma\ra q\bar{q}$ and $q\ra\gamma q$ are shifted
and absorbed by the corresponding quark density in the photon ($f_{q/\gamma}$)
or by the quark-to-photon fragmentation function ($D_{\gamma/q}$), respectively
(for precise definitions and details see \cite{Aurenche:1984hc,jan}).

It is worth mentioning 
that there was a discussion whether
for the photoproduction of isolated photons in $e^+e^-$ collisions
the conventional factorization breaks down, and whether the cross section
is an infrared safe quantity  
\cite{Berger:1995cc,Aurenche:1996ng,Catani:1998yh}.
In principle these questions could as well  
occur for the photoproduction of isolated photons in $ep$ collisions.
However we do not deal with this problem
because it arises from $2\rightarrow 3$ processes 
involving the parton-to-photon fragmentation, which are absent
in our calculation.
We checked this explicitly and found that all the singularities in
our calculations for the isolated photon production 
are canceled or factorized, as in the case of non-isolated photon
production, and the cross sections are well defined 
\cite{Krawczyk:1998it,Krawczyk:2001tz,Zembrzuski:2003nu} (see 
also~\cite{Gordon:1995km,Fontannaz:2001ek,Gordon:1998yt,Fontannaz:2001nq},
\cite{Fontannaz:2003yn}-\cite{Gehrmann-DeRidder:2000ce}).

\begin{figure}[h]
\vskip 4cm\relax\noindent\hskip -2cm
       \relax{\includegraphics{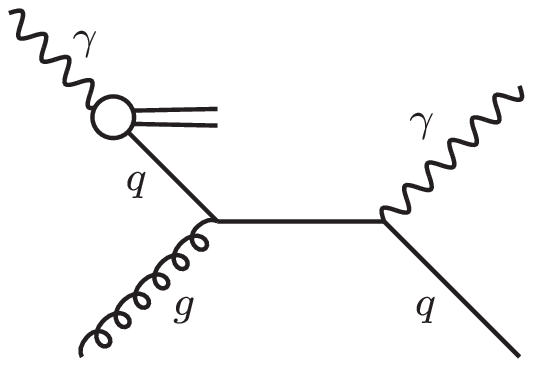}}
\vskip 1cm\relax\noindent\hskip 4cm
       \relax{\includegraphics{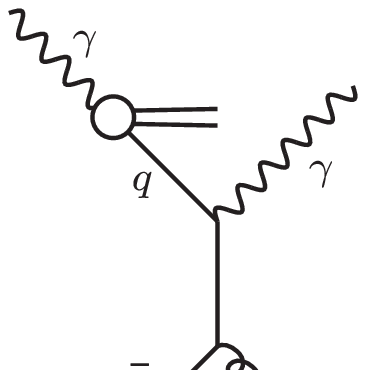}}
\vskip 2cm\relax\noindent\hskip -2cm
       \relax{\includegraphics{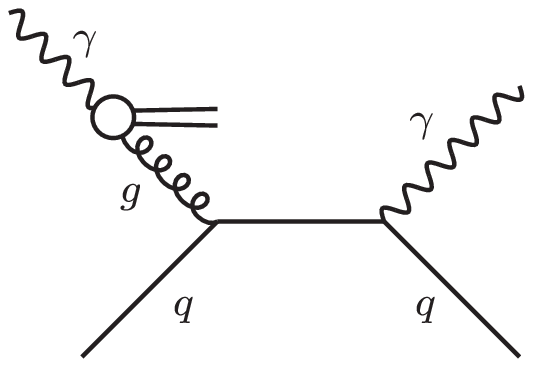}}
\begin{picture}(0,0)
\put(340,10){+ crossed diagrams}
\end{picture}
\vspace{0cm}
\caption{\small\sl The processes with the resolved initial photon.}
\label{figsingi} 
\end{figure}

\begin{figure}[ht]
\vskip 4cm\relax\noindent\hskip -2.5cm
       \relax{\includegraphics{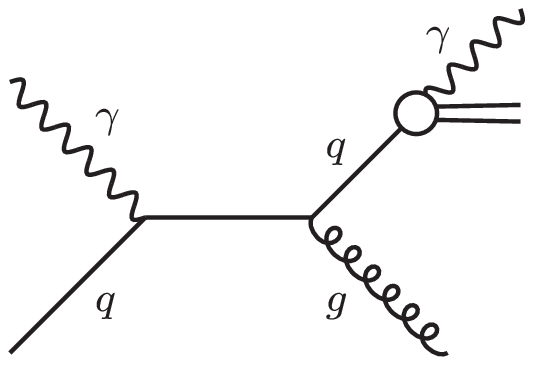}}
\vskip 1cm\relax\noindent\hskip 3.5cm
       \relax{\includegraphics{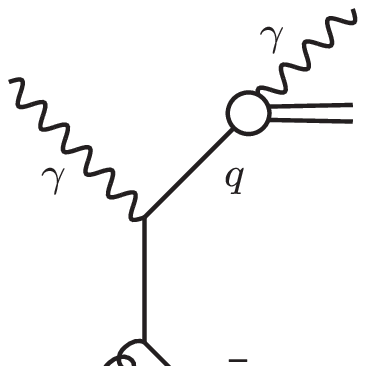}}
\vskip 2cm\relax\noindent\hskip -2.5cm
       \relax{\includegraphics{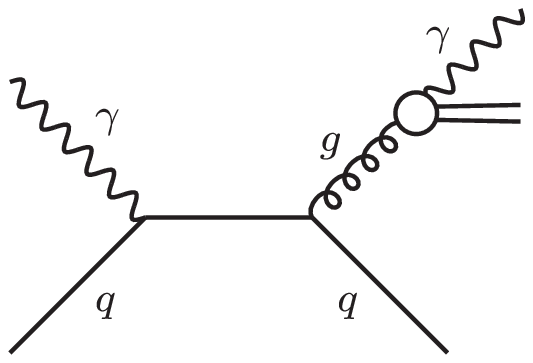}}
\begin{picture}(0,0)
\put(340,10){+ crossed diagrams}
\end{picture}
\vspace{0cm}
\caption{\small\sl The processes with the parton-to-photon
fragmentation.}
\label{figsingf}
\end{figure}

The processes involving the parton densities in the {\sl resolved}
initial photon are shown in Fig.
\ref{figsingi}. In these processes a parton (quark or gluon) from the photon 
interacts with a parton arising from the proton yielding the {\sl direct}
final photon and a parton. Fig. \ref{figsingf} shows the processes
with the fragmentation of a parton into the final photon. Here 
the {\sl direct} initial photon interacts with a parton from the proton
leading to a production of two partons; 
one of these partons produces the final photon
in the 
%\PP 
%long-distance 
{\sl fragmentation} process. The name {\sl direct} photons
stand for the photons participating directly in the hard partonic
process.


%%%%%%%%%%%%%%%%%%%%%%%%%%%%%%%%%%%%%%%%%%%%%%%%%%%%%%%%%%%%%%%%%%%%%%%%%%
\subsection{Counting of orders. Alternative set of diagrams}\label{Cnlo:disc}
%%%%%%%%%%%%%%%%%%%%%%%%%%%%%%%%%%%%%%%%%%%%%%%%%%%%%%%%%%%%%%%%%%%%%%%%%%

The hard partonic processes in Figs. \ref{figsingi} and \ref{figsingf}
are of order $\mathcal{O}$$(\alpha\alpha_S)$. They are convoluted
with the corresponding parton densities or fragmentation functions
which are proportional to the electromagnetic coupling constant $\alpha$.
These distributions contain the logarithmic dependence
on the factorization/renormalization scale $\mu$: 
$f_{\gamma}\sim\alpha\ln\mu^2/\Lambda_{QCD}^2$ and
$D_{\gamma}\sim\alpha\ln\mu^2/\Lambda_{QCD}^2$.
The logarithm behaves like the inverse
of the strong coupling, $\ln\mu^2/\Lambda_{QCD}^2\sim 1/\alpha_S$ and
many authors (e.g. 
\cite{Gordon:1995km,Fontannaz:2001ek,Gordon:1998yt,Fontannaz:2001nq,
Fontannaz:2003yn,Heinrich:2003vx,Fontannaz:2004qv})
treat $f_{\gamma}$ and $D_{\gamma}$ as being of order 
$\mathcal{O}$$(\alpha/\alpha_S)$. This conclusion arises
from the {\sl evolution equation} and the  
{\sl renormalization group equation}, as it shown in \cite{Vogt:1999mu}.
If so, then the processes shown in Figs. \ref{figsingi}, \ref{figsingf}
are of order $\mathcal{O}$$(\alpha\alpha_S\cdot\alpha/\alpha_S)$=
$\mathcal{O}$$(\alpha^2)$, i.e. identical with the order of the Born process
(Fig. \ref{figborn}).
Note however, that the logarithmic behavior of the parton densities
in the photon or the parton-to-photon fragmentation functions
arises from the pure electromagnetic coupling $\gamma\ra q\bar{q}$
and $q\ra\gamma q$, respectively \cite{Walsh:1973mz}, 
%\PP 
as already discussed in Introduction. 
From this point of view the considered distributions
should be rather treated as the quantities of order
$\mathcal{O}$$(\alpha)$ since the logarithm $\ln\mu^2/\Lambda_{QCD}^2$
coming from the electromagnetic interaction is not related to the
strong coupling constant,
see also \cite{Krawczyk:1990nq,Krawczyk:1998it,Krawczyk:2001tz} 
and for more detailed discussion in \cite{Chyla:1999mw}.
Such a counting, $f_{\gamma}$ and $D_{\gamma}\sim$
$\mathcal{O}$$(\alpha)$, leads to an assignment
of the $\mathcal{O}$$(\alpha^2\alpha_S)$ order to the processes
shown in Figs. \ref{figsingi}, \ref{figsingf}.

The different counting of orders in the strong coupling leads
to the different sets of processes included in our NLO calculation
for the prompt photon production at HERA
\cite{Krawczyk:1998it,Krawczyk:2001tz,Zembrzuski:2003nu} 
in comparison with other NLO calculations.
They include in addition (beside the $\mathcal{O}$$(\alpha_S)$ corrections to
the Born cross section)
the $\mathcal{O}$$(\alpha_S)$ corrections to the processes
with the resolved initial photon \underline{or}
parton-to-photon fragmentation
\cite{Gordon:1995km,Fontannaz:2001ek,Gordon:1998yt,Fontannaz:2001nq},
and the $\mathcal{O}$$(\alpha_S)$ corrections to the processes
with both the resolved initial photon \underline{and} 
parton-to-photon fragmentation
\cite{Gordon:1995km,Fontannaz:2001ek,Fontannaz:2001nq}~\footnote{All these
QCD corrections have been calculated (most of them twice) in 
\cite{AurenchePLB140}-\cite{Gordon:1994wu}}.
These corrections are not included in our NLO calculation
since in our approach they should be taken into account
together with the next-to-next-to-leading (NNLO) order corrections to
the Born process.

It should be emphasized that the $\mathcal{O}$$(\alpha_S)$ corrections
to the Born process contain (after the factorization procedure)
an explicit logarithmic dependence
on the scale $\mu$, $\ln s/\mu^2$. This dependence is
compensated by $\ln\mu^2/\Lambda_{QCD}^2$ from $f_{\gamma}$ and
$D_{\gamma}$ if the $\mathcal{O}$$(\alpha_S)$ corrections to the
Born process are
included in the cross section consistently with the corresponding 
contributions involving parton densities in the photon or parton-to-photon
fragmentation functions. Similarly, the 
$\mathcal{O}$$(\alpha_S)$ corrections to the processes with the
resolved initial photon or the fragmentation into the final photon
should be, from our point of view, 
taken into account \underline{together} with the
$\mathcal{O}$$(\alpha_S^2)$ corrections to the Born process
in order to compensate the dependence on the $\mu^2$ in the hadronic 
cross section.

One can argue that, since the other authors 
\cite{Gordon:1995km,Fontannaz:2001ek,Gordon:1998yt,Fontannaz:2001nq}
include more diagrams, their calculation
is more accurate. It is true that including more diagrams usually
improves the quality of QCD predictions 
but it is not obvious that it is always the case 
%\PP
(see also the discussion in Secs. \ref{results3} and \ref{results5}).

It is worth mentioning that there are also other authors who
treat the processes involving photons in a similar way as we do.
The authors of \cite{Glover:1993xc} 
(see also \cite{Glover:1992sf,Glover:1994th})
present the NLO calculation for the prompt photon production
in the $e^+e^-$ collision with the quark-to-photon fragmentation. 
They include the fragmentation
function at the same order as the $\mathcal{O}$$(\alpha_S)$ corrections,
since the fragmentation is in a close relation with the corresponding
collinear configuration in $\mathcal{O}$$(\alpha_S)$ corrections
(compare Fig. \ref{fact-frag}). It allows to cancel
an explicit dependence on the scale $\mu^2$.
A similar NLO calculation for the prompt photon production
in the deep inelastic $ep$ events (with $Q^2> 10$ GeV$^2$)
at the HERA collider is presented in 
\cite{Gehrmann-DeRidder:2000ce,Gehrmann-DeRidder:1999wy,
Gehrmann-DeRidder:1999yu}. 
In this calculation the 
diagrams shown in Figs. \ref{figborn}-\ref{fig23} and
\ref{figsingf} are included without the $\mathcal{O}$$(\alpha_S)$ corrections
to the processes involving the fragmentation
(the contributions due to the resolved initial photon are negligible
for large $Q^2$). The cancellation (to a large extent) of the
$\mu^2$ dependence is therein also stressed.

Another kind of arguments is presented in \cite{Kunszt:1992ab}, where 
the NLO calculation for the prompt photon production in the 
$e^+e^-$ collision is investigated. The authors of \cite{Kunszt:1992ab}
claim that the 
parton-to-photon fragmentation, despite being of order
$\mathcal{O}$$(\alpha/\alpha_S)$ for the non-isolated photon production, 
should be counted as the quantity of order
$\mathcal{O}$$(\alpha)$ for the isolated
final photon, since the isolation itself is a correction
of order $\mathcal{O}$$(\alpha_S)$.

Counting $f_{\gamma}$ and $D_{\gamma}$ as
being of order $\mathcal{O}$$(\alpha/\alpha_S)$ allows for a self-consistent
expansion of physical quantities (structure functions, cross sections)
in powers of $1/\ln\mu^2/\Lambda_{QCD}^2$ (with 
$\alpha_S\sim 1/\ln\mu^2/\Lambda_{QCD}^2+...$) \cite{Vogt:1999mu}.
On the other hand, our approach,
where $f_{\gamma}$, $D_{\gamma}\sim$$\mathcal{O}$$(\alpha)$,
allows for a consistent counting of powers of $\alpha_S$
(since the logarithm $\ln\mu^2/\Lambda_{QCD}^2$ in $f_{\gamma}$ and 
$D_{\gamma}$ it is not $1/\alpha_S$,
see also \cite{Krawczyk:1990nq,Chyla:1999mw}).

The photon is a very special particle: being the point-like object 
it sometimes exhibits hadronic-like ``structure''. 
This double nature of the photon leads to more complicated description
within the QCD than for purely hadronic processes, 
and further study is needed to clarify what is the 
proper organization of the QCD perturbative 
series for processes involving photons. 
%\PP 
This should include among others also the calculation of the
$\mathcal O$$(\alpha_s^2)$ corrections to the Born process.

%%%%%%%%%%%%%%%%%%%%%%%%%%%%%%%%%%%%%%%%%%%%%%%%%%%%%%%%%%%%%%%%%%%%%%%%%%
\subsection{Full set of diagrams included in analysis.
Resolved $\gamma$ {\underline{and}} fragmentation into $\gamma$,
box diagram}\label{Cnlo:other}
%%%%%%%%%%%%%%%%%%%%%%%%%%%%%%%%%%%%%%%%%%%%%%%%%%%%%%%%%%%%%%%%%%%%%%%%%%
%\PP 

Besides the diagrams discussed in Sec. \ref{CnloB}, we include 
in our numerical analysis also other diagrams, 
%\PP these are 
namely the diagrams with the 
resolved $\gamma$ {\underline{and}} the
fragmentation into $\gamma$ and the box diagram. 
In Figs. \ref{figdoub}, \ref{figdoub2} the processes with the resolved 
initial photon \underline{and} 
the fragmentation into the final photon are shown.
\begin{figure}[t]
\vskip 4cm\relax\noindent\hskip -3.5cm
       \relax{\includegraphics{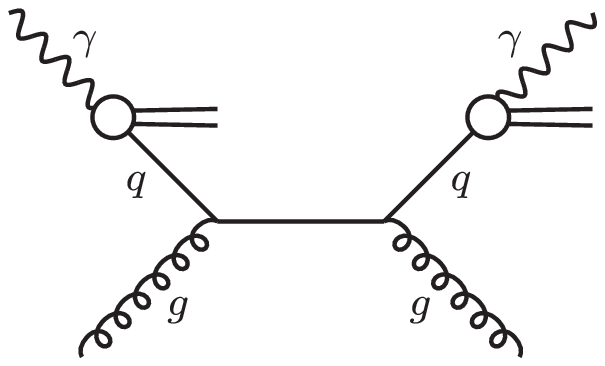}}
\vskip -0.5cm\relax\noindent\hskip 3cm
       \relax{\includegraphics{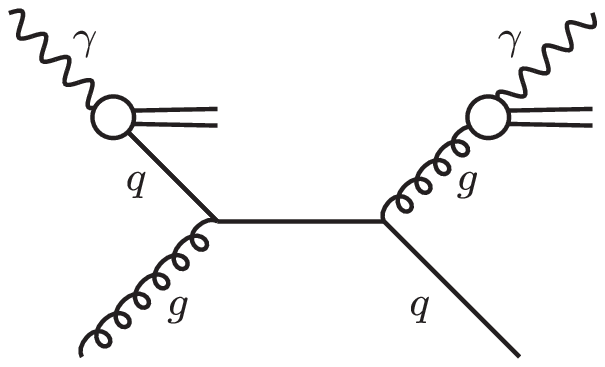}}
\vskip 3.5cm\relax\noindent\hskip -3.5cm
       \relax{\includegraphics{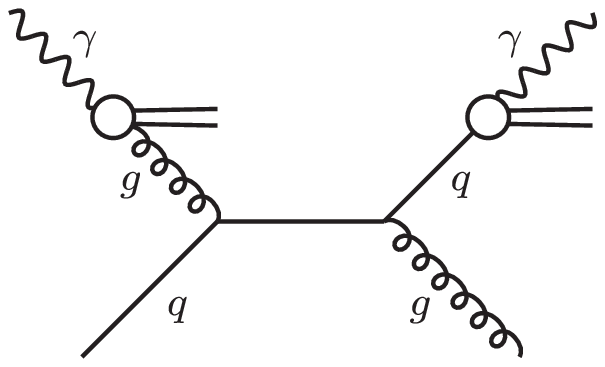}}
\vskip -0.5cm\relax\noindent\hskip 3cm
       \relax{\includegraphics{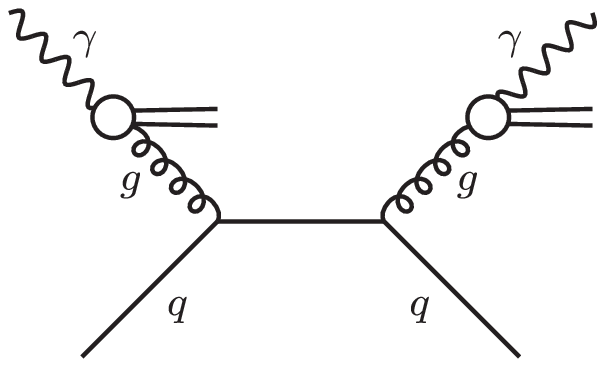}}
\vskip 3.5cm\relax\noindent\hskip -3.5cm
       \relax{\includegraphics{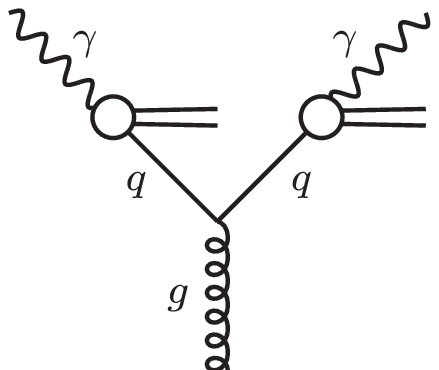}}
\vskip -1.cm\relax\noindent\hskip 3cm
       \relax{\includegraphics{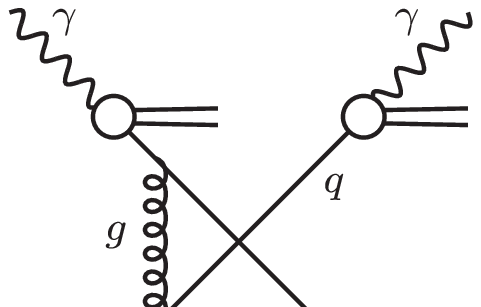}}
\vskip 4cm\relax\noindent\hskip 0.5cm
       \relax{\includegraphics{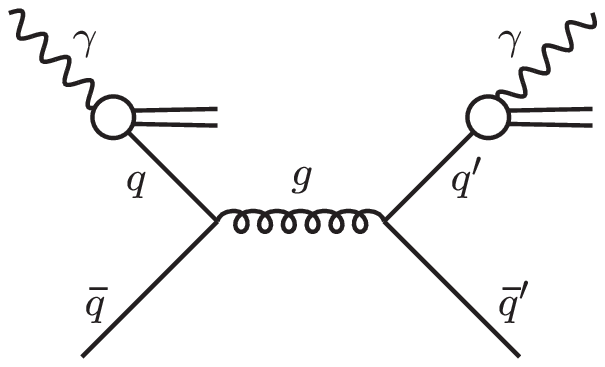}}
\vskip 3.5cm\relax\noindent\hskip -3.5cm
       \relax{\includegraphics{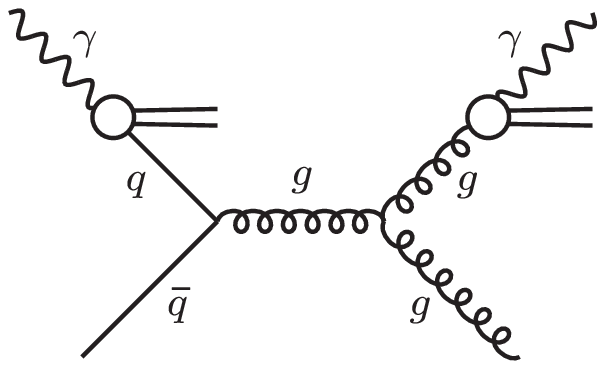}}
\vskip -0.5cm\relax\noindent\hskip 3cm
       \relax{\includegraphics{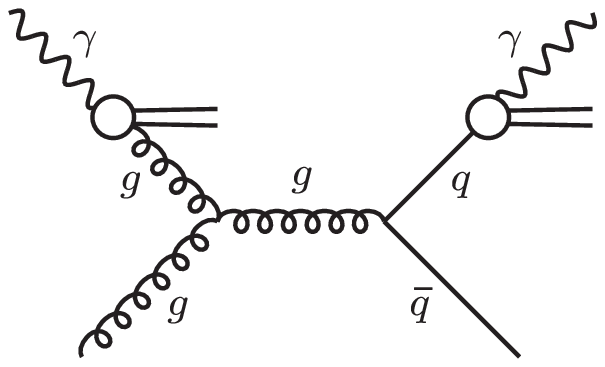}}
\vspace{0cm}
\caption{\small\sl The processes involving the resolved initial photon
and the parton-to-photon fragmentation.}
\label{figdoub}
\end{figure}
\begin{figure}[t]
\vskip 4cm\relax\noindent\hskip -4.5cm
       \relax{\includegraphics{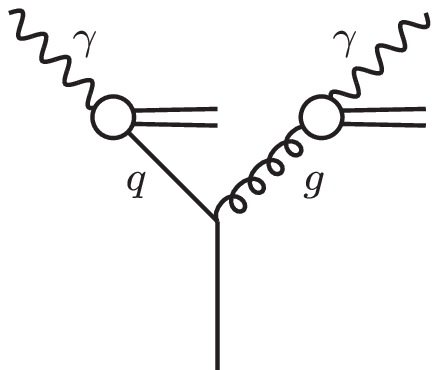}}
\vskip -0.5cm\relax\noindent\hskip 0.4cm
       \relax{\includegraphics{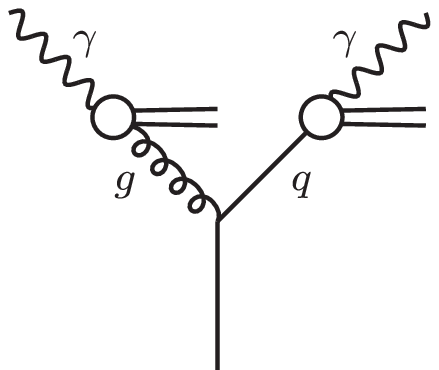}}
\vskip -0.5cm\relax\noindent\hskip 5.4cm
       \relax{\includegraphics{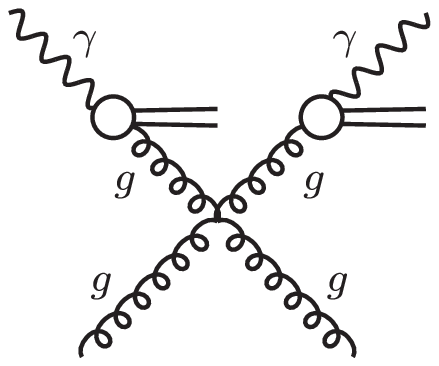}}
\vskip 5cm\relax\noindent\hskip -1.cm
       \relax{\includegraphics{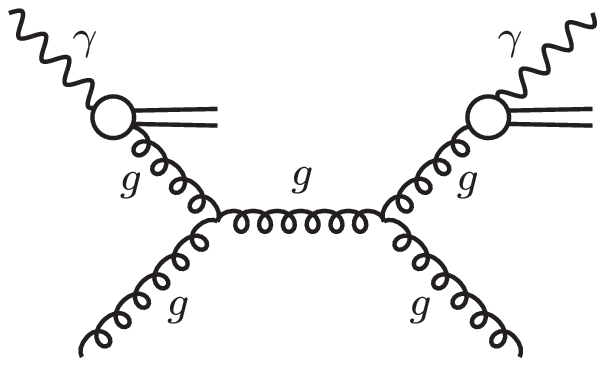}}
\vskip -2cm\relax\noindent\hskip 5.4cm
       \relax{\includegraphics{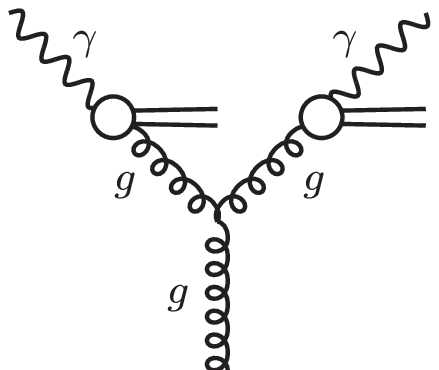}}
\vspace{1.5cm}
\caption{\small\sl Continuation of Fig.~\ref{figdoub}.}
\label{figdoub2}
\end{figure}
If we take $f_{\gamma}$, $D_{\gamma}\sim$$\mathcal{O}$$(\alpha)$,
all these processes are of order $\mathcal{O}$$(\alpha^2\alpha_S^2)$
and strictly speaking
go beyond the NLO accuracy of our calculations. Nevertheless
we include them, since they were taken into account in most of
existing NLO calculations for the (non-isolated or isolated)
prompt photon production at HERA and are found to be important
\cite{Duke:1982bj,Aurenche:1984hc},
\cite{Krawczyk:1990nq}-\cite{Aurenche:1992sb},
\cite{Gordon:1994sm}-\cite{Fontannaz:2003yn}.

\begin{figure}[h]
\vskip 5.3cm\relax\noindent\hskip 0.2cm
       \relax{\includegraphics{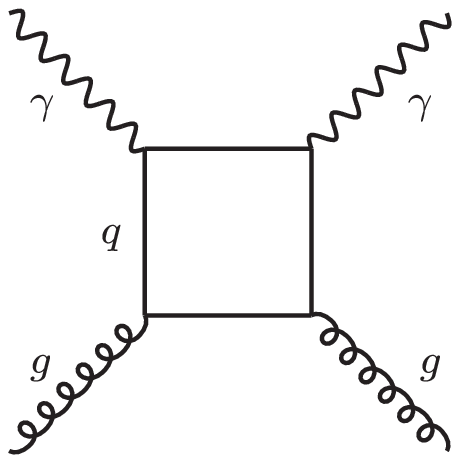}}
\begin{picture}(0,0)
\put(290,65){+ crossed diagrams}
\end{picture}
\vspace{0cm}
\caption{\small\sl The box diagram.}
\label{figbox}
\end{figure}

Finally, we take into account the photon-by-gluon scattering,
so called box process, 
$\gamma g\ra\gamma g$, shown in Fig. \ref{figbox} \cite{Combridge:1980sx}.
The box process is also of order $\mathcal{O}$$(\alpha^2\alpha_S^2)$,
i.e. beyond the accuracy of our calculation. 
However it is a very special process as it is 
the lowest order contribution to the photon-by-gluon
scattering, it contains no
singularities, and no parton densities in the photon or fragmentation
functions are involved. We think that including the box process
does not introduce any additional ambiguities  in the
summation of partonic cross sections. The box contribution
to the reaction $ep\ra e\gamma X$ or $\gamma p\ra\gamma X$ was
considered in previous calculations and it is known to be large
\cite{Fontannaz:1982et,Aurenche:1984hc},
\cite{Bawa:1988qs}-\cite{Aurenche:1992sb}, \cite{Gordon:1994sm}.
It is included in the calculations for the isolated photon
production at HERA 
by authors K\&Z \cite{Krawczyk:1998it,Krawczyk:2001tz,Zembrzuski:2003nu}
and FGH \cite{Fontannaz:2001ek,Fontannaz:2001nq}
but it is omitted by GV \cite{Gordon:1995km}
and LG \cite{Gordon:1998yt}.

To summarize, in our NLO calculation we take into account the
following contributions:
\begin{itemize}
\item the Born contribution (Fig. \ref{figborn}),
\item the $\mathcal{O}$$(\alpha_S)$ corrections to the Born diagram,
%\PP 
including the  $\gamma g\ra \gamma q \bar{q}$ process 
(Figs. \ref{figvirt}-\ref{fig23}), 
\item the processes with the resolved initial photon (Fig. \ref{figsingi})
\underline{or} with the fragmentation into the final
photon (Fig. \ref{figsingf}),
\item the processes with the resolved initial photon \underline{and} 
the fragmentation into the final photon (Figs. \ref{figdoub}, \ref{figdoub2}),
\item the box process (Fig. \ref{figbox}).
\end{itemize}

%%%%%%%%%%%%%%%%%%%%%%%%%%%%%%%%%%%%%%%%%%%%%%%%%%%%%%%%%%%%%%%%%%%%%%%%%
\section{Calculation details}\label{det}
%%%%%%%%%%%%%%%%%%%%%%%%%%%%%%%%%%%%%%%%%%%%%%%%%%%%%%%%%%%%%%%%%%%%%%%%%

We perform NLO QCD calculations in the modified Minimal Subtraction
($\overline{\rm MS}$) renormalization scheme \cite{Bardeen:1978yd}.
The fac\-tor\-ization/renor\-malization scales in parton densities and 
fragmentation functions are assumed being equal to the
renormalization scale in the strong coupling constant and are denoted
as $\mu$. As a reference we take $\mu$ equal to the transverse
momentum~\footnote{By the transverse momentum, $p_T$,
we mean the component of the momentum perpendicular to 
momenta of the initial particles.} (transverse energy) of the final photon,
$\mu=p_T^{\gamma}=E_T^{\gamma}$. For a comparison $\mu=E_T^{\gamma}/2$ 
and $\mu=2E_T^{\gamma}$ will be also considered.

The quark masses are neglected in the calculation
and the number of active flavors 
is assumed to be $N_f$=4 or, for a comparison, $N_f$=3 or 5.
The cross section is proportional to the electric 
charge of the quark in the fourth power, so
the contribution of the bottom quark ($e_b$=-1/3) is expected
to be much smaller than the contributions of the up and charm quarks
($e_u$=$e_c$=2/3), and the predictions for $N_f$=4 and 5 should not
differ considerably. On the other hand, the differences between
the results obtained using $N_f$=4 and 3 can be large.

The two-loop coupling constant $\alpha_s$ is applied in the form
\be
\alpha_S(\mu^2)={{12 \pi}\over {(33-2N_f)\ln(\mu^2/\Lambda_{QCD}^2)}}
[1-{{6(153-19N_f)}\over (33-2N_f)^2} {{\ln[\ln(\mu^2/\Lambda_{QCD}^2)]}
\over{\ln(\mu^2/\Lambda_{QCD}^2)}}].
\label{alfas}
\ee
To obtain the QCD parameter $\Lambda_{QCD}$ appearing in the strong coupling 
constant (\ref{alfas}), we use the world average of $\alpha_S$ at the scale 
$M_Z$ (the mass of the $Z^0$ gauge boson) \cite{Bethke:2002rv}:
\be
\label{1183}
\overline{\alpha_S} (M_Z^2) = 0.1183 \pm 0.0027.
\ee
In order to 
minimize theoretical and experimental uncertainties the above
$\overline{\alpha_S} (M_Z^2)$ value was determined in \cite{Bethke:2002rv}
from precise data based on NNLO analyses only;
the data given at scales different than $M_Z$ were 
extrapolated to the $M_Z$ scale using the four-loop coupling.
Although we use the two-loop expression (\ref{alfas}),
we apply (\ref{1183})
as the best estimation of the true value of $\alpha_S (M_Z)$.
We take the number of active flavours $N_f$=3, 4 or 5
at scales $\mu < m_c$, $m_c<\mu < m_b$ and $\mu < m_b$, respectively,
with the following charm, bottom and $Z$ masses: $m_c=1.5$ GeV, $m_b=4.7$ GeV,
$M_Z=91.2$ GeV \cite{Bethke:2002rv}.
Assuming that $\alpha_S (\mu^2)$ is a continuous function
at $\mu=m_c$ and $\mu=m_b$ the obtained $\Lambda_{QCD}$ parameters are:
$\Lambda_{QCD}$=0.386, 0.332 and 0.230 GeV
for $N_f$=3, 4 and 5, respectively. 
%The corresponding LO parameter for $N_f$=4 is $\Lambda_{QCD}$=0.123 GeV.
The above $\Lambda_{QCD}$ values are used in numerical calculations 
discussed in the next chapters (Chapters~\ref{non}, 
\ref{isol}, \ref{jet})~\footnote{Note
that in Chapters~\ref{small} and \ref{Sglu}
%slightly 
different numbers are used for 
consistency with the published results.}.

In the calculations we apply the Gl\"uck-Reya-Vogt (GRV)
parton densities in the proton \cite{Gluck:1995uf}
and photon \cite{Gluck:1992ee}, and the GRV
fragmentation functions \cite{Gluck:1993zx}. For a comparison 
we also use other parametrizations, namely
Martin-Ro\-berts-Stir\-ling\--Thorne (MRST98) \cite{Martin:1998sq},
(MRST99) \cite{Martin:1999ww}, (MRST2002) \cite{Martin:2002aw}, 
CTEQ4M \cite{Lai:1996mg}, CTEQ6M \cite{Pumplin:2002vw}, 
Au\-ren\-che\--Chiap\-pet\-ta\--Fon\-tan\-naz-Guillet-Pilon
(ACFGP) \cite{Aurenche:1992sb}, 
Aurenche-Guillet-Fon\-tan\-naz
(AFG) \cite{Aurenche:1994in} and (AFG02) \cite{Fontannaz:2002nu}, 
Gordon-Storrow (GS) \cite{Gordon:1997pm},
Cornet-Jankowski-Kraw\-czyk\--Lor\-ca (CJ\-KL) \cite{Cornet:2002iy},
Duke-Owens (DO) \cite{Duke:1982bj},
Bourhis-Fontannaz-Guillet (BFG) \cite{Bourhis:1997yu} and
Gl\"uck-Reya-Stratmann (GRS) \cite{Gluck:1994tv}.

Following experimental analysis \cite{Breitweg:1997pa}-\cite{unknown:2004uv},
we consider the photoproduction of the photon at HERA with transverse
momentum (transverse energy) higher than 5 GeV. In such processes
the emission of the final large-$p_T$ photon directly from the electron
(Bethe-Heitler process) is negligible  \cite{ula} and we omit 
it~\footnote{For transverse energy higher than 5 GeV the momentum transfer, 
-$t$ (see Eq. (\ref{mand})), is higher than 25 GeV$^2$.
Note that for $Q^2_{max}$ = 1 GeV$^2$ and lower values of
the momentum transfer, 4$\le -t \le 10$ GeV$^2$,
the contribution of the Bethe-Heitler process may be no-negligible
in some kinematic regions, see Fig. 6 in Ref. \cite{Hoyer:2000mb}.}
(this process is neglected in all existing calculations for the 
photoproduction of isolated photons at HERA
\cite{Gordon:1995km}-\cite{Fontannaz:2003yn}).

The results presented in next sections are obtained in NLO QCD
with use of the GRV set of parametrizations 
with $\mu =E_T^{\gamma}$, $N_f=4$ and
$\Lambda_{QCD}$=0.332 GeV
unless stated otherwise.

%%%%%%%%%%%%%%%%%%%%%%%%%%%%%%%%%%%%%%%%%%%%%%%%%%%%%%%%%%%%%%%%%%%%%%%%%
\section{Equivalent photon approximation}\label{Cnlo:epa}
%%%%%%%%%%%%%%%%%%%%%%%%%%%%%%%%%%%%%%%%%%%%%%%%%%%%%%%%%%%%%%%%%%%%%%%%%

\begin{figure}[b]
\vskip 4.5cm\relax\noindent\hskip 0cm
       \relax{\includegraphics{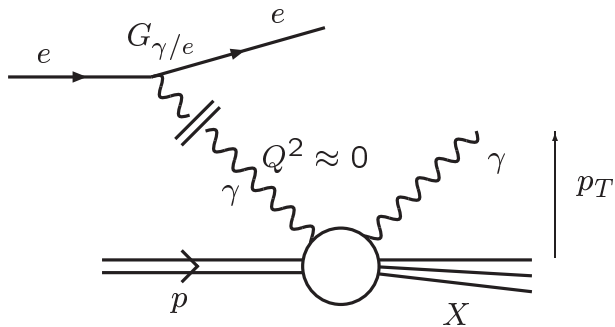}}
\vspace{0cm}
\caption{\small\sl The factorization of the $ep\ra e\gamma X$ reaction.}
\label{figepa.slide}
\end{figure}

Our aim is to consider the production of photons with large transverse momentum
in the electron-proton scattering in processes in which the electron 
is scattered at a small angle. In such events the mediating photon is almost 
on-shell, $Q^2\approx 0$ (photoproduction events), and  
the cross section can be calculated using the 
equivalent photon (Weizs\"{a}cker-Williams) 
approximation~\cite{vonWeizsacker:1934sx,Williams:1934ad} 
(see also e.g.~\cite{Budnev:1974de,Frixione:1993yw,Nisius:2000cv}).
In this approximation the differential cross section for the $ep$ collision
is related
to the corresponding differential cross section for the $\gamma p$ collision; 
in case of the considered herein reaction, $ep\ra e\gamma X$
(or $ep\ra e\gamma ~jet ~X$), 
we get (see Fig.~\ref{figepa.slide}):
\be\label{epa}
d\sigma^{ep\ra e\gamma (jet) X} = \int G_{\gamma /e}(y)
d\sigma^{\gamma p\ra \gamma (jet) X} dy ~,
\ee
where $y$ is the fraction of the initial electron momentum carried 
by the exchanged photon,
and $G_{\gamma /e}$ stands for the flux of the real photons emitted from 
the electron.
\\
We use the photon spectrum in the form \cite{Frixione:1993yw}:
\be
G_{\gamma/e}(y)={\alpha\over 2\pi} \{ {1+(1-y)^2\over y}
\ln [ {Q^2_{max}(1-y)\over m_e^2 y^2}]
- ~ {2\over y}(1-y-{m_e^2y^2\over Q^2_{max}}) \} ~,
\ee
with $m_e$ being the electron mass. In the numerical calculations
the maximal photon virtuality $Q^2_{max}$ = 1 GeV$^2$ is assumed,
what is a typical value for the 
recent photoproduction measurements at the HERA 
collider~\cite{Breitweg:1997pa}-\cite{unknown:2004uv}.
The above formula describes the spectrum of equivalent real
(transversally polarized) photons. We do not take into account 
longitudinally polarized
photons and the interference between longitudinally and transversally 
polarized photons, since they give a very small contribution 
%\PP 
in the kinematic regions which we study~\cite{ula,Jezuita-Dabrowska:bp}.

\chapter{Inclusive photon production}\label{non}

The inclusive production of the photon with a large transverse momentum
in the $\gamma p$ or $ ep$ collision at HERA and other colliders was the 
subject of detailed studies in the literature 
\cite{Bjorken:1969ja}, \cite{Tu:1979vg}-\cite{Gordon:1994sm}, 
and herein only the main aspects of the calculation of 
the cross section including NLO QCD corrections are briefly
discussed. The main cross section formula and explicit expressions for 
various contributions are given in Sec. \ref{Snon:x} and in the 
Appendices.

\section{Cross section formulae}\label{Snon:x}

The differential cross section for the $\gamma p\ra \gamma X$ 
(or $\gamma p\ra \gamma ~jet ~X$) process has the following form:
\bea
d\sigma^{\gamma p\ra \gamma (jet) X}=
\sum_{q,\bar{q}} \int dx f_{q/p}(x,\mu^2) d\sigma^{\gamma q\ra\gamma q} +
\int dx f_{g/p}(x,\mu^2) d\sigma^{\gamma g\ra\gamma g}
%\nonumber 
\label{cross1}
\\
+ \sum_{a=q,\bar{q},g} \int dx_{\gamma}
\sum_{b=q,\bar{q},g} \int dx 
f_{a/\gamma}(x_{\gamma},\mu^2)
f_{b/p}(x,\mu^2) 
d\sigma^{ab\ra\gamma d} 
%\nonumber 
\label{cross2}
\\
+ \sum_{b=q,\bar{q},g} \int dx 
\sum_{c=q,\bar{q},g} \int {dz\over z^2}
f_{b/p}(x,\mu^2) 
D_{\gamma /c}(z,\mu^2)
d\sigma^{\gamma b\ra cd} 
%\nonumber 
\label{cross3}
\\
+ \sum_{a=q,\bar{q},g} \int dx_{\gamma}
\sum_{b=q,\bar{q},g} \int dx 
\sum_{c=q,\bar{q},g} \int {dz\over z^2}
f_{a/\gamma}(x_{\gamma},\mu^2)
f_{b/p}(x,\mu^2)
\nonumber \\ \cdot 
D_{\gamma /c}(z,\mu^2) \raisebox{5mm}{} \raisebox{-5mm}{}
d\sigma^{ab\ra cd}
\label{cross4}
\\
+ \sum_{q,\bar{q}} \int dx \left\{
f_{q/p}(x,\mu^2) \left[ d\sigma^{\gamma q\ra\gamma q}_{\alpha_S}
+d\sigma^{\gamma q\ra\gamma qg}_{\alpha_S}\right]
%\\
+ f_{g/p}(x,\mu^2) d\sigma^{\gamma g\ra\gamma q\bar{q}}_{\alpha_S} 
\right\},
\label{cross5}
\eea
where $\gamma q\ra\gamma q$, $\gamma g\ra\gamma g$ etc. are various partonic 
processes described in the previous chapter.
The functions $f_{a/\gamma}$, 
$f_{b/p}$, and $D_{\gamma /c}$ stand for the parton distributions
in the photon and proton, and the parton-to-photon fragmentation function,
respectively. The corresponding longitudinal-momentum fractions carried
by partons in the proton and photon are denoted as $x$ and $x_{\gamma}$.
Similarly, for the parton-to-photon fragmentation 
the variable $z$ is introduced.
The $\mu$ scale is the factorization/renormalization scale
related to the physical hard scale of the considered
process.

The contribution of processes with the direct both initial and final
photon are given on the right-hand side in (\ref{cross1})
(see Figs. \ref{figborn}, \ref{figbox}).
The expressions (\ref{cross2}-\ref{cross4}) stand for contributions
of processes with the resolved photons or/and the fragmentation
into the photon (Figs. \ref{figsingi}-\ref{figdoub2}). 
The formulae for the partonic cross sections, 
$d\sigma^{\gamma q\ra\gamma q}$ etc., appearing in
(\ref{cross1}-\ref{cross4}) are listed in Appendix \ref{lox}.
The $\mathcal{O}$$(\alpha_S)$ corrections to the
Born process are included in (\ref{cross5}),
where $d\sigma^{\gamma q\ra\gamma q}_{\alpha_S}$,
$d\sigma^{\gamma q\ra\gamma qg}_{\alpha_S}$ and
$d\sigma^{\gamma g\ra\gamma q\bar{q}}_{\alpha_S}$ 
are the partonic cross sections for the processes shown in Figs. 
\ref{figvirt}-\ref{fig23}.

The $\mathcal{O}$$(\alpha_S)$ corrections to the Born process 
were first numerically calculated for the inclusive photon production 
in \cite{Duke:1982bj}, and the compact analytical formulae 
can be found in \cite{Aurenche:1984hc,jan}.
These corrections contain the infrared
and/or collinear singularities. The collinear singularities are present in
the squared amplitudes of the $2\ra 3$
processes (Figs.~\ref{figreal},~\ref{fig23}),
and are factored out and absorbed by the corresponding parton densities
in the proton or photon (Fig. \ref{fact-phot}), 
and by the parton-to-photon fragmentation functions (Fig.~\ref{fact-frag}).
The infrared singularities are due to the soft gluon emission or
the soft gluon exchange and cancel when the real gluon corrections
(Fig.~\ref{figreal}) are added to the virtual gluon corrections
(Fig.~\ref{figvirt}) in agreement with the Bloch-Nordsieck 
theorem \cite{BlochNordsieck} (see \cite{Aurenche:1984hc,jan} for the
calculation details).
The corresponding singular-free partonic cross sections 
$d\sigma^{\gamma q\ra\gamma q}_{\alpha_S}+
d\sigma^{\gamma q\ra\gamma qg}_{\alpha_S}$ and
$d\sigma^{\gamma g\ra\gamma q\bar{q}}_{\alpha_S}$ 
for the inclusive photon production are given in Eqs.~(\ref{eq:non:k}) 
and (\ref{eq:non:kp}) in Appendix \ref{xsec:w1}. 

To obtain the predictions for the inclusive photon
production one needs to perform the integrations of the cross section
(\ref{cross1}-\ref{cross5}) within the whole range of the
fractional momenta: $0<x<1$, $0<x_{\gamma}<1$, $0<z<1$.
The $\Theta$-functions in
(\ref{eq:non:k}) and (\ref{eq:non:kp}), and the $\delta$-function
in (\ref{eq.lox}) ensure that in fact the $x$, $x_{\gamma}$ and $z$
close to zero do not contribute to the cross section.

\section{Numerical results}\label{S:incl:res}

The results for the inclusive (non-isolated) photon production at the 
HERA collider are obtained with the parameters and parton distributions
given in Sec. \ref{det}. The initial electron and proton energies
corresponding to energies at HERA are taken $E_e=27.6$ GeV 
and $E_p=920$ GeV, respectively \cite{unknown:2004uv}.

\begin{figure}[t]
\vskip 24.5cm\relax\noindent\hskip -2cm
       \relax{\includegraphics{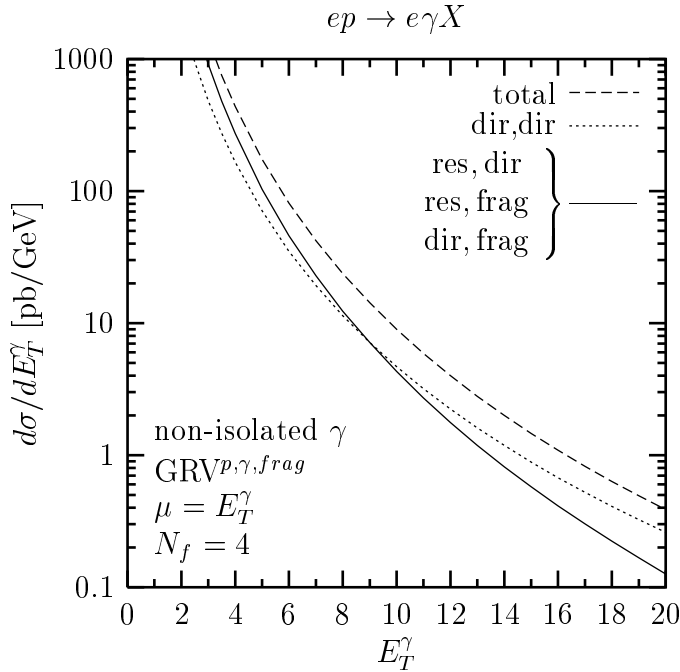}}
\vspace{-16cm}
\caption{\small\sl The cross section $d\sigma /dE_T^{\gamma}$
for the inclusive $\gamma$ production at the HERA collider (dashed line).
The contributions of processes with the direct initial and final $\gamma$
(dotted line) and processes with the resolved initial $\gamma$ or/and
the fragmentation into final $\gamma$ (solid line) are shown separately.}
\label{fig.nonpg}
\end{figure}

In Fig.~\ref{fig.nonpg} the NLO cross section $d\sigma /dE_T^{\gamma}$ 
for the inclusive photon production in the $ep$ scattering is presented
(dashed line). The contribution of the processes with the resolved
initial photon or/and the parton-to-photon fragmentation (solid line)
and the contribution of the processes with the direct both photons 
(dotted line) are shown as well.
The cross section decreases by three orders of magnitude when $E_T^{\gamma}$
increases from 4 GeV to 20 GeV, and obviously the most important 
contribution to the total cross section 
is coming from the lowest $E_T^{\gamma}$ region.
The processes with the direct initial and final photons, i.e. Born
plus $\mathcal{O}$$(\alpha_S)$ 
corrections and the box contribution (Figs. \ref{figborn}-\ref{fig23},
\ref{figbox}), dominate in the cross section for large 
$E_T^{\gamma}\ge 9$ GeV. For lower $E_T^{\gamma}$ the processes with 
the resolved initial photon or/and the fragmentation into final photon
(Figs. \ref{figsingi}-\ref{figdoub2}) constitute more
than a half of the cross section.

\begin{figure}[t]
\vspace{24.3cm}
\vskip 0cm\relax\noindent\hskip -2cm
       \relax{\includegraphics{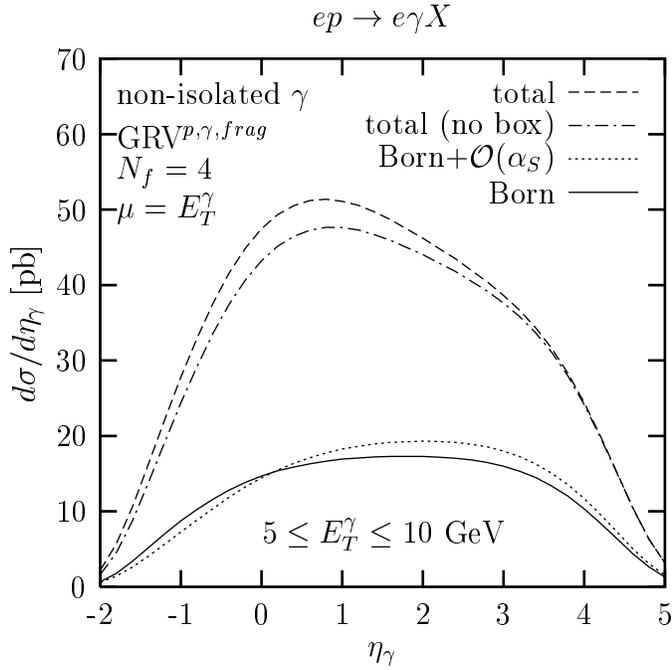}}
\vspace{-16.cm}
\caption{\small\sl The cross section $d\sigma /d\eta^{\gamma}$ 
for the inclusive $\gamma$ production at the HERA collider with
(dashed line) and without (dashed-dotted line) the box contribution.
The contributions of the Born process (solid line) and the Born process 
with the $\mathcal{O}$$(\alpha_S)$ corrections (dotted line) are shown.}
\label{fig.noneg1}
\end{figure}

\begin{figure}[b]
\vspace{23.5cm}
\vskip 0cm\relax\noindent\hskip -2cm
       \relax{\includegraphics{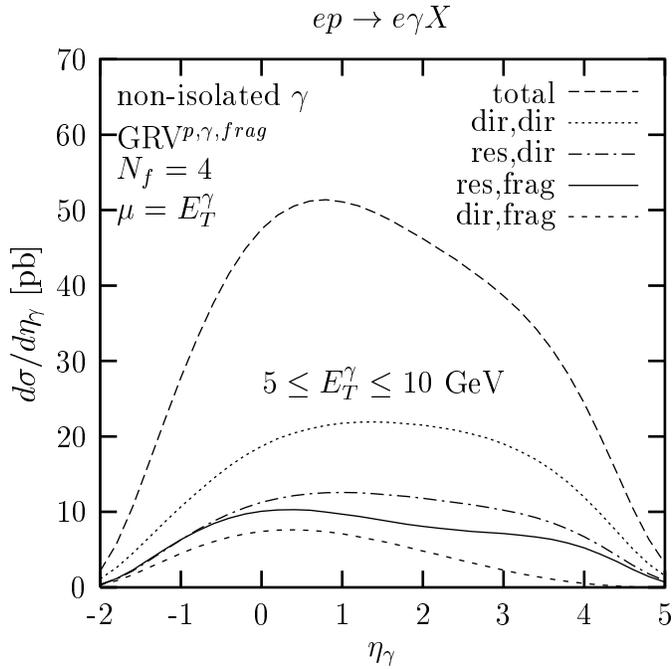}}
\vspace{-16.cm}
\caption{\small\sl The cross section $d\sigma /d\eta^{\gamma}$ for the
inclusive $\gamma$ production at the HERA collider 
(dashed line) and the contributions
of various types of processes with:
the direct initial $\gamma$ and the fragmentation into 
the final $\gamma$ (short-dashed line), 
the resolved initial $\gamma$ and the fragmentation into the final $\gamma$ 
(solid line), the resolved initial and the direct final $\gamma$ 
(dashed-dotted line), and direct both $\gamma$'s (dotted line).} 
\label{fig.noneg2}
\end{figure}

The importance of individual contributions to the inclusive photon 
cross section, integrated over 5 GeV $<E_T^{\gamma}<$ 10 GeV, is illustrated 
in Tab.~\ref{tab1} (the first row)
and in Figs. \ref{fig.noneg1} and \ref{fig.noneg2}, where the distributions
of the final photon rapidity, $\eta^{\gamma}$, are shown.
In this $E_T^{\gamma}$ range the integrated (``total'') 
NLO cross section is equal to 241 pb. 
Processes other than the lowest order (Born) one give all together the 
contribution almost two times
larger than the cross section for the Born process alone.
The $\mathcal{O}$$(\alpha_S)$ corrections to the Born process
are relatively small, negative at $\eta_{\gamma}<0$
and positive at $\eta_{\gamma}>0$ (Fig. \ref{fig.noneg1}), 
and constitute only 2\% of the total cross section (Tab.~\ref{tab1}). 
The box contribution constitutes 6\% of the cross section
integrated over all $\eta_{\gamma}$ (Tab.~\ref{tab1})
and it is even larger, 7-12\%, 
for the differential cross section in the range $-1<\eta_{\gamma}<1$
(Fig. \ref{fig.noneg1}), where the data 
\cite{Breitweg:1999su,unknown:2004uv} exists
(see Chapters \ref{small}, \ref{isol}, \ref{jet}).
Although the box diagram, $\gamma g\rightarrow\gamma g$, 
is of order ${\mathcal{O}}(\alpha^2\alpha_s^2)$,
it gives relatively large contribution 
mainly due to a large gluonic content of the proton at small $x$.

The contribution of processes of order ${\mathcal{O}}(\alpha^2\alpha_s^2)$
with the resolved initial photon \underline{and} 
the quark-to-photon fragmentation is large, being of order 
20\% (Tab.~\ref{tab1}).
It is worth noticing that this contribution
is build from many, relatively small, individual terms 
(Figs. \ref{figdoub}, \ref{figdoub2}).
The parton-to-photon fragmentation with the direct or resolved initial
photon (Figs. \ref{figsingi}-\ref{figdoub2}) gives in sum a contribution 
of 31\% (Tab.~\ref{tab1}) or 32-39\% for the central rapidity
range $-1<\eta_{\gamma}<1$ (Fig. \ref{fig.noneg2}),
and this contribution is similar to the Born one (36\%). 
Note that this part of the cross section
involving poorly known fragmentation functions, $D_{\gamma}$,
is much suppressed by implementing of experimental cuts, 
%\PP this is a main issue of  
see Chapter \ref{small}.

In the whole range of $\eta_{\gamma}$ the processes with the resolved
initial and direct final photons constitute about 25\% of the cross 
section~\footnote{This contribution is even enhanced if experimental 
constrains are taken into account (Chapter \ref{small}).}
(Tab.~\ref{tab1}, Fig. \ref{fig.noneg2}). 
It makes the cross section sensitive 
to the photon structure function \cite{Duke:1982bj},
\cite{Aurenche:1984hc}-\cite{Aurenche:1992sb},
\cite{Gordon:1994sm}-\cite{Fontannaz:2004qv}, 
\cite{Krawczyk:1997zv,Krawczyk:1999eq}
and in particular to the gluonic content of the photon, what will be
shortly discussed in Chapter \ref{Sglu}.

%%%%%%%%%%%%%%%%%%%%%%%%%%%%%%%%%%%%%%%%%%%%%%%%%%%%%%%%%%%%%%%%%%%%%%%
\chapter{Isolated photon production. Small cone approximation}\label{small}
%%%%%%%%%%%%%%%%%%%%%%%%%%%%%%%%%%%%%%%%%%%%%%%%%%%%%%%%%%%%%%%%%%%%%%%

In order to compare QCD predictions with data, one should consider
such physical quantities, which are actually measured
or are as close to the measured ones as possible.
Since in experimental analyses 
%\PP 
in order to reduce backgrounds the observed photon is often required 
to be {\sl isolated} from hadrons 
\cite{Breitweg:1997pa}-\cite{unknown:2004uv}, similar
isolation of the photon produced in the $ep$ scattering
was included in the QCD calculations. It was done for the first time by
Gordon and Vogelsang~\cite{Gordon:1995km,Gordon:1998yt}
and Gordon~\cite{Gordon:1998yt}.
The second independent calculation was performed by
Krawczyk and Zembrzuski \cite{Krawczyk:1998it,Krawczyk:2001tz}.
Then the calculation of Fontannaz, Guillet, Heinrich,
\cite{Fontannaz:2001ek,Fontannaz:2001nq}
and a new analysis of Zembrzuski and Krawczyk \cite{Zembrzuski:2003nu}
were presented.
In this chapter we discuss
the QCD predictions for the isolated photon production
presented in the paper of Krawczyk and Zembrzuski
\cite{Krawczyk:2001tz}~\footnote{This is an extended and updated version 
of \cite{Krawczyk:1998it}.}.

%%%%%%%%%%%%%%%%%%%%%%%%%%%%%%%%%%%%%%%%%%%%%%%%%%%%%%%%%%%%%%%%%%%%%%%
\section{Isolation restrictions}\label{small.isol}
%%%%%%%%%%%%%%%%%%%%%%%%%%%%%%%%%%%%%%%%%%%%%%%%%%%%%%%%%%%%%%%%%%%%%%%

The final photon is ``isolated'' if  
the sum of hadronic transverse energy within a cone of radius $R$ 
around the photon is smaller than the photon 
transverse energy multiplied by a small parameter 
$\epsilon$~\cite{Breitweg:1997pa}-\cite{unknown:2004uv}:
\be\label{eps}
\sum_{hadrons} E_T^{hadron} < \epsilon E_T^{\gamma} .
\ee
The cone is defined in the rapidity and azimuthal angle phase space, namely
\be\label{R}
\sqrt{(\eta^{hadron}-\eta^{\gamma})^2+(\phi^{hadron}-\phi^{\gamma})^2} < R.
\ee
The isolation constraint allows to reduce the background 
from neutral mesons ($\pi^0$, $\eta$), which decay into two photons
and from photons radiated by final state hadrons.
It suppresses considerably the contribution of processes involving 
the parton fragmentation into the photon.
For comparison with data the
same isolation restriction (\ref{eps}) is taken into
account in theoretical 
calculations~\cite{Gordon:1995km}-\cite{Fontannaz:2004qv}. 

Note, that a different type of isolation which removes from the cross section
the whole fragmentation component is advocated by 
Frixione~\cite{Frixione:1998jh}. We do not apply this type of
isolation, since so far it is not applied in experimental analyses 
of the prompt photon production at the HERA collider.

%%%%%%%%%%%%%%%%%%%%%%%%%%%%%%%%%%%%%%%%%%%%%%%%%%%%%%%%%%%%%%%%%%%%%%%
%\section{Analytical results for small cone approximation}\label{Sapprox}
\section{Small cone approximation}\label{Sapprox}
%%%%%%%%%%%%%%%%%%%%%%%%%%%%%%%%%%%%%%%%%%%%%%%%%%%%%%%%%%%%%%%%%%%%%%%
%\PP
%%%%%%%%%%%%%%%%%%%%%%%%%%%%%%%%%%%%%%%%%%%%%%%%%%%%%%%%%%%%%%%%%%%%%%%
%\subsection{Definition}
%%%%%%%%%%%%%%%%%%%%%%%%%%%%%%%%%%%%%%%%%%%%%%%%%%%%%%%%%%%%%%%%%%%%%%%

The simplest way to calculate the differential cross section
for an isolated photon production, $d\sigma_{isol}$, is to calculate
the difference of the differential cross section for a non-isolated 
(inclusive) photon production,
$d\sigma_{non-isol}$, and a subtraction term,
$d\sigma_{sub}$
~\cite{Berger:1990es,Gordon:1994ut,Gluck:1994iz,Gordon:1995km}:
\bea
d\sigma_{isol}=d\sigma_{non-isol}-d\sigma_{sub}.
\eea
The calculation of $d\sigma_{non-isol}$ was discussed in the previous chapter.
The subtraction term corresponds to the 
cuts which are opposite to the isolation 
cuts, i.e. within a cone of radius $R$ around the final photon
the total transverse energy of hadrons should be {\underline{higher}} 
than the photon transverse 
energy multiplied by $\epsilon$,
\be\label{epssub}
\sum_{hadrons} E_T^{hadron} > \epsilon E_T^{\gamma} .
\ee

In this chapter we apply the subtraction term $d\sigma_{sub}$
calculated in an approximate way. This approximation bases on the 
assumption that an angle $\delta$ between the final photon and a parton 
inside the cone of radius $R$ is small~\cite{Gordon:1994ut,Gordon:1995km}.
It allows to simplify the calculations considerably
and leads to the compact analytical expressions for all relevant
matrix elements involved in $d\sigma_{sub}$. 
Note that in this approximation the maximal value of the angle $\delta$ 
is proportional to the radius $R$: 
\bea
\delta \le R/cosh(\eta^{\gamma}) = R E_{T}^{\gamma} / E^{\gamma} ,
\eea
where $\eta^{\gamma}$, $E^{\gamma}$ and $E_{T}^{\gamma}$ stand for the photon 
rapidity, energy and transverse energy, respectively.
We stress that the above small angle (cone) approximation is used 
\underline{only} in calculation of the 
$\mathcal{O}$$(\alpha_S)$ 
corrections in the subtraction cross section $d\sigma_{sub}$, while
other contributions to $d\sigma_{sub}$ as well as $d\sigma_{non-isol}$ 
are obtained in an exact way.

%isolation cone is narrow

%%%%%%%%%%%%%%%%%%%%%%%%%%%%%%%%%%%%%%%%%%%%%%%%%%%%%%%%%%%%%%%%%%%%%%%
\section{Analytical results}
%%%%%%%%%%%%%%%%%%%%%%%%%%%%%%%%%%%%%%%%%%%%%%%%%%%%%%%%%%%%%%%%%%%%%%%
%\PP

The calculations include partonic processes with two ($2\ra 2$) or three
($2\ra 3$) particles in the final state:
\be\label{223}
ab\ra cd
\ee
or
\be\label{23eq}
ab\ra\gamma d_1d_2 ,
\ee
where $a$ is the photon or a parton originating from the photon, 
$b$ is a parton from the proton, $c$ stand for the final photon 
or a parton which decays into the photon in the fragmentation process, 
and $d_{(i)}$ denotes quarks and/or gluons. 
For the isolated final photon the summation in Eqs.~(\ref{eps}, \ref{epssub}) 
runs over the $c$-parton remnant and over the $d_i$-partons, 
if they are inside the cone (\ref{R}). 

In the partonic $2\ra 2$ processes with a direct final photon 
($c=\gamma$),
\be\label{22nofrag}
ab\ra\gamma d ,
\ee 
the photon is isolated by definition, so they give no
contribution to the subtraction term.
If the final photon comes from the fragmentation process ($c\ne\gamma$),
\be
ab\ra c d ,
\ee 
it takes the $z$-fraction of the
$c$-parton transverse energy, $E_{T}^{\gamma} = z E_{T}^{c}$.
The hadronic remnant of the $c$-parton takes the fraction of transverse energy
equal to $(1-z) E_{T}^{c}$, and the photon is isolated if
\be
E_T^{\gamma} = z E_T^c > \epsilon (1-z) E_T^c 
\ee
or
\be
z > 1 / (1+\epsilon)
\label{zeps} .
\ee
The subtraction term for $2\ra 2$ processes can be obtained from 
cross sections (\ref{cross3}, \ref{cross4}) involving fragmentation
functions integrated 
over $z$, with $z$-values fulfilling inequality opposite to (\ref{zeps}),
i.e. $z<1/(1+\epsilon)$. So we have
\bea
%d\sigma^{\gamma p\ra \gamma X}_{2\ra 2,sub}=
d\sigma_{frag, sub}^{\gamma p\rightarrow\gamma X}=
%\makebox[9cm]{}
%\nonumber \\
\sum_{b=q,\bar{q},g} \int\limits_0^1 dx 
\sum_{c=q,\bar{q},g} \int\limits_0^{1/(1+\epsilon)}{dz\over z^2}
f_{b/p}(x,\mu^2) 
D_{\gamma /c}(z,\mu^2) \raisebox{5mm}{} \raisebox{-5mm}{}
d\sigma^{\gamma b\ra cd}
\nonumber \\
+\sum_{a=q,\bar{q},g} \int\limits_0^1 dx_{\gamma}
%\nonumber \\ \cdot
\sum_{b=q,\bar{q},g} \int\limits_0^1 dx 
\sum_{c=q,\bar{q},g} \int\limits_0^{1/(1+\epsilon)} {dz\over z^2}
f_{a/\gamma}(x_{\gamma},\mu^2)
%\nonumber \\ \cdot
f_{b/p}(x,\mu^2) 
\nonumber \\ \cdot
D_{\gamma /c}(z,\mu^2)
d\sigma^{ab\ra cd} .
\label{subLO}
\eea
The total subtraction cross section consists of 
$d\sigma_{frag, sub}^{\gamma p\rightarrow\gamma X}$
and $\mathcal{O}$$(\alpha_S)$ corrections:
\bea
d\sigma_{sub}^{\gamma p\rightarrow\gamma X} = 
d\sigma_{frag, sub}^{\gamma p\rightarrow\gamma X} +
d\sigma_{\alpha_S, sub}^{\gamma p\rightarrow\gamma X} .
\eea
The virtual gluon corrections do not contribute
to the subtraction,
since they come from  $2\ra 2$ processes (Fig.~\ref{figvirt}),
where the direct final photon
is isolated by definition. 
The cross section 
$d\sigma_{\alpha_S, sub}^{\gamma p\rightarrow\gamma X}$
includes only contributions of $2\ra 3$ processes (\ref{23eq}).
In these processes the photon and two partons are produced. One
parton, say $d_1$, enters the cone of radius R around the photon
(\ref{R}), 
and its transverse energy should be higher 
than the photon's transverse energy multiplied by $\epsilon$ (\ref{epssub}).
There are three types of such processes, namely:
%\newline
%$\bullet$ $\gamma q\rightarrow\gamma q + g$ (with a quark inside the cone),
%\newline
%$\bullet$ $\gamma q\rightarrow\gamma g + q$ (with a gluon inside the cone),
%\newline
%$\bullet$ $\gamma g\rightarrow\gamma q + \bar{q}$ 
%(with a quark or antiquark inside the cone).
%\newline
\bea
\bullet \gamma q\rightarrow\gamma q + g {\rm 
~(with ~a ~quark ~inside ~the ~cone)},
\label{2plus1a}
\eea
\bea
\bullet \gamma q\rightarrow\gamma g + q 
{\rm ~(with ~a ~gluon ~inside ~the ~cone)},
\label{2plus1b}
\eea
\bea
\bullet \gamma g\rightarrow\gamma q + \bar{q} 
{\rm ~(with ~a ~quark ~or ~antiquark ~inside ~the ~cone)}.
\label{2plus1c}
\eea
%\end{itemize}
The subtraction cross section for $\mathcal{O}$$(\alpha_S)$ corrections
%$d\sigma_{\alpha_S, sub}^{\gamma p\rightarrow\gamma X}$
has a general form:
\bea
d\sigma_{\alpha_S, sub}^{\gamma p\rightarrow\gamma X} = 
\sum_{i=1}^{2 N_f}\int\limits_0^1 
\Theta \left( {v(1-w)\over 1-v+vw} - \epsilon \right)
\left[ f_{q_i/p}(x,\mu^2) 
d\sigma_{sub}^{\gamma q_i\rightarrow\gamma q_i + g} + 
\right.
\makebox[0cm]{} \nonumber \\ \left.
f_{q_i/p}(x,\mu^2)
d\sigma_{sub}^{\gamma q_i\rightarrow\gamma g + q_i} +
f_{g/p}(x,\mu^2)
d\sigma_{sub}^{\gamma g\rightarrow\gamma q_i + \bar{q}_i} \right] dx,
\label{eqcor}
\eea
where the variables $v$ and $w$ are defined in Appendix~\ref{Anot2}.
These variables are related to the final photon momentum and
to the initial momenta, and depend on the fractional momenta $y$ and $x$.
To fulfill the condition (\ref{epssub}) the integration is performed 
over $x$ according to $\Theta$ function with
\be\label{eps1}
v(1-w)/(1-v+vw) > \epsilon ,
\ee
since in the small cone approximation the transverse energy of the 
parton $d_1$ 
is related to the photon transverse energy in the following way:
\be\label{eps2}
E_T^{d_1} = E_T^{\gamma} v(1-w)/(1-v+vw) .
\ee

The analytical results for the $\mathcal{O}$$(\alpha_S)$ corrections 
contributing to the subtraction term (\ref{eqcor}) are presented below. 
All collinear singularities are shifted into the fragmentation functions 
$D_{\gamma/c}$ (Fig.~\ref{fact-frag}) according to the standard 
factorization procedure, 
%\PP 
as discussed in Introduction
(for details  for the non-isolated photon production see also
\cite{Aurenche:1984hc,jan}).
The infrared singularities do not appear in this calculations. 
This is due to the fact that there are no virtual gluon
corrections, the gluons emitted inside the cone can not be too soft 
being constrained by Eqs. (\ref{eps1}, \ref{eps2}),
while the gluon emitted outside 
the isolation cone can not be soft, because its transverse momentum has to
balance transverse momenta of the photon and the parton moving parallel
to the photon.
The cross sections for partonic processes (\ref{2plus1a}-\ref{2plus1c}), 
denoted in (\ref{eqcor}) by
$d\sigma_{sub}^{\gamma b\ra\gamma d_1+d_2}$,
are given by 
following expressions:
\bea\label{small1}
E^{\gamma}{d\sigma_{sub}^{\gamma q_i\rightarrow\gamma q_i + g}\over
d^3p^{\gamma}} = \Theta (1-w)
{\alpha_{em}^2\alpha_se_{q_i}^4\over \pi\hat{s}^2} C_F
{(1-v+vw)^2+(1-v)^2\over (1-v+vw)^2(1-v)} \cdot P ,
\eea
\bea\label{small2}
E^{\gamma}{d\sigma_{sub}^{\gamma q_i\rightarrow\gamma g + q_i}\over
d^3p^{\gamma}} = \Theta (1-w)
{\alpha_{em}^2\alpha_se_{q_i}^4\over \pi\hat{s}^2}C_F 
{(R\cdot E_T^{\gamma})^2\over \hat{s}}\cdot
%\makebox[3cm]{}
\nonumber 
\eea
\bea 
\makebox[1cm]{}\cdot
{(1-v) {[}(1-v+vw)^2+(vw)^2{]}\over 
(1-v+vw)^5v(1-w) (vw)^2}
%\nonumber \\ \cdot
{[}1+(1-v+vw)^4+v^4(1-w)^4{]} ~,
\eea
\bea\label{small3}
E^{\gamma}{d\sigma_{sub}^{\gamma g\rightarrow\gamma q_i + \bar{q}_i}\over
d^3p^{\gamma}} = \Theta (1-w)
{\alpha_{em}^2\alpha_se_{q_i}^4\over \pi\hat{s}^2} \,{1\over 2} \,
{(vw)^2+(1-v)^2\over (1-v+vw)vw(1-v)} \cdot P ,
\eea
where $C_F=4/3$ and
\bea\label{small4}
P = {1+v^2(1-w)^2\over 1-v+vw} \ln
{\left( {(R^2\cdot E_T^{\gamma})^2 v^2(1-w)^2\over \mu^2} \right) } +1 .
\eea
The function $P$ (\ref{small4}) which appears in Eqs. 
(\ref{small1}) and (\ref{small3}) was previously presented e.g. in
\cite{Gordon:1994ut}, where it was used in calculations for the
prompt photon production in the hadron-hadron collisions.
Other terms on the right-hand side in Eqs. (\ref{small1}) and (\ref{small3})
correspond to the partonic cross sections for the $2\ra 2$ processes,
$\gamma q\ra q g$ or $\gamma g\ra q\bar{q}$, respectively.
The expression (\ref{small2}) has been 
%\PP 
derived by us and presented for the first time
in the paper of Krawczyk and Zembrzuski \cite{Krawczyk:2001tz}.
The derivation of formulae (\ref{small1}-\ref{small3}) is briefly
discussed in Appendices \ref{xsec:col1} and \ref{xsec:col4}.

%%%%%%%%%%%%%%%%%%%%%%%%%%%%%%%%%%%%%%%%%%%%%%%%%%%%%%%%%%%%%%%%%%%%%%%
%\subsection{Discussion}
%%%%%%%%%%%%%%%%%%%%%%%%%%%%%%%%%%%%%%%%%%%%%%%%%%%%%%%%%%%%%%%%%%%%%%%
These results are obtained with the assumption that the angle between
the photon and the parton inside the cone is small 
and only the leading terms are kept.
In $d^3\sigma_{sub}^{\gamma q_i\rightarrow\gamma q_i + g}$ (\ref{small1})
and $d^3\sigma_{sub}^{\gamma g\rightarrow\gamma q_i + \bar{q}_i}$
(\ref{small3}) the leading contribution is ${\mathcal{O}} (\ln R) + const$
and all terms of order ${\mathcal{O}} (R^2)$ or higher are neglected
(terms of order ${\mathcal{O}} (R)$ do not appear). In
$d^3\sigma_{sub}^{\gamma q_i\rightarrow\gamma g + q_i}$ (\ref{small2})
there are no logarithmic ${\mathcal{O}} (\ln R)$ or constant terms. 
For small $\epsilon$, a soft gluon can be emitted into the isolation cone
in the process ${\gamma q_i\rightarrow\gamma g + q_i}$, 
leading to a large ${\mathcal{O}} (R^2)$ term in (\ref{small2}),
and therefore this contribution can not be neglected. 
%\cite{Gordon:1994ut}

The small cone approximation was previously used to obtain the predictions
for the prompt photon production at the hadron-hadron \cite{Gordon:1994ut}
as well as at the electron-proton \cite{Gordon:1995km} reactions.
Then the exact (i.e. without the assumption that the cone
is small) calculations were performed for both the hadron-hadron
\cite{Gordon:1996ug} and the electron-proton \cite{Gordon:1998yt} collisions.
The author of \cite{Gordon:1996ug} compared the approximated and 
exact results and concluded that the applied approximation was reliable
for the hadronic processes. Moreover, he mentioned in \cite{Gordon:1998yt}
that it is also an accurate technique for including isolation
effects in the cross section for the prompt photon production
in the electron-proton scattering. We have performed a 
detailed comparison between the
predictions obtained for this process
with the approximated and with the exact implementation
of the isolation restrictions, see Sec. \ref{Sisol:ae}. 
 
%%%%%%%%%%%%%%%%%%%%%%%%%%%%%%%%%%%%%%%%%%%%%%%%%%%%%%%%%%%%%%%%%%% 
\section{Numerical results and discussion}\label{S:small:res} 
%%%%%%%%%%%%%%%%%%%%%%%%%%%%%%%%%%%%%%%%%%%%%%%%%%%%%%%%%%%%%%%%%%%
 
We perform the numerical calculations for the isolated 
photon production at the $ep$ scattering using
the HERA collider energies: $E_e$=27.5 GeV and 
$E_p$=820 GeV~\cite{Breitweg:1999su}, and applying the parton
densities specified in Sec. \ref{det}. In this section we use the parameter
$\Lambda_{QCD}$=0.365, 0.320 and 0.220 GeV, fitted by us 
to the experimental value of $\alpha_s(M_Z) = 0.1177$~\cite{Biebel:1999zt},
for $N_f$=3, 4 and 5, respectively.
 
%%%%%%%%%%%%%%%%%%%%%%%%%%%%%%%%%%%%%%%%%%%%%%%%%%%%%%%%%%%%%%%%%%%%%
\subsection{Effects of isolation}\label{results1} 
%%%%%%%%%%%%%%%%%%%%%%%%%%%%%%%%%%%%%%%%%%%%%%%%%%%%%%%%%%%%%%%%%%%%%
 
\begin{figure}[ht]   
\vskip 25cm\relax\noindent\hskip -2cm
       \relax{\includegraphics{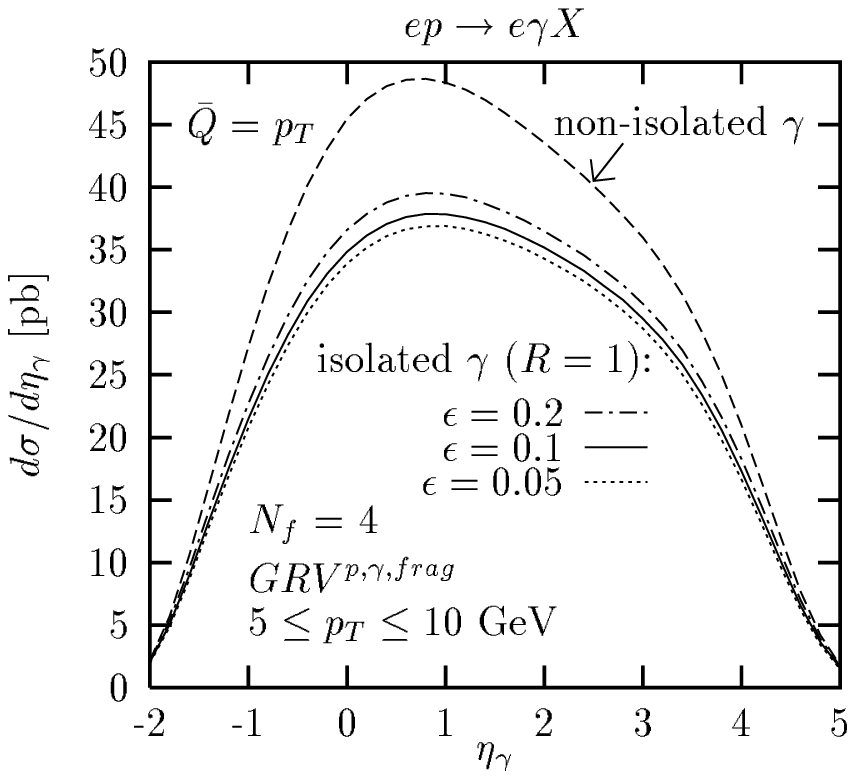}}
\vspace{-17.5cm}
\caption{\small\sl The differential cross section $d\sigma/d\eta^{\gamma}$ 
for the prompt $\gamma$ photoproduction at HERA
as a function of the photon rapidity $\eta^{\gamma}$
for non-isolated photon (dashed line) and isolated photon
with $R = 1$ and $\epsilon =$ 0.05 (dotted line), 0.1 (solid line)
and 0.2 (dashed-dotted line). The photon transverse energy,
$E_T^{\gamma}\equiv p_T$, is taken in the range
$5 \le E_T^{\gamma} \le 10$ GeV. The factorization/renormalization scale
$\mu\equiv \bar{Q} = E_T^{\gamma}$ is used.}
\label{Ffig7a} 
%\end{figure}   
%\begin{figure}[hb]   
\vskip 25cm\relax\noindent\hskip -2cm
       \relax{\includegraphics{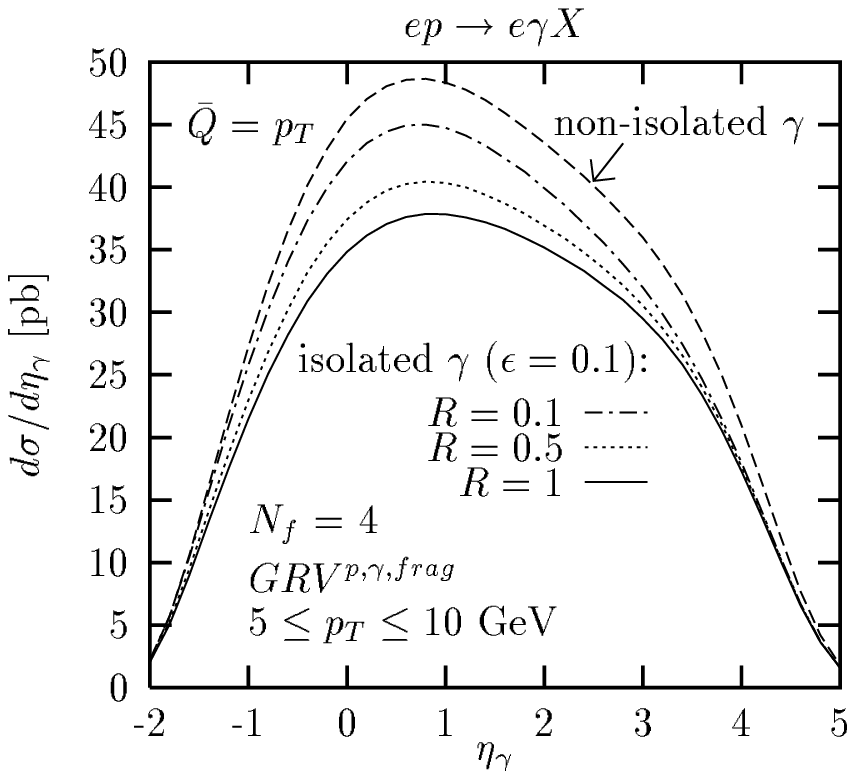}}
\vspace{-17.5cm}
\caption{\small\sl As in Fig.~\ref{Ffig7a} for
non-isolated photon (dashed line) and isolated photon
with $\epsilon = 0.1$ and $R =$ 0.1 (dashed-dotted line), 0.5 (dotted line)
and 1 (solid line).}
\label{Ffig7b}
\end{figure}

The results for the isolated photon production are presented
in Figs.~\ref{Ffig7a} and \ref{Ffig7b} in comparison with the results for
the non-isolated photon production in the $ep$ collision. 
The presented differential cross sections $d\sigma /d\eta^{\gamma}$ are 
obtained for $5 \le E_T^{\gamma} \le 10$ GeV using 
various isolation cone parameters, $\epsilon$ and $R$.
The isolation cut suppresses the cross section by above 10\% in the whole 
range of the photon rapidities shown in Figs.~\ref{Ffig7a}, 
\ref{Ffig7b}. For $\epsilon$=0.1 and $R=1$
the suppression is 17-23\% at rapidities $-1.5<\eta^{\gamma}\le 4$.
This large effect is not very sensitive to the value of $\epsilon$:
changing the value by a factor of 2 from $\epsilon =0.1$ to
$\epsilon =0.2$ or to $\epsilon =0.05$ varies the results for isolated photon
by about $4\%$ (Fig.~\ref{Ffig7a}).
The dependence on $R$ is stronger but also not very large: 
when the
$R$ value is changed by a factor of 2 (from 1 to 0.5) 
the results increase by about $7\%$
(Fig.~\ref{Ffig7b}).

The suppression due to the isolation imposed on the photon is seen
in Tab.~\ref{tab2}, for the individual contributions and for the
sum of all contribution. 
The results for both the non-isolated and isolated photons 
are shown in the first and the second row, respectively.
As expected, the cross section for processes 
with fragmentation into final photon is strongly suppressed: 
due to the isolation it is lowered by a factor of 5. 
At the same time the $\mathcal{O}$$(\alpha_S)$ corrections to the Born 
diagram increase
from 4.8 pb for the non-isolated $\gamma$ to 13.1pb for the isolated $\gamma$, 
i.e. the contribution of this corrections to the subtraction
cross section, $d\sigma_{sub}$, is negative.
The requirement of isolation does not modify contributions of other 
processes since they are of $2\ra 2$ type and
involve only the direct final photons.
The subtraction cross section, being a sum of negative 
$\mathcal{O}$$(\alpha_S)$ corrections
and fragmentation contributions, is of course positive. Finally, the
total cross section for isolated final photon is lower, by $20\%$,
than for the non-isolated one.

In further calculations discussed in this work
the values $R=1$ and $\epsilon =0.1$ are assumed,
as in analyses of the H1 and ZEUS 
Collaborations~\cite{Breitweg:1997pa}-\cite{unknown:2004uv}.

%%%%%%%%%%%%%%%%%%%%%%%%%%%%%%%%%%%%%%%%%%%%%%%%%%%%%%%%%%%%%%%%%%%%%%%
\subsection{Effects of other  cuts}\label{results1b} 
%%%%%%%%%%%%%%%%%%%%%%%%%%%%%%%%%%%%%%%%%%%%%%%%%%%%%%%%%%%%%%%%%%%%%%%
%\PP

\begin{figure}[t]  
\vskip 25cm\relax\noindent\hskip -2cm
       \relax{\includegraphics{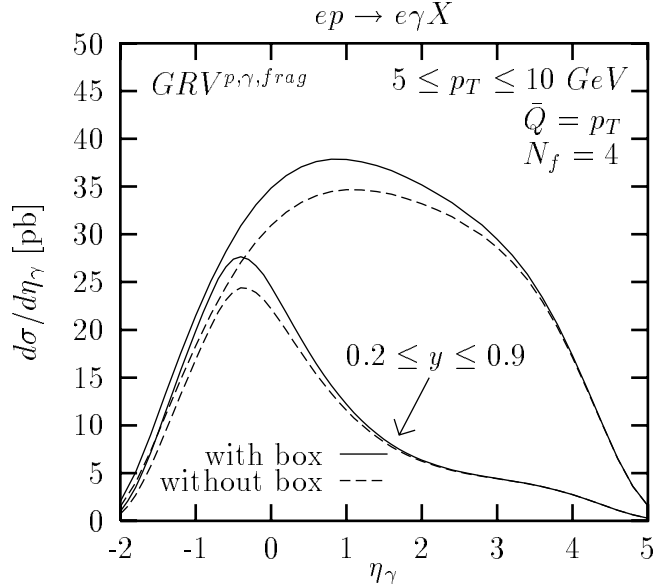}}
\vspace{-17.5cm}
\caption{\small\sl The differential cross section $d\sigma/d\eta^{\gamma}$ 
for the photoproduction of the isolated $\gamma$ ($\epsilon = 0.1$, $R = 1$) 
at HERA as a function of the photon rapidity $\eta^{\gamma}$ 
with (solid lines) and without (dashed lines) the box contribution. 
The results are obtained with imposed $y$-cut ($0.2\leq y\leq 0.9$)
and without this cut. The photon transverse energy,
$E_T^{\gamma}\equiv p_T$, is assumed in the range
$5 \le E_T^{\gamma} \le 10$ GeV.}
\label{Ffig8}
\end{figure}

In order to compare the results obtained by us with the data we take into 
account the isolation restrictions, as well as other kinematic
cuts imposed in experimental analyses of the prompt photon events 
at the HERA collider. In this section we 
%\PP 
study effects of cuts applied 
in the analysis of the ZEUS Collaboration~\cite{Breitweg:1999su}, where
the initial $\gamma$ fractional momentum and the final $\gamma$
transverse energy and rapidity are required to be in the range:
$0.2\le y\le 0.9$, $5 \le E_T^{\gamma} \le 10$ GeV and
$-0.7\leq\eta^{\gamma}\leq 0.9$, respectively.
The influence of the limited $y$ range is shown in Fig.~\ref{Ffig8}.
The cross section (solid line) is strongly reduced, 
by 30-85\%, in the positive rapidity 
region. At negative rapidities the change due to the $y$-cut is weaker: 5-10\% 
at $-1.2<\eta^{\gamma}\le -0.4$ and 10-30\% at other negative rapidities.
Separately we show the results obtained without including the box 
subprocess (dashed line). The box diagram 
contributes mainly in the rapidity region between -1 and 3. After 
imposing the $y$-cut it is important in the narrower region, 
from -1 to 1. The influence of the $y$-cut can be 
read also from the third row in Tab.~\ref{tab2}:
the Born contribution is reduced 3.5 times in comparison with
the results obtained without the $y$-cut (the second row), 
while other contributions are suppressed less, roughly by a factor of 2.

\begin{figure}[t]  
\vskip 25cm\relax\noindent\hskip -2cm
       \relax{\includegraphics{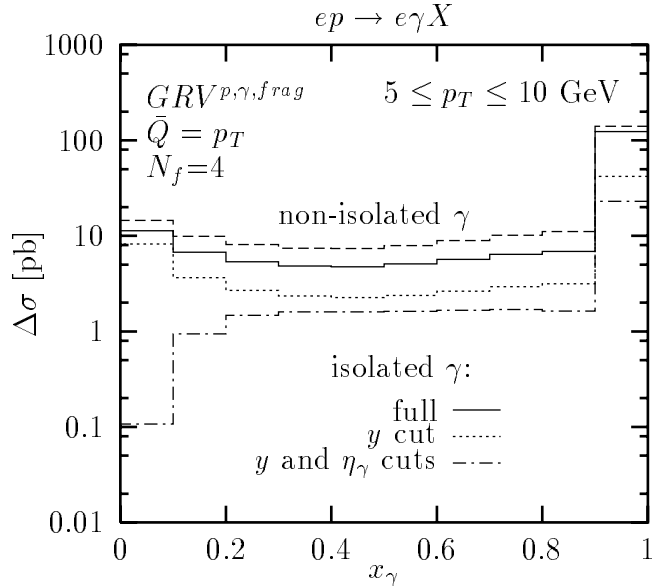}}
\vspace{-17.5cm}
\caption{\small\sl The cross section in $x_{\gamma}$ bins of the length 0.1
for the photon transverse energy in the range $5 \le E_T^{\gamma} \le 10$ GeV
($E_T^{\gamma}\equiv p_T$).
The results for the non-isolated $\gamma$ integrated over the whole range 
of $y$ and $\eta^{\gamma}$ are shown with the dashed line. 
The solid line represents results integrated over the whole 
range of $y$ and $\eta^{\gamma}$ for the isolated $\gamma$ 
with $\epsilon = 0.1$ and $R = 1$.
Results with additional cuts in the
isolated $\gamma$ cross section are shown with: dotted line
($0.2\leq y\leq 0.9$) and dashed-dotted line ($0.2\leq y\leq 0.9$,
$-0.7\leq\eta^{\gamma}\leq 0.9$).}
\label{Ffig9}
\end{figure}

The results obtained for the isolated photon with the $y$-cut and in addition 
with the cut on the final photon rapidity, $-0.7\le\eta^{\gamma}\le 0.9$, 
are presented in the last row of Tab.~\ref{tab2}. The restriction on 
$\eta^{\gamma}$ decreases the contributions of all subprocesses
approximately by a factor of 2, except for the contribution of
processes with resolved initial $\gamma$ and fragmentation into final $\gamma$
which is reduced almost 3 times.

The influence of various 
%\PP 
cuts on the $x_{\gamma}$-distribution is illustrated in Fig.~\ref{Ffig9}.
This distribution is scheme-dependent, nevertheless it is important 
to study the resolved photon contributions in the chosen renormalization
scheme. Such an analysis was performed for the first time in our
paper \cite{Krawczyk:2001tz}.
Note that $x_{\gamma}<1$ and $x_{\gamma}=1$ correspond to processes with 
the resolved or direct initial photon, respectively. So, the cross section
in the last bin 
in Fig.~\ref{Ffig9}, $0.9<x_{\gamma}\le 1$, contain the contributions of
both types of processes, while the cross sections in other bins,
$x_{\gamma}\le 0.9$, consists of the resolved photon contribution alone.
In particular we see in Fig.~\ref{Ffig9}
that the isolation and the energy cut reduce considerably
the cross section at large and medium $x_{\gamma}$, while the cross section
at $x_{\gamma}$ below 0.1
is reduced less. On the other hand, the small $x_{\gamma}$
contributions are strongly, by two orders of magnitude, 
diminished by the photon rapidity cut.
This shows that measurements in the central $\eta^{\gamma}$ region 
($-0.7\le\eta^{\gamma}\le 0.9$)
are not too sensitive to the parton densities in the photon
at small $x_{\gamma}$ region.

%%%%%%%%%%%%%%%%%%%%%%%%%%%%%%%%%%%%%%%%%%%%%%%%%%%%%%%%%%%%%%%%%%%%%%%
\subsection{Dependence on the choice of the 
renormalization scale}\label{results3} 
%%%%%%%%%%%%%%%%%%%%%%%%%%%%%%%%%%%%%%%%%%%%%%%%%%%%%%%%%%%%%%%%%%%%%%%

\begin{figure}[b]  
\vskip 25cm\relax\noindent\hskip -2cm
       \relax{\includegraphics{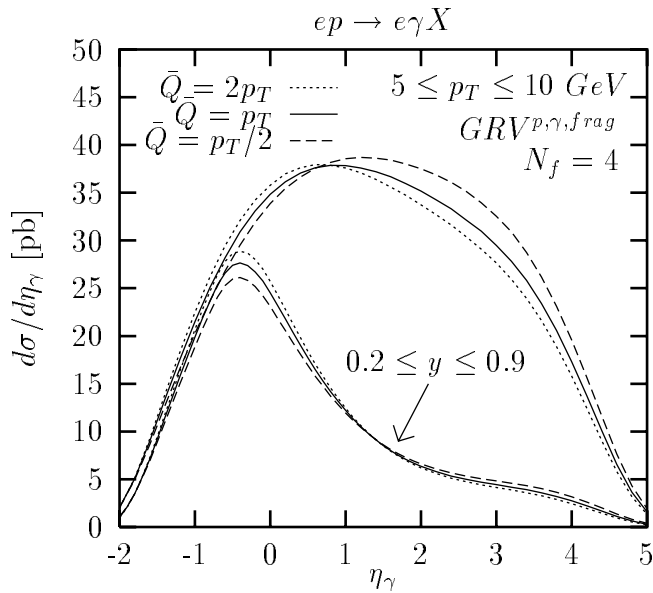}}
%\vskip 7.5cm\relax\noindent\hskip -2cm
%       \relax{\special{psfile=fig13b.ps}}
\vspace{-17.5cm}
\caption{\small\sl The cross section $d\sigma/d\eta^{\gamma}$ for
$5\le E_T^{\gamma}\equiv p_T \le 10$ GeV and 
$\mu\equiv\bar{Q} = E_T^{\gamma}/2$ (dashed line), $E_T^{\gamma}$ (solid line)
and $2\cdot E_T^{\gamma}$ (dotted line). The results obtained without and
with the $y$-cut, $0.2\leq y\leq 0.9$, are shown.}
\label{Ffig13}
\end{figure}

In order to estimate the contribution due to missing higher order 
terms, the influence of the choice of the renormalization/factorization
scale $\mu$ 
is studied for the $\eta^{\gamma}$ distribution.
In Fig.~\ref{Ffig13} the results obtained
with and without the $y$-cut are shown.
When changing $\mu$ from $E_T^{\gamma}$ to 2$\cdot E_T^{\gamma}$ 
($E_T^{\gamma}$/2) the cross section
increases (decreases) at rapidities below $\sim 1$
and decreases (increases) at higher rapidity values. 
Only at high rapidities,
$\eta^{\gamma} > 3$, the dependence on a choice 
of the scale is strong, above 10\%, however here
the cross section is relatively small.
In the wide kinematic range,  
$-2 < \eta^{\gamma} < 2$, the relative differences between the results
obtained (with and without the $y$-cut)
for $\mu = E_T^{\gamma}$ and  
for $\mu = 2\cdot E_T^{\gamma}$ or $E_T^{\gamma}/2$ 
are small and do not exceed 6\%.
Around the maximum of the cross section at rapidities 
$-1\le\eta^{\gamma}\le 0$ these differences are 4-6\%. 
This small sensitivity of the results to the change of the scale is 
important since it may indicate that the contribution from neglected NNLO
and higher order terms is not significant.
This is important for a comparison 
%\PP   
(see below, and in Sec. 5.5.3 and 6.2.4) with other NLO calculations
using other set of diagrams and, in particular, including the
$\mathcal{O}$$(\alpha_S)$ corrections to the processes with the resolved
photon which are not included in our calculation.

Note that individual contributions are strongly dependent
on the choice of $\mu$, e.g. results for the
processes with the resolved photon vary by
$\pm$10-20\% at rapidities $\eta^{\gamma}\le 1$
while the $\mathcal{O}$$(\alpha_S)$ corrections to the Born
diagram vary by a factor
of a few and in some kinematic ranges they even change sign (not shown).
The results are much more stable when the sum of the resolved photon processes
and the $\mathcal{O}$$(\alpha_S)$ corrections to the Born process is included
(compare the discussion in Sec. \ref{Cnlo:disc}). 

%%%%%%%%%%%%%%%%%%%%%%%%%%%%%%%%%%%%%%%%%%%%%%%%%%%%%%%%%%%%%%%%%%%%%%%
\subsection{Comparison with the ZEUS data}\label{results2}
%%%%%%%%%%%%%%%%%%%%%%%%%%%%%%%%%%%%%%%%%%%%%%%%%%%%%%%%%%%%%%%%%%%%%%%

The results of the first measurements of the isolated prompt photon 
cross section (with no jet requirements)
in photoproduction events at the HERA collider
have been published in \cite{Breitweg:1999su}. Below we 
%\PP 
compare of our predictions with these data.

\subsubsection{\boldmath $E_T^{\gamma}$, $N_f$} 

Fig.~\ref{Ffig10}a shows the distribution of the photon transverse energy,
$E_T^{\gamma}$, for various number of active massless quarks, $N_f$.
The predictions for $N_f=4$ and 5 are in agreement with most of the
experimental points but tend to lie slightly below the data.
The difference between predictions for $N_f=4$ and $N_f=5$ is very
small due to the fourth power of electric charge characterizing  
processes with two photons.
The contribution of the charm quark is an order of magnitude larger
than the contribution of beauty and the predictions for $N_f=3$
are significantly below the predictions for $N_f=4$ and below the data,
confirming the need to include the charm contribution in the calculations.

\subsubsection{\boldmath $\eta^{\gamma}$, $N_f$} 

\begin{figure}[t]  
\vskip 25cm\relax\noindent\hskip -2cm
       \relax{\includegraphics{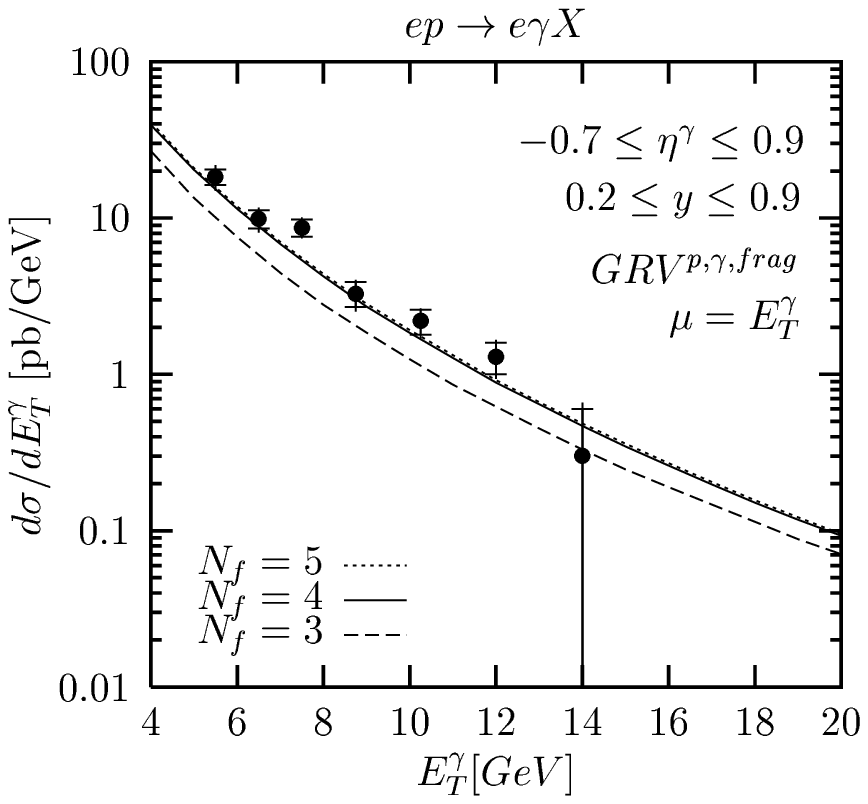}}
\vskip 7.8cm\relax\noindent\hskip -2cm
%       \relax{\special{psfile=fig10b.ps}}
       \relax{\includegraphics{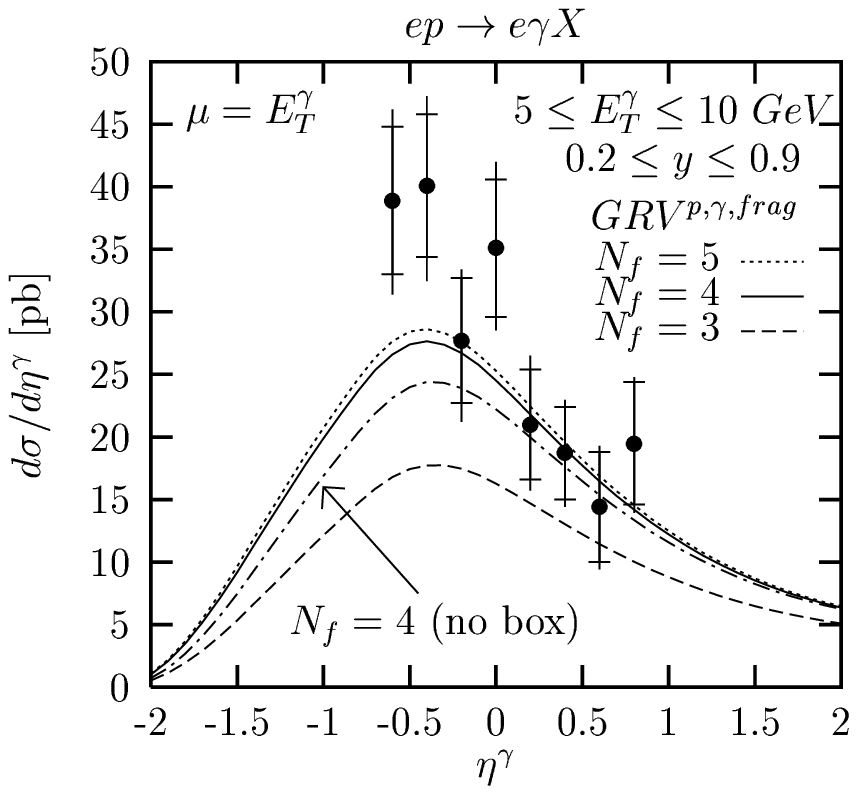}}
\vspace{-17.2cm}
\caption{\small\sl The cross section $d\sigma/dE_T^{\gamma}$ (a) and
$d\sigma/d\eta^{\gamma}$ (b) 
with various numbers of active massless flavors:
$N_f$ = 3 (dashed lines), 4 (solid lines) and 5 (dotted lines),
compared to the ZEUS data~\cite{Breitweg:1999su}.
The result without the box contribution
is also shown for $N_f$ = 4 (dot-dashed line)}
\label{Ffig10}
\end{figure}

A similar comparison between the predictions and the data, now for the 
distribution of the photon rapidity, $\eta^{\gamma}$,  
is shown in Fig.~\ref{Ffig10}b. A good description of the 
data is obtained for $N_f$=4, 5 in the rapidity region 
$0.1\le\eta^{\gamma}\le 0.9$. For $-0.7\le\eta^{\gamma}\le 0.1$ our 
predictions lie mostly below the experimental points. 
Note that this 
disagreement between predicted and measured cross sections is also observed 
for calculations of Gordon \cite{Gordon:1998yt} and of
Fontannaz, Guillet and Heinrich \cite{Fontannaz:2001ek}
as well as for Monte Carlo 
simulations~\cite{Breitweg:1999su,Bussey2001}.
In order to find the source of such a disagreement various checks have
been performed, discussed below. 

\subsubsection{\boldmath $\eta^{\gamma}$, integrated $\sigma$, 
$\mathcal O$$(\alpha^2\alpha_S^2)$} 

First, we study the size of 
$\mathcal{O}$$(\alpha^2\alpha_S^2)$ contributions included by us.
In Fig.~\ref{Ffig10}b we present separately an effect due to the box 
subprocess (for $N_f=4$). One can see that the box term enhances considerably 
the cross section in the measured rapidity region. Its contribution to the 
integrated cross section is equal to 10\%, see Tab. \ref{tab2} (fourth line). 
The processes with resolved photon and fragmentation also give a
no-negligible contribution, although roughly two times smaller than the 
box one (Tab. \ref{tab2}, fourth line).
We conclude, that taking into account of these 
$\mathcal{O}$$(\alpha^2\alpha_s^2)$ contributions improves the description 
of the data.

\subsubsection{\boldmath $\eta^{\gamma}$, $f_{\gamma}$, $\mu$}

\begin{figure}[ht]  
\vskip 25cm\relax\noindent\hskip -2cm
       \relax{\includegraphics{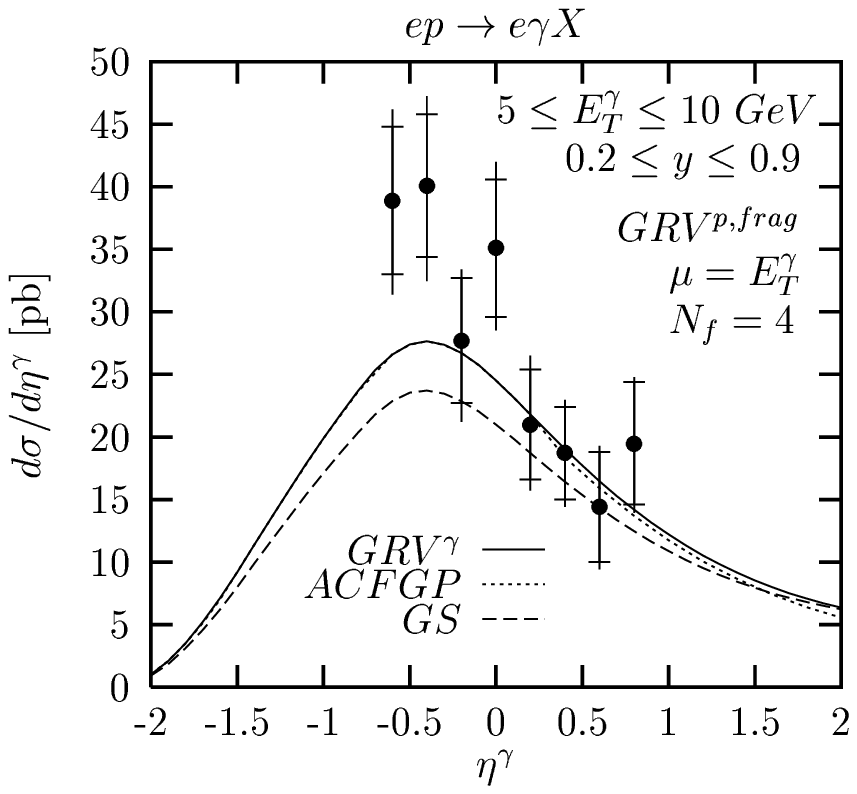}}
\vskip 8cm\relax\noindent\hskip -2cm
       \relax{\includegraphics{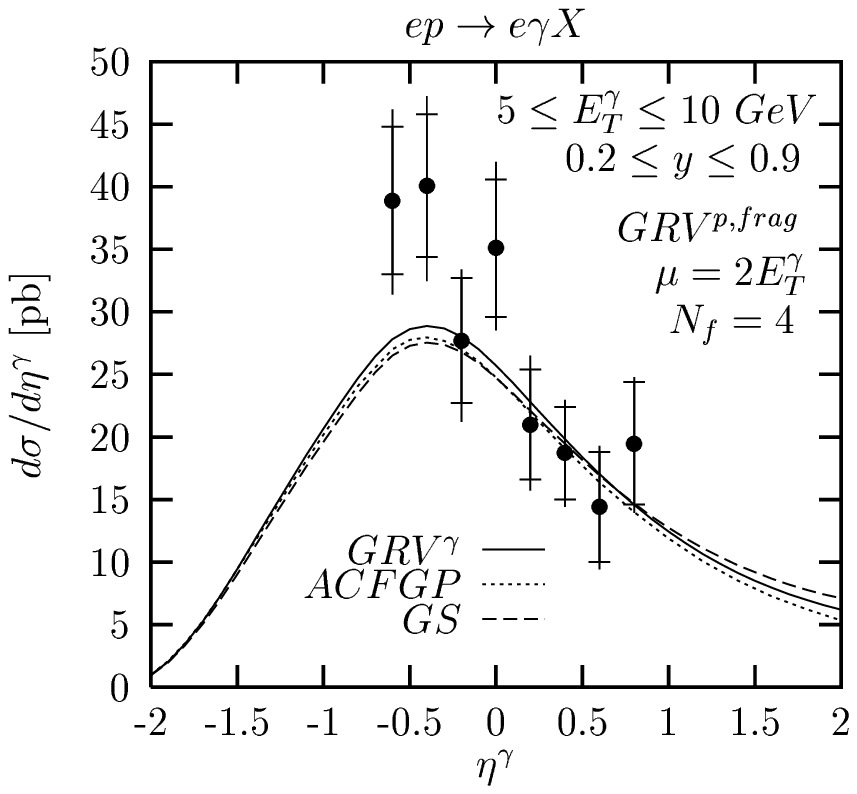}}
\vspace{-17.cm}
\caption{\small\sl The cross section $d\sigma /d\eta^{\gamma}$ in the
range $5<E_T^{\gamma}<10$ GeV for the 
renormalization/factorization scale $\mu = E_T^{\gamma}$ (a)
and $\mu = 2\cdot E_T^{\gamma}$ (b). The 
GRV$^{\gamma}$ \cite{Gluck:1992ee} (solid line), ACFGP$^{\gamma}$ 
\cite{Aurenche:1992sb} (dotted line) and GS$^{\gamma}$
\cite{Gordon:1997pm} (dashed line) parton densities in the photon were
used together with GRV$^p$ densities in the proton \cite{Gluck:1995uf}
and GRV$^{frag}$ fragmentation \cite{Gluck:1993zx}.}
\label{Ffig11}
\end{figure}

Next, we study the dependence on the parton densities.
The predictions obtained using three different NLO parton densities in the 
photon (ACFGP$^{\gamma}$~\cite{Aurenche:1992sb}, 
GRV$^{\gamma}$~\cite{Gluck:1992ee} and 
GS$^{\gamma}$~\cite{Gordon:1997pm}) are presented for $N_f=4$
in Fig.~\ref{Ffig11}a ($\mu = E_T^{\gamma}$) and in Fig.~\ref{Ffig11}b 
($\mu = 2 E_T^{\gamma}$) together with the ZEUS data~\cite{Breitweg:1999su}.
The results based on ACFGP$^{\gamma}$ and GRV$^{\gamma}$ 
parametrizations differ by less than 4\%
at rapidities $\eta^{\gamma} < 1$ (at higher $\eta^{\gamma}$ the differences 
are bigger), and both give a good description of the 
data in the rapidity range
$0.1\le\eta^{\gamma}\le 0.9$ 
(for $\mu = E_T^{\gamma}$ and $\mu = 2 E_T^{\gamma}$). 
For $-0.7\le\eta^{\gamma}\le 0.1$ none of the predictions is in
agreement with the measured cross section.

\begin{figure}[ht]  
\vskip 15.5cm\relax\noindent\hskip 0.cm
       \relax{\includegraphics{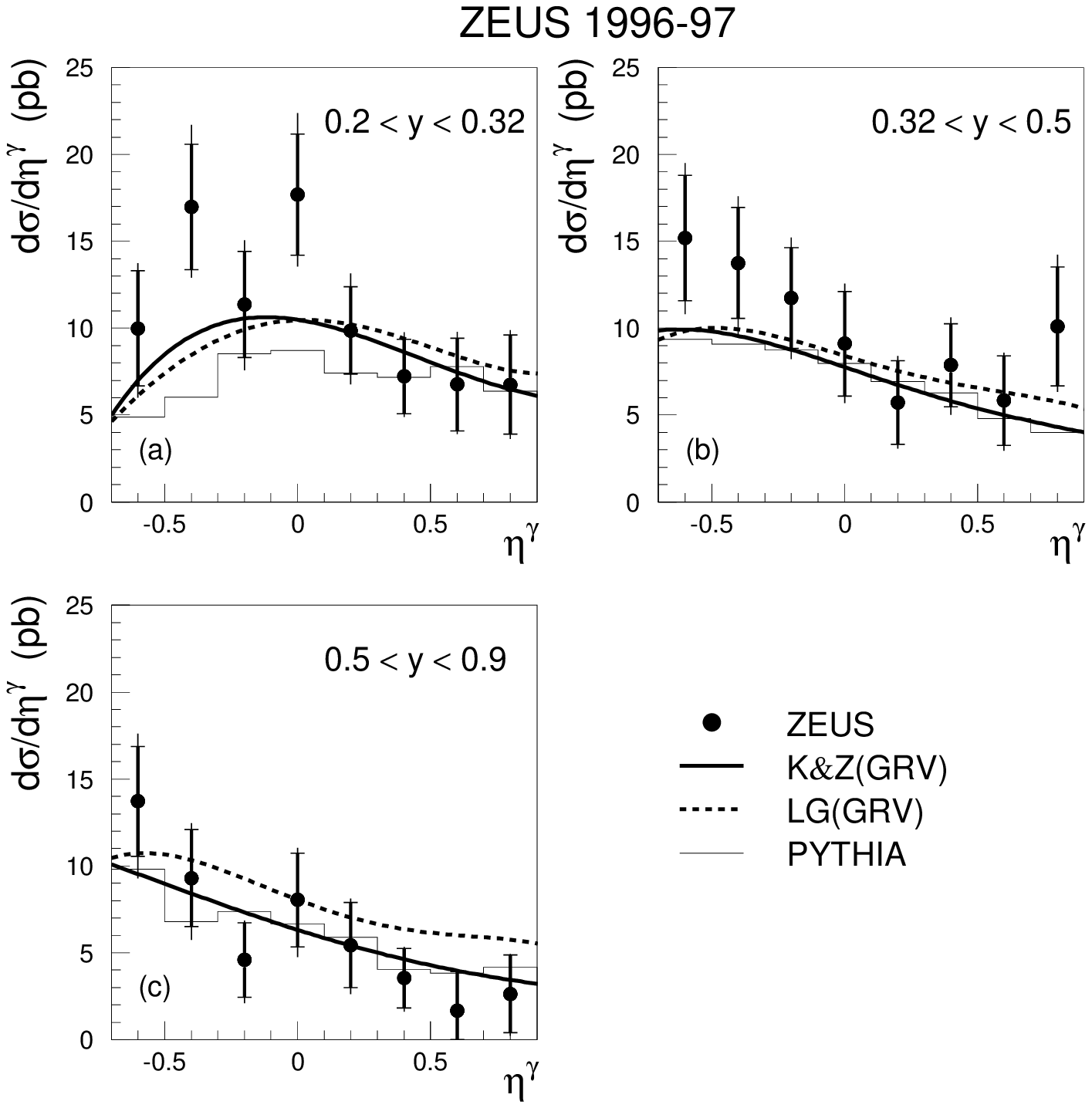}}
\vspace{-1.cm}
\caption{\small\sl The cross section $d\sigma /d\eta^{\gamma}$ for three
ranges of $y$: $0.2<y<0.32$ (a), $0.32<y<0.5$ (b) and $0.5<y<0.9$ (c).
The predictions of K\&Z \cite{Krawczyk:2001tz} and
LG \cite{Gordon:1998yt}
are compared with the ZEUS Collaboration data \cite{Breitweg:1999su}.
The figure is taken from \cite{Breitweg:1999su}.}
\label{Fslide.3eta}
\end{figure}

For $\mu$=$E_T^{\gamma}$ (Fig.~\ref{Ffig11}a) the GS$^{\gamma}$ 
distribution leads to 
results considerably below ones obtained using ACFGP$^{\gamma}$ and 
GRV$^{\gamma}$ densities,
especially in the rapidity region from roughly -1 to~1, where the
differences are up to 14\%.
This difference between the GS$^{\gamma}$ and other 
considered herein parton parametrizations is 
mainly due to their different treatment of the 
charm quark in the photon, namely
in the GS$^{\gamma}$ approach the charm quark is absent
for $\mu^2$ below 50 GeV$^2$.
Since we take $5\leq \mu = E_T^{\gamma}\leq 10$ GeV, and the most important 
contribution to the cross section arises from the lower $E_T^{\gamma}$ region 
(see Fig.~\ref{Ffig10}a), the $\mu^2$ value 
usually lies below the GS$^{\gamma}$ charm quark threshold.
As a consequence, the 
predictions based on GS$^{\gamma}$ have strongly suppressed 
contribution
of subprocesses involving charm from the photon - contrary
to GRV$^{\gamma}$ and ACFGP$^{\gamma}$ 
predictions where the charm threshold is assumed at lower $\mu^2$.

All the considered parton parametrizations give similar description of the data
when the scale is changed to $\mu=2E_T^{\gamma}$, see Fig.~\ref{Ffig11}b. 
Here the calculation corresponds to $\mu^2$
which is always above 50 GeV$^2$ and 
the charm density in the GS$^{\gamma}$ parametrization is non-zero,
as in other parametrizations.

\subsubsection{\boldmath $\eta^{\gamma}$, $f_p$, $D_{\gamma}$}

We have also studied the dependence of our results on a choice of 
the parton densities in the proton and parton fragmentation
into the photon (not shown). 
The MRST1998$^p$ (set ft08a)~\cite{Martin:1998sq} and 
CTEQ4M$^p$~\cite{Lai:1996mg} NLO densities in the proton
give predictions lower than GRV$^p$ \cite{Gluck:1995uf} NLO densities 
by 4-7\% and 4-6\% at negative rapidities.
At positive rapidities the cross sections calculated using 
MRST1998$^p$, CTEQ4M$^p$ and GRV$^p$
vary among one another by less than 4\%. 
Results for the isolated photon production are not too sensitive to the
parton-to-photon fragmentation function. 
For rapidity ranging from -1 to 4 the cross section
obtained with DO$^{frag}$(LO)~\cite{Duke:1982bj} 
fragmentation function is $2-4\%$ lower than
the cross section based on GRV$^{frag}$(NLO)~\cite{Gluck:1993zx} 
parametrization. 
Only at minimal ($\eta^{\gamma}<-1$)
and maximal ($4<\eta^{\gamma}$) rapidity values
this difference is larger, being at a level of $4-8\%$.
All these parton densities in the proton and fragmentation functions lead
to similar descriptions of the data.

%\subsubsection{ Relative energy bins}
\subsubsection{\boldmath $\eta^{\gamma}$, $y$}

In Fig.~\ref{Fslide.3eta} our predictions are compared to the ZEUS data
divided into three ranges of the initial photon fractional momentum $y$. 
%\PP relative energy of initial photons.. 
This allows to establish that
the above discussed discrepancy between the data and the predictions for 
$\eta^{\gamma} < 0.1$ is coming mainly from the low, $0.2<y<0.32$,
and medium, $3.2<y<0.5$, $y$ region.
In the high $y$ region, $0.5<y<0.9$, a good agreement is obtained.

\subsubsection{summary of comparison with data}

To summarize, there are discrepancies between the predictions and the
ZEUS data \cite{Breitweg:1999su} at $\eta^{\gamma}<0.1$ and $y<0.5$.
The variation of theoretical parameters and parton densities
do not improve the description of the data. Note that there are
also discrepancies between predictions and the new H1 data 
for the prompt photon production at HERA \cite{unknown:2004uv} 
(Chapter \ref{isol}). So far the source of these discrepancies is not 
established.

Finally, it is worth mention that in Ref. \cite{SungWonLee:thesis}
there was performed an analysis, in which our computer program
for the prompt photon production at HERA \cite{Krawczyk:2001tz}
was modified: the parton densities in the photon for $x_{\gamma}>0.8$
were multiplied by a factor of 2 or 3. For doubled $f_{\gamma}$
at $x_{\gamma}>0.8$ the description of the data was much improved,
and for tripled $f_{\gamma}$ at $x_{\gamma}>0.8$ a good agreement with
the ZEUS data was found in the whole range of $\eta^{\gamma}$. 
This agreement is probably accidental, but it can also indicate 
a need of improvement of the parton densities in the photon 
at $x_{\gamma}$ close to 1 \cite{SungWonLee:thesis}.

%%%%%%%%%%%%%%%%%%%%%%%%%%%%%%%%%%%%%%%%%%%%%%%%%%%%%%%%%%%%%%%%%%%%%%%
\subsection{Comparison with other QCD predictions (LG)}\label{results5}
%%%%%%%%%%%%%%%%%%%%%%%%%%%%%%%%%%%%%%%%%%%%%%%%%%%%%%%%%%%%%%%%%%%%%%%

The calculation of L.~Gordon (LG) \cite{Gordon:1998yt} 
for the photoproduction of isolated photons at HERA contains the
$\mathcal{O}$$(\alpha_S)$ corrections to the
processes with the resolved initial photon, which are not included in our
calculation (Chapter \ref{nlo}).
On the other hand, we include the box diagram (Fig. \ref{figbox})
neglected
in \cite{Gordon:1998yt}. The $\mathcal{O}$$(\alpha_S^2)$ diagrams
with resolved initial $\gamma$ and fragmentation into final $\gamma$,
as well as other diagrams shown in Figs. \ref{figborn}-\ref{fig23} and 
\ref{figsingi}-\ref{figdoub2},
are taken into account in both calculation.

The LG calculation applies the MRST$^p$(NLO) \cite{Martin:1998sq}
parton distributions in the proton which give predictions 4\% lower
on average than GRV$^p$(NLO) \cite{Gluck:1995uf} para\-metri\-zation
used by us (Sec. \ref{results2}). Both calculations use
GRV$^{\gamma}$ \cite{Gluck:1992ee} and GS$^{\gamma}$ \cite{Gordon:1997pm}
parton densities in the photon. Since the GS$^{\gamma}$ densities
lead to worse description of the data than the GRV$^{\gamma}$ ones
(see Sec.~\ref{results2}), below we discuss the 
predictions obtained using the GRV$^{\gamma}$ para\-metri\-zation.

\begin{figure}[th]  
\vskip 8.5cm\relax\noindent\hskip 3.cm
       \relax{\includegraphics{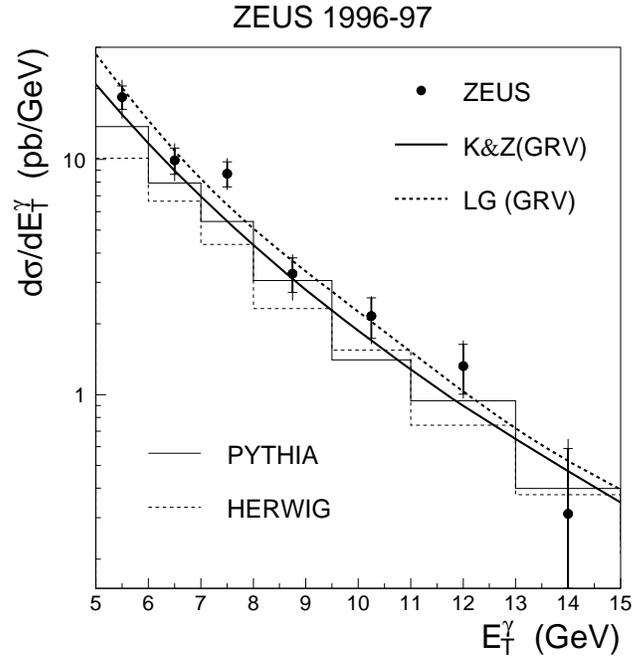}}
\vspace{-0.2cm}
\caption{\small\sl The cross section $d\sigma /dE_T^{\gamma}$
for the photoproduction of isolated photons in the kinematic range
$0.2<y<0.9$ and $-0.7\leq\eta^{\gamma}\leq 0.9$.
The predictions of K\&Z \cite{Krawczyk:2001tz} and
LG \cite{Gordon:1998yt} using GRV$^{\gamma}$ \cite{Gluck:1992ee}
parton distributions in the photon, are compared with 
the ZEUS Collaboration data \cite{Breitweg:1999su}.
The Monte Carlo simulations are also shown.
The figure is taken from \cite{Breitweg:1999su}.}
\label{Fslide.dpt}
\end{figure}

\begin{figure}[ht]  
\vskip 16.cm\relax\noindent\hskip -0.cm
       \relax{\includegraphics{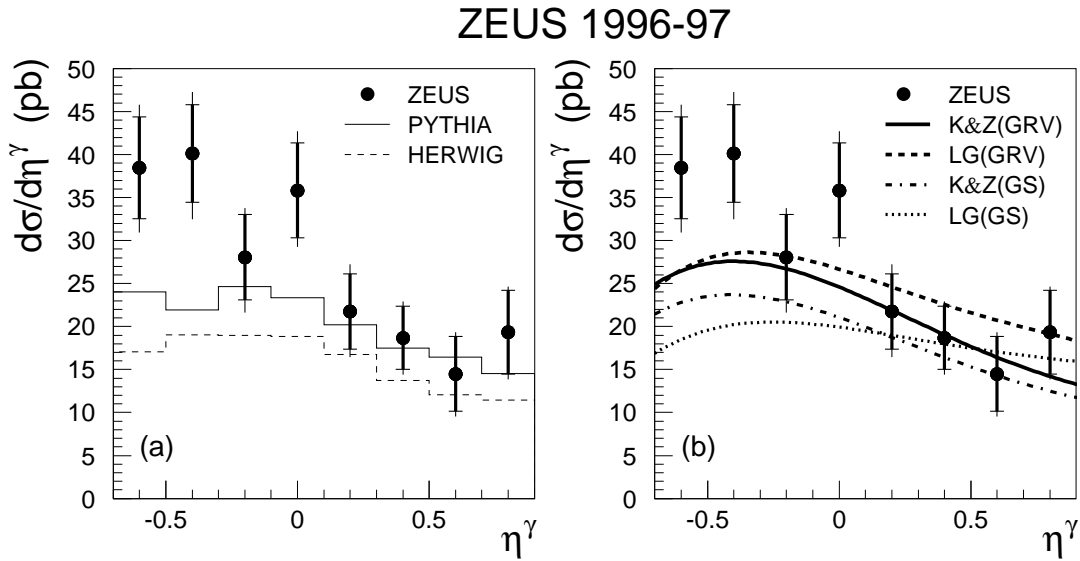}}
\vspace{-8.cm}
\caption{\small\sl As in Fig. \ref{Fslide.dpt} for the 
$d\sigma /d\eta^{\gamma}$ cross section in the range 
$5< E_T^{\gamma}<10$ GeV. In addition, the predictions
of K\&Z \cite{Krawczyk:2001tz} and LG \cite{Gordon:1998yt} 
using GS$^{\gamma}$ \cite{Gordon:1997pm} parton densities in the photon
are shown. The figure is taken from \cite{Breitweg:1999su}.}
\label{Fslide.eta}
\end{figure}

The comparison between the predictions of
Krawczyk and Zembrzuski (K\&Z) \cite{Krawczyk:2001tz}, predictions
of Gordon (LG) \cite{Gordon:1998yt} 
%(both with the GRV$^{\gamma}$ parametrization)
and the 
%\PP 
data was performed by the ZEUS Collaboration \cite{Breitweg:1999su}.
Results of this analysis 
are shown in Figs. \ref{Fslide.3eta}, \ref{Fslide.dpt} and \ref{Fslide.eta}
\footnote{In the comparison between LG \cite{Gordon:1998yt}
and K\&Z \cite{Krawczyk:2001tz} we consider the predictions 
presented in the ZEUS paper \cite{Breitweg:1999su}, because the 
numerical results presented in the original paper \cite{Gordon:1998yt}
are given for different kinematic ranges.}.

The LG predictions for $d\sigma /dE_T^{\gamma}$ cross section are 
higher than ours in the whole presented range of transverse 
momentum, $5\le E_T^{\gamma}\le 15$ GeV (Fig. \ref{Fslide.dpt}). 
Both calculations agree with the data within large experimental
errors, although the K\&Z predictions tend to lie slightly below
the data.

For $d\sigma /d\eta^{\gamma}$
the biggest differences are at $\eta^{\gamma}$=0.9 where the LG results 
are about 35\% higher than K\&Z (Fig. \ref{Fslide.eta}b). 
The differences decrease towards negative rapidity 
values and are negligible at $-0.7\le\eta^{\gamma} < -0.5$.
The average difference in the range $-0.7\le\eta^{\gamma} < 0.9$
is about 10\%.
Both predictions are in disagreement with data at $\eta^{\gamma} < 0.1$.
At $\eta^{\gamma} > 0.1$ K\&Z agrees with the measured cross section
but LG tends to lie slightly to high.

For $y$ range limited to low values only, 0.2 $< y <$ 0.32, the LG cross 
section is higher than ours by up to 20\% at positive $\eta^{\gamma}$, 
while at negative $\eta^{\gamma}$ it is lower by up to 10\%.
(Fig. \ref{Fslide.3eta}a). In this $y$ range, as well as for medium
range with 0.32 $< y <$ 0.5 (Fig. \ref{Fslide.3eta}b), 
the K\&Z and LG calculations lead to
similar discrepancies with data at 
$\eta^{\gamma}$ \begin{minipage}[t]{10pt} \raisebox{3pt}{$<$} \makebox[-17pt]{}
\raisebox{-3pt}{$\sim$}\end{minipage} 0.1.
For large $y$ values, 0.5 $< y <$ 0.9,
where our predictions agree with data, the LG results are higher than 
ours by up to 70\% at $0.7<\eta^{\gamma}< 0.9$ and does not describe the data 
well (Fig. \ref{Fslide.3eta}c).

%For $\mu=E_T^{\gamma}$ the GS$^{\gamma}$ 
%parton distributions in the photon lead to K\&Z results lower by 
%11-14\% than those obtained with GRV$^{\gamma}$ densities 
%at $\eta^{\gamma}$ between -1 and 1 (see Sec.~\ref{results2}). 
%In the LG calculation \cite{Gordon:1998yt} 
%this difference is even twice larger. The LG predictions
%obtained using the GS$^{\gamma}$ parametrization lie up to 20\%
%below (30\% above) predictions of K\&Z also using GS$^{\gamma}$ 
%in the rapidity range
%$-0.7\le\eta^{\gamma}\le 0.2$ ($0.2\le\eta^{\gamma}\le 0.9$).

To summarize, the differences between both calculations including
different sets of diagrams are of order 10\%. However there are 
some kinematic ranges where the differences are larger
up to 70\% for 0.5 $< y <$ 0.9 and
$0.7<\eta^{\gamma}< 0.9$. 
The experimental errors are to large to conclude
which calculation describe better the ZEUS Collaboration data
presented in \cite{Breitweg:1999su}.
Nevertheless, for large $y$ region, 0.5 $< y <$ 0.9, our predictions 
give better description of the data.

There is also another NLO calculation for the photoproduction of
isolated photons at HERA presented by
Fontannaz, Guillet and Heinrich (FGH) \cite{Fontannaz:2001ek}.
We compare our predictions with predictions of FGH in Secs. 
\ref{Sisol:fgh} and \ref{fgh}, where we use the same parton distributions 
as FGH do.
Let us only note here, that in the range considered by the ZEUS
group \cite{Breitweg:1999su}, $0.2< y <0.9$ and
$-0.7<\eta^{\gamma}< 0.9$,
the FGH predictions obtained using
MRST1999\hspace{0.2mm}$^{p}$ \cite{Martin:1999ww}, 
AFG$^{\gamma}$ \cite{Aurenche:1994in} and BFG$^{frag}$ \cite{Bourhis:1997yu}
parametrizations are about 10\% higher on average 
than our predictions obtained
using GRV$^{p,\gamma,frag}$ \cite{Gluck:1995uf,Gluck:1992ee,Gluck:1993zx}
parametrizations. However, for the large $y$ region, 0.5 $< y <$ 0.9, and
for $\eta^{\gamma}$ in the narrowed range $0.7<\eta^{\gamma}< 0.9$
the FGH predictions are about 50\% higher.

The dependence of our predictions on the 
renormalization/factorization scale $\mu$ is about
$\pm 5\%$ in the kinematic range considered by the ZEUS Collaboration
\cite{Breitweg:1999su} and it is true also for 0.5 $< y <$ 0.9.
So, we could expect that the size
of higher order contributions is not large
%\PP 
in this particular region of $y$ as well.
However we found that 
 the difference between K\&Z and LG (FGH) is up to 
70\% (50\%)
for $0.5< y <0.9$ and $0.7<\eta^{\gamma} < 0.9$.
To understand the origin of this large difference we should know
the dependence of the LG and FGH predictions on the scale $\mu$
in this particular range,
which however was not presented in \cite{Gordon:1998yt}
and \cite{Fontannaz:2001ek}
\footnote{The dependence on the choice of the scale $\mu$ was studied in
\cite{Gordon:1998yt,Fontannaz:2001ek} for the \underline{total} cross sections,
i.e. cross sections integrated within considered therein kinematic
ranges. The total cross section of FGH for the prompt photon
production vary by about $\pm 4$\%. 
The total cross section of LG for the prompt photon \underline{plus jet}
production vary by $\pm 3$\%.}.
If this dependence was large, it could be a confirmation of our
supposition that the $\mathcal{O}$$(\alpha_S)$ corrections to the
processes with resolved photons should be included 
\underline{together} with the $\mathcal{O}$$(\alpha_S^2)$ corrections
to the Born process in next-to-next-to-leading order
(NNLO) accuracy and, that they should not
be included in NLO to avoid possible uncertainties due to the choice
of $\mu$.

%%%%%%%%%%%%%%%%%%%%%%%%%%%%%%%%%%%%%%%%%%%%%%%%%%%%%%%%%%%%%%%%%%%%%%%
\chapter{Isolated photon production. Exact implementation of isolation
cuts}\label{isol}
%%%%%%%%%%%%%%%%%%%%%%%%%%%%%%%%%%%%%%%%%%%%%%%%%%%%%%%%%%%%%%%%%%%%%%%

In the small cone approximation, discussed in the previous section,
one obtains simple analytical formulae which are
easy to implement in numerical calculations for
the isolated photon production. The accuracy of this approximation
is good~\cite{Gordon:1994ut,Gordon:1995km,Gordon:1996ug,Gordon:1998yt}, 
however 
the method has one drawback: it does not allow to obtain any
predictions for the rapidity or transverse energy of the jet (or jets)
produced together with the photon. This is because the formulae 
(\ref{small1}-\ref{small4}) and (\ref{eq:non:k},\ref{eq:non:kp})
used in the small cone approximation are 
integrated over all allowed momenta of final particles other than the photon.

In order to compare predictions with the existing data 
not only for the photon observed in the final state but also
for the $ep\ra e\gamma ~jet ~X$ process,
there is a need of more general technique allowing to consider
both the photon and the jet momenta in any kinematic range for which
the data are taken. An introduction of such a technique, based on the 
division ({\sl slicing}) of the three-body phase space
\cite{Ellis:1980wv}-\cite{Klasen:1997br}, 
\cite{Fontannaz:2001ek,Gordon:1998yt,Fontannaz:2001nq}
is presented below  and was discussed in the paper of 
Zembrzuski and Krawczyk \cite{Zembrzuski:2003nu}.
Note, that our method of the division of the phase space 
\cite{Zembrzuski:2003nu} slightly differs from methods applied in other papers.

The use of the {\sl phase space slicing} allows to calculate the cross section
in an exact way, i.e. with no assumption that the isolation cone is small.
We take some cross section and matrix elements formulae from the literature and
derive all the missing expressions, see Secs. \ref{Sisol:x}-\ref{isol:a}
and Appendices \ref{lox}-\ref{xsec:}.
The comparison between exact predictions for the 
$ep\ra e\gamma X$ cross section
and the predictions based on the small cone approximation,
as well as the comparison with the new H1 data \cite{unknown:2004uv}
for the $ep\ra e\gamma X$ reaction,
are presented in Sec.~\ref{pssres}.

The slicing of the phase space is applied to obtain predictions for the
$ep\ra e\gamma ~jet ~X$ cross section, and comparing them with
corresponding H1 data \cite{unknown:2004uv}, in 
Chapter~\ref{jet}.

~ \\ ~ \\ ~

%%%%%%%%%%%%%%%%%%%%%%%%%%%%%%%%%%%%%%%%%%%%%%%%%%%%%%%%%%%%%%%%%%%%%%%
\section{Cross section formulae}\label{Sisol:x}
%%%%%%%%%%%%%%%%%%%%%%%%%%%%%%%%%%%%%%%%%%%%%%%%%%%%%%%%%%%%%%%%%%%%%%%

The general formula for the isolated photon (or isolated photon + jet)
cross section is obtained from Eqs. (\ref{cross1}-\ref{cross5})
integrated over the whole range of $x$ and $x_{\gamma}$ and over $z$
within the region given by Eq. (\ref{zeps}), namely:
\bea
d\sigma^{\gamma p\ra \gamma (jet) X}=
\sum_{q,\bar{q}} \int\limits_0^1 dx f_{q/p}(x,\mu^2) d\sigma^{\gamma q\ra\gamma q} +
\int\limits_0^1 dx f_{g/p}(x,\mu^2) d\sigma^{\gamma g\ra\gamma g}
%\nonumber 
\label{cross1i}
\\
+ \sum_{a=q,\bar{q},g} \int\limits_0^1 dx_{\gamma}
\sum_{b=q,\bar{q},g} \int\limits_0^1 dx 
f_{a/\gamma}(x_{\gamma},\mu^2)
f_{b/p}(x,\mu^2) 
d\sigma^{ab\ra\gamma d}
%\nonumber 
\label{cross2i}
\\
+ \sum_{b=q,\bar{q},g} \int\limits_0^1 dx 
\sum_{c=q,\bar{q},g} \int\limits_{1/1+\epsilon}^1 {dz\over z^2}
f_{b/p}(x,\mu^2) 
D_{\gamma /c}(z,\mu^2)
d\sigma^{\gamma b\ra cd}
%\nonumber 
\label{cross3i}
\\
+ \sum_{a=q,\bar{q},g} \int\limits_0^1 dx_{\gamma}
\sum_{b=q,\bar{q},g} \int\limits_0^1 dx 
\sum_{c=q,\bar{q},g} \int\limits_{1/1+\epsilon}^1 {dz\over z^2}
f_{a/\gamma}(x_{\gamma},\mu^2)
f_{b/p}(x,\mu^2) 
\nonumber 
\\ \cdot
D_{\gamma /c}(z,\mu^2)
d\sigma^{ab\ra cd} \raisebox{5mm}{} \raisebox{-5mm}{}
%\makebox[0.5cm]{}
\label{cross4i}
\\
+ \sum_{q,\bar{q}} 
\int\limits_0^1 dx \left\{ f_{q/p}(x,\mu^2) 
\left[ d\sigma^{\gamma q\ra\gamma q}_{\alpha_S}
+d\sigma^{\gamma q\ra\gamma qg}_{\alpha_S}\right]
+ f_{g/p}(x,\mu^2) d\sigma^{\gamma g\ra\gamma q\bar{q}}_{\alpha_S} \right\},
\label{cross5i}
\eea
where the partonic cross sections in (\ref{cross1i}-\ref{cross4i})
are given by Eqs. (\ref{eq.lox}-\ref{Combridge}).
Note, that these partonic cross sections are not integrated over
the photon momentum, so they allow to obtain various types of
differential cross sections (e.g. $d\sigma /d\eta^{\gamma}$ or
$d\sigma /dE_{T}^{\gamma}$) in any range of the final photon 
or final parton momenta. This is due to the fact that 
in $2\ra 2$ processes the parton momentum 
is determined, in the centre-of-mass system, by the final photon momentum.
The isolation is taken into account simply through the lower limit
of the integration over $z$.

An inclusion of the isolation in calculations of
the $\mathcal{O}$$(\alpha_S)$ corrections (\ref{cross5i})
is more complicated, since one needs to restrict the final momenta 
of two partons, and to take care of a cancellation and factorization 
of singularities.

%%%%%%%%%%%%%%%%%%%%%%%%%%%%%%%%%%%%%%%%%%%%%%%%%%%%%%%%%%%%%%%%%%%%%%%
\section{Division of phase space}\label{pss}
%%%%%%%%%%%%%%%%%%%%%%%%%%%%%%%%%%%%%%%%%%%%%%%%%%%%%%%%%%%%%%%%%%%%%%%

The $\mathcal{O}$$(\alpha_S)$ corrections (\ref{cross5i}) 
include $2\ra 3$ processes
(\ref{23eq}) and the virtual gluon corrections. The squared matrix
elements of these processes contain infrared and collinear singularities
%\PP we collect them in Appendix \ref{xsec:}.
\cite{Aurenche:1984hc,jan}, see Appendix \ref{xsec:}.
In order to obtain numerical predictions for the isolated
photon production 
%\PP
with arbitrary kinematic cuts,
one needs to isolate these singularities,
and cancel or factorize them in such a way, that an integration
over any range of final momenta is still possible
\cite{Fontannaz:2001ek,Gordon:1998yt}.
The standard method to achieve this goal is to divide the phase space
(Appendix \ref{A3b}) into a few parts
\cite{Ellis:1980wv}-\cite{Klasen:1997br}, 
\cite{Fontannaz:2001ek,Gordon:1998yt,Fontannaz:2001nq}, and herein this 
technique is adopted.

Let us assume that 
$\theta_{ji}$ is an angle between the momentum of the final parton $d_i$
(see Eq. (\ref{23eq}))
and the momentum of an initial particle $j$, where $j=e,~p$
\footnote{The initial particles $a$ and $b$ in Eq. (\ref{23eq}) move
parallel to the initial electron and proton, respectively.}.
Let us also define the distance, $R_{\gamma i}$, between the 
$d_i$ parton and the final photon:
\be
R_{\gamma i}=\sqrt{(\eta_{d_i}-\eta_{\gamma})^2+(\phi_{d_i}-\phi_{\gamma})^2}.
\ee
We define the parts of the phase space in the following way:
\begin{itemize}
\item Part 1. In the first part the variable $w$ 
(see Appendix \ref{Anot})
is in the range $w_{cut} \le w \le 1$, where $w_{cut}$ 
is an arbitrary parameter close to 1: $0<1-w_{cut}\ll 1$. 
This region of the phase space contains
all types of NLO corrections: the virtual gluon exchange, the real gluon
emission, and the process $\gamma g\ra\gamma q\bar{q}$.
The virtual gluon exchange is a $2\ra 2$ process, while the other
ones are of $2\ra 3$. However, for $w$ close to 1, $w_{cut} < w \le 1$,
the two final partons in the $2\ra 3$ processes are almost collinear
or/and one of the final partons is soft.
For $w_{cut}$ sufficiently close to 1 the kinematics of
the $2\ra 3$ processes is almost the same as in the $2\ra 2$ case:
\be
\gamma b\ra\gamma d \makebox[1cm]{,}d\equiv d_1 + d_2,
\ee
where the ``particle'' $d$ has four-momentum equal to the sum
of four-momenta of the $d_i$-partons, and its mass is almost zero.
The additional restriction $1-w_{cut} < \epsilon$ is sufficient to ensure 
that if parton $d_1$ or $d_2$ is inside the isolation cone then its
energy is smaller than $\epsilon E_{\gamma}$, so the final photon is isolated.
\end{itemize}
The other parts of the phase space, Parts 2-5 described below, 
are defined for $w<w_{cut}$, and contain only $2\ra 3$ diagrams.
\begin{itemize}
\item Part 2. In this region $w<w_{cut}$ and 
$\theta_{cut} > min (\theta_{e 1}, \theta_{e 2})$,
where $\theta_{cut}$ is a small arbitrary parameter, $\theta_{cut}\ll 1$.
Here, one of the final partons has the 
momentum almost collinear to the momentum
of the initial electron and it, for sufficiently small $\theta_{cut}$,  
does not enter the isolation cone around the photon.
The second parton has a large transverse momentum balancing the
photon transverse momentum, so the final photon is isolated.
\item Part 3. Here $w<w_{cut}$ and 
$\theta_{cut} > min (\theta_{p 1}, \theta_{p 2})$.
One of the final partons has momentum almost collinear to the momentum
of the proton, and, like in Part 2, the final photon is isolated.
\item Part 4. Here $w<w_{cut}$ and 
$R_{cut} > min (R_{\gamma 1}, R_{\gamma 2})$,
where $R_{cut}$ is a small parameter, $R_{cut}\ll1$.
In this case one of the final partons, say $d_1$, is almost collinear 
to the photon, and the photon is isolated if
the transverse momenta obey the inequality (\ref{eps}): 
$E_{T d_1}<\epsilon E_{T \gamma}$. 
\item Part 5. The last part is defined as the region in which there
are no collinear configurations: $w<w_{cut}$, 
$\theta_{cut} < min (\theta_{e 1}, \theta_{e 2})$,
$\theta_{cut} < min (\theta_{p 1}, \theta_{p 2})$, and
$R_{cut} < min (R_{\gamma 1}, R_{\gamma 2})$.
\end{itemize}
The $\mathcal{O}$$(\alpha_S)$ corrections (\ref{cross5i}) contain 
contributions from all parts of the phase space:
\bea 
\left[ d\sigma^{\gamma q\ra\gamma q}_{\alpha_S}
+d\sigma^{\gamma q\ra\gamma qg}_{\alpha_S}\right]=
\left[ d\sigma^{\gamma q\ra\gamma q}_{\alpha_S (1)}
+d\sigma^{\gamma q\ra\gamma qg}_{\alpha_S (1)}\right]+
\sum_{i=2,3,4,5} d\sigma^{\gamma q\ra\gamma qg}_{\alpha_S (i)}
\label{cross5rr}
\\
d\sigma^{\gamma g\ra\gamma q\bar{q}}_{\alpha_S}=
\sum_{i=1,2,3,4,5} d\sigma^{\gamma g\ra\gamma q\bar{q}}_{\alpha_S (i)}
\makebox[4.1cm]{}
\label{cross523}
\eea
In numerical calculations for Part 1 the formulae (\ref{eq:non:k}) and
(\ref{eq:non:kp}) (the same as for the inclusive photon case) can be
used with the requirement $w>w_{cut}$.
%To take into account the isolation of the final photon, 
%the additional cut (\ref{eps1}) has to be included.
In these formulae all the soft gluon singularities present in the virtual 
gluon and real gluon corrections are canceled, and all the
collinear singularities are factorized into the parton densities.

Parts 2-5 contain no contribution from the virtual gluon exchange,
and no soft gluon singularities. 
Analytical expressions for cross sections corresponding to
Parts 2, 3, and 4 are derived in Appendix \ref{xsec:}, and 
%\PP 
will be presented in next section.

The cross section in Part 5 has no singularities at all, 
and one can perform an exact numerical integration over
final four-momenta in any kinematic range, including isolation 
restrictions and other cuts.
For this purpose the general formulae (\ref{ps3E}) and
(\ref{eq:xsec:m2}-\ref{xeq:og})
with $\varepsilon =0$ are used.

The presented method of the division of the phase space differs from the 
methods applied in other calculations 
for the photoproduction of isolated photons at HERA
\cite{Gordon:1998yt,Fontannaz:2001ek,Fontannaz:2001nq},
since we use the cut-off parameter $w_{cut}$ not used before.
%\PP 
Our method allows to obtain relatively the simplest
analytical singular-free formulae in all parts of the phase space,
%\PP{\bf\large na czym polega???}
%\footnote{As it is described above, in Part 1 we just take the same formula
%as for the non-isolated photon production. 
%The unintegrated over momenta formula
%used in Part 5 is also the same as for the non-isolated photon case.
%In Parts 2-4 the relatively simple formulae for collinear configurations
%are used.},
as all the singularities due to the soft gluons (or gluons collinear
to the final quark) are collected in one part of the phase space,
i.e. in Part 1, and this part is chosen this way that the formulae for
the non-isolated photon production can be used.
The unintegrated over momenta formulae
used in Part 5 are also the same as for the non-isolated photon case,
and in Parts 2-4 the relatively simple expressions for collinear configurations
are applied.

The drawback of our method is that the numerical calculations are not 
stable if the integration over final momenta in Part 5 is performed
in the laboratory $ep$ frame or in the rest frame of the initial
photon and parton. However, we found that the calculations were sufficiently 
stable if the rest frame of the final partons was chosen instead.

The predictions should not depend on a choice of unphysical cut-off parameters,
$1-w_{cut}$, $\theta_{cut}$ and $R_{cut}$, if they are small enough. 
On the other hand, they can not be too small, since for very low values
of these parameters large numerical errors occur. 
The results presented in Secs. \ref{pssres} and 
\ref{Sjet:res} are obtained
with $1- w_{cut}=\theta_{cut}=R_{cut}=0.01$ where the angle $\theta_{cut}$ 
is defined in the centre of mass of the initial photon and the initial parton
(despite the fact that the numerical calculation is performed in other frame).
We have checked that the change of the predictions due to the variation
of these parameters is negligible, below 1\%, if they are taken
in the range $10^{-4} \le 1 - w_{cut} \le 0.03$, 
$10^{-4} \le \theta_{cut} \le 0.05$ and
$10^{-4} \le R_{cut} \le 0.05$.

%%%%%%%%%%%%%%%%%%%%%%%%%%%%%%%%%%%%%%%%%%%%%%%%%%%%%%%%%%%%%%%%%%%%%%%
\section{Analytical results}\label{isol:a}
%%%%%%%%%%%%%%%%%%%%%%%%%%%%%%%%%%%%%%%%%%%%%%%%%%%%%%%%%%%%%%%%%%%%%%%
Below we present analytical results for cross sections
for $2\ra 3$ processes including various collinear configurations
described in Sec. \ref{pss}. These results are obtained 
(for the details see Appendices \ref{A3b}, \ref{xsec:})
using the standard factorization procedure, which for the inclusive
(non-isolated) photon production in the $ep$ scattering was discussed
in Ref. \cite{Aurenche:1984hc,jan}. 
All the collinear
singularities appearing in the calculations are factorized
into parton densities in the photon (Part 2), proton (Part 3),
and into fragmentation functions (Part 4).

Some of the presented below ``splitting functions'' $P$ have been already 
presented in Ref. \cite{Gordon:1994ut}, where they were used
to calculate in the small cone approximation
the cross section for the isolated photon production
in the hadron-hadron collision (see also e.g. 
\cite{Furman:1981kf,Giele:1991vf,Giele:1993dj}). Nevertheless,
we find it 
%\PP 
useful to present herein compact analytical formulae
for all the relevant
cross sections for the isolated photon production in the $ep$
scattering.
% using an unified notation.
The partonic cross sections for the $2\ra 3$ processes in the parts
of the phase space labeled in Sec. \ref{pss} as Part 2, 3 and 4
have the form: 
\bea
d\sigma^{\gamma g\ra\gamma q\bar{q}}_{\alpha_S (2)}
=
\Theta(s+t+u)\int\limits_0^1
d\sigma^{qg\ra\gamma q}(\xi s,\xi t,u) 
P_{\gamma\ra q\bar{q}}\Bigl (\xi,yE_e(1-\xi)\theta_{cut} \Bigl ) d\xi ,
\label{co1}
\eea\bea
d\sigma^{\gamma q\ra\gamma qg}_{\alpha_S (2)}
=
\Theta(s+t+u)\int\limits_0^1
d\sigma^{q\bar{q}\ra\gamma g}(\xi s,\xi t,u) 
P_{\gamma\ra q\bar{q}}\Bigl (\xi,yE_e(1-\xi)\theta_{cut} \Bigl )  d\xi ,
\label{co2}\eea\bea
d\sigma^{\gamma q\ra\gamma q g}_{\alpha_S (3)}
=
\Theta(s+t+u)\int\limits_0^1
d\sigma^{\gamma q\ra\gamma q}(\xi s,t,\xi u) 
P_{q\ra qg}\Bigl (\xi,xE_p(1-\xi)\theta_{cut} \Bigl ) d\xi ,
\label{co3}\eea\bea
d\sigma^{\gamma g\ra\gamma q\bar{q}}_{\alpha_S (3)}
=
\Theta(s+t+u)\int\limits_0^1
d\sigma^{\gamma q\ra\gamma q}(\xi s,t,\xi u) 
P_{g\ra q\bar{q}}\Bigl (\xi,xE_p(1-\xi)\theta_{cut} \Bigl )  d\xi ,
\label{co4}\eea\bea
d\sigma^{\gamma q\ra\gamma q g}_{\alpha_S (4)}
=\Theta({s+t+u\over t+u}+\epsilon)\cdot
%\nonumber
\label{theta1}
\eea
\bea
\makebox[2cm]{}\cdot\Theta(s+t+u)\int\limits_0^1
d\sigma^{\gamma q\ra qg}(s,{t\over \xi}, {u\over \xi}) 
P_{q\ra\gamma q}\Bigl (\xi,E_T^{\gamma}(1-\xi)R_{cut} \Bigl )  {d\xi\over \xi^2} ,
\label{co5}\eea\bea
d\sigma^{\gamma g\ra\gamma q \bar{q}}_{\alpha_S (4)}
=\Theta({s+t+u\over t+u}+\epsilon)\cdot
%\nonumber
\label{theta2}
\eea
\bea
\makebox[2cm]{}\cdot\Theta(s+t+u)\int\limits_0^1
d\sigma^{\gamma g\ra q\bar{q}}(s,{t\over \xi}, {u\over \xi}) 
P_{q\ra\gamma q}\Bigl (\xi,E_T^{\gamma}(1-\xi)R_{cut} \Bigl )  {d\xi\over \xi^2} ,
\label{co6}
\eea
where $\Theta({s+t+u\over t+u}+\epsilon)$ includes the isolation restrictions
(\ref{eps}), the partonic cross sections for the $2\ra 2$ processes 
are given in Appendix \ref{lox} and the functions $P$ are given by
\bea
P_{q\ra\gamma q}(\xi,E_{\perp})={\alpha\over 2\pi} e_q^2
\left\{ {1+(1-\xi)^2\over \xi} \ln {E_{\perp}^2\over\mu^2} + \xi\right\},
\eea\bea
P_{\gamma\ra q\bar{q}}(\xi,E_{\perp})=3 {\alpha\over 2\pi} e_q^2
\left\{ [\xi^2+(1-\xi)^2] \ln {E_{\perp}^2\over\mu^2} + 1\right\},
\eea\bea
P_{q\ra qg}(\xi,E_{\perp})= {4\over 3}{\alpha_S\over\alpha e_q^2} 
P_{q\ra\gamma q}(1-\xi,E_{\perp}),
\eea\bea
P_{g\ra q\bar{q}}(\xi,E_{\perp})= {1\over 6}{\alpha_S\over \alpha e_q^2} 
P_{\gamma\ra q\bar{q}}(\xi,E_{\perp}).
\label{co10}
\eea
The variable $\xi$ in the functions $P_{a\ra bc}(\xi,E_{\perp})$ stands for 
the fraction of the energy of the particle $a$ taken by the particle $b$:
$E_b = \xi E_a$. The variable $E_{\perp}$ is the maximal possible
(for given $\theta_{cut}$ or $R_{cut}$) momentum of the particle
$c$ perpendicular to the momentum of the particle $a$:
$E_{\perp}= max |\vec{p}_c\times \vec{p}_a|/|\vec{p}_a|$.
The results (\ref{co1}-\ref{co10}) are obtained with the assumption that
$\theta_{cut}$ and $R_{cut}$ are small, see Appendices \ref{A3b} and 
\ref{xsec:} for details.

Note, that the integration over $\xi$ in Eqs. (\ref{co1}-\ref{co6})
is trivial as the $2\ra 2$
partonic cross sections, $d\sigma^{ab\ra cd}(s,t,u)$,
contain the $\delta (s+t+u)$ function, see Eq. (\ref{eq.lox}).
The expressions (\ref{co5}) and (\ref{co6}) are equivalent to the
expressions (\ref{small1}) and (\ref{small3}) for the subtraction 
terms used in the small cone approximation (Sec. \ref{Sapprox}).

The cut-off parameter $R_{cut}$ is chosen in a Lorentz-invariant form
(Sec. \ref{pss}), but the parameter $\theta_{cut}$ depends on the
choice of the frame of reference. It can be defined in any 
frame of reference provided that the energies $yE_e$ and $xE_p$
in Eqs. (\ref{co1}-\ref{co4}) are given in the same frame.
In numerical calculations we use the expressions (\ref{co1}-\ref{co4})
with $\theta_{cut}$ defined in the centre of mass system of the
initial photon and the initial parton, where $yE_e=xE_p$.

%%%%%%%%%%%%%%%%%%%%%%%%%%%%%%%%%%%%%%%%%%%%%%%%%%%%%%%%%%%%%%%%%%%%%%%
\section{Numerical results and discussion}\label{pssres}
%%%%%%%%%%%%%%%%%%%%%%%%%%%%%%%%%%%%%%%%%%%%%%%%%%%%%%%%%%%%%%%%%%%%%%%

The predictions for the photoproduction of the prompt photons
at the HERA collider are obtained with parameters 
and parton densities specified in Sec. \ref{det}.
For a comparison with the H1 Collaboration data,
kinematic ranges used in \cite{unknown:2004uv} are applied:
the fraction of the electron energy transferred to the photon
is restricted to the range $0.2<y<0.7$, the final photon rapidity
and transverse energy are taken in the limits
$-1<\eta^{\gamma}<0.9$ and $5<E_T^{\gamma}<10$~GeV, 
and the initial energies are $E_e=27.6$ GeV and $E_p=920$ GeV.
The final photon is isolated with parameters $\epsilon =0.1$ and $R=1$.

%%%%%%%%%%%%%%%%%%%%%%%%%%%%%%%%%%%%%%%%%%%%%%%%%%%%%%%%%%%%%%%%%%%%%%%
\subsection{Comparison between exact and approximated results}
\label{Sisol:ae}
%%%%%%%%%%%%%%%%%%%%%%%%%%%%%%%%%%%%%%%%%%%%%%%%%%%%%%%%%%%%%%%%%%%%%%%

\begin{figure}[t]
\vskip 25cm\relax\noindent\hskip -2cm
       \relax{\includegraphics{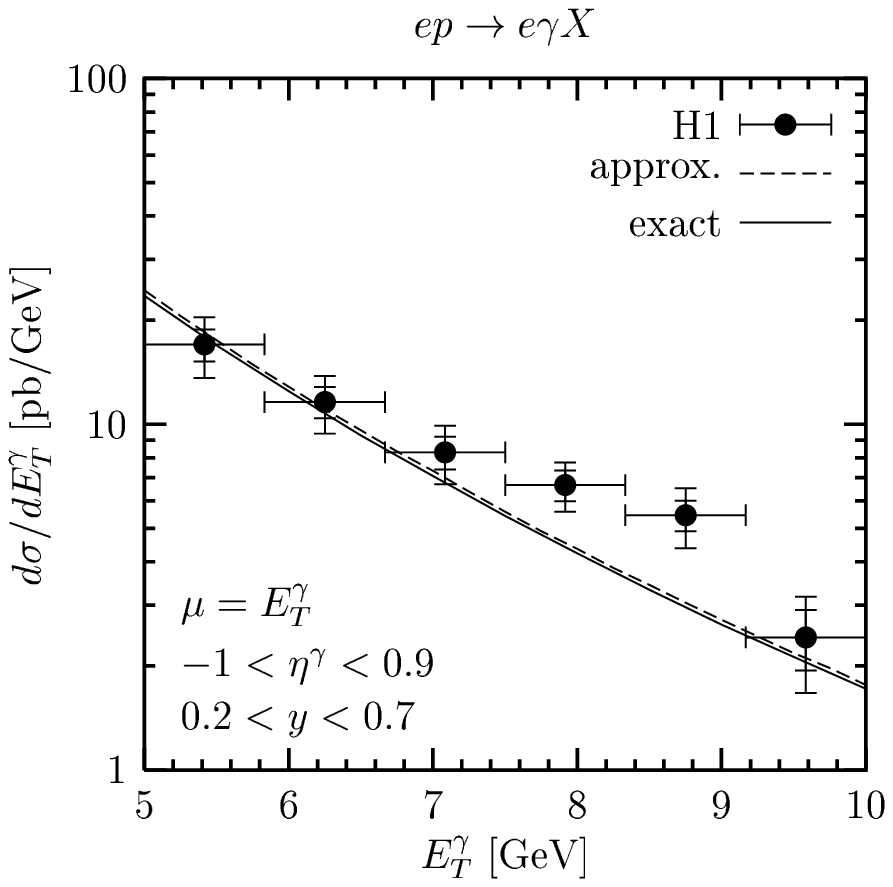}}
\vspace{-16.cm}
\caption{\small\sl The comparison between exact predictions (solid line),
predictions obtained in the small cone approximation (dashed line)
and the H1 \cite{unknown:2004uv} data for $d\sigma /dE_T^{\gamma}$ 
cross section.
The GRV$^{p,\gamma,frag}$ 
\cite{Gluck:1995uf,Gluck:1992ee,Gluck:1993zx} 
parton densities are used}
\label{fignojet.ptgamma}
%\end{figure}
%\vspace*{7.5cm}
%\begin{figure}[ht]
\vskip 25cm\relax\noindent\hskip -1.9cm
       \relax{\includegraphics{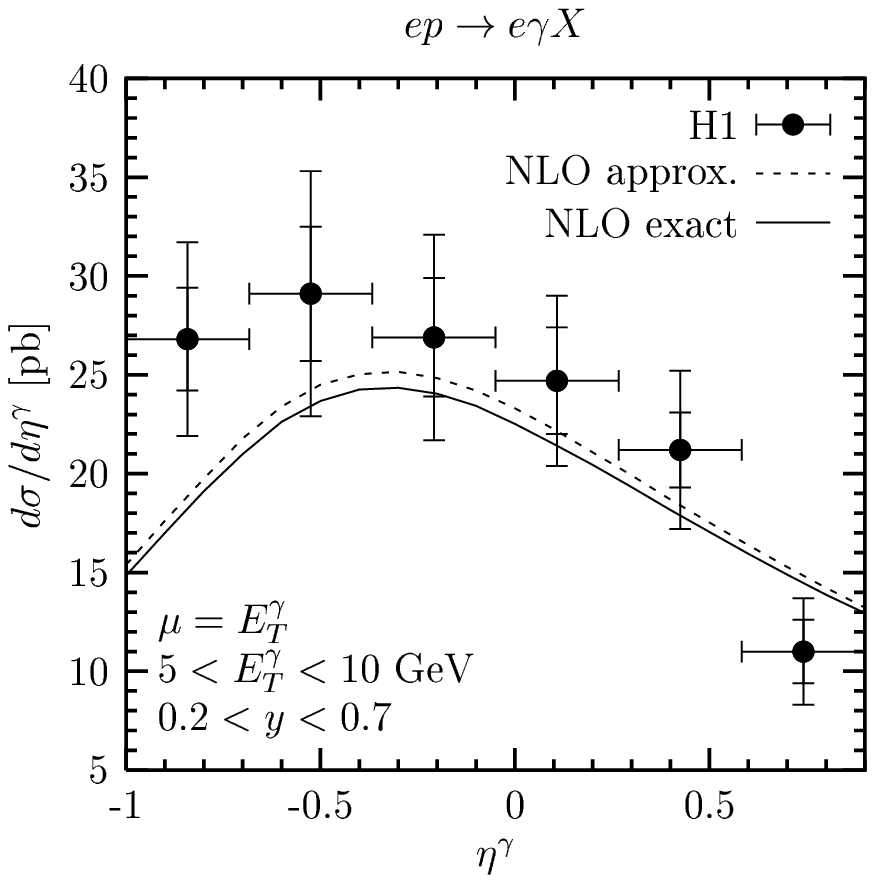}}
\vspace{-16.cm}
\caption{\small\sl As in Fig.~\ref{fignojet.ptgamma} for
$d\sigma /d\eta_{\gamma}$ cross section.}
\label{fignojet.etagamma}
\end{figure}

In Figs. \ref{fignojet.ptgamma} and \ref{fignojet.etagamma}
the differential cross sections $d\sigma /dE_T^{\gamma}$ and
$d\sigma /d\eta^{\gamma}$ are presented. The exact 
predictions~\footnote{By exact predictions we mean the predictions 
obtained  using the technique described in Sec. \ref{pss}
with no assumption that the isolation cone is small.}
(solid lines) 
are compared with the predictions obtained in the small cone approximation 
(dashed lines) and with the new H1 data \cite{unknown:2004uv}
(see also \cite{Lemrani:thesis}).
The differences between both predictions are about 3\%, so the
small cone approximation works well, despite the fact that the isolation
cone of radius $R=1$ is not a small one.

The comparison between the two predictions are also presented
in Tab. \ref{tab3} for individual contributions and total cross sections.
The Born, box and resolved photon contributions
are the same in both calculations, since the approximation
was applied to $\mathcal{O}$$(\alpha_S)$ 
corrections only (Sec. \ref{Sapprox}). 
Nevertheless the difference between exact and approximated 
$\mathcal{O}$$(\alpha_S)$ 
corrections is very large. We have found that for the isolated photon
production the $\mathcal{O}$$(\alpha_S)$ 
corrections to the Born process obtained using the small cone
approximation are 67\% higher than the corresponding contribution
obtained using the division of the phase space
\footnote{For non-isolated photon there is no difference
between both types of calculations.}.
We conclude that the small cone approximation gives
a very rough estimation of the order of magnitude of the 
$\mathcal{O}$$(\alpha_S)$ corrections,
however the total cross section is estimated quite well, since
the $\mathcal{O}$$(\alpha_S)$
corrections give relatively small (5\%) contribution to the
total cross section (Tab. \ref{tab3}).

%%%%%%%%%%%%%%%%%%%%%%%%%%%%%%%%%%%%%%%%%%%%%%%%%%%%%%%%%%%%%%%%%%%%%%%
\subsection{Comparison with the H1 data}\label{Sisol:h1}
%%%%%%%%%%%%%%%%%%%%%%%%%%%%%%%%%%%%%%%%%%%%%%%%%%%%%%%%%%%%%%%%%%%%%%%

The comparison with the new H1 Collaboration 
data \cite{unknown:2004uv} is presented
in Figs. \ref{fignojet.ptgamma} and \ref{fignojet.etagamma}
and we see that
the predictions tend to lie below the data for $\eta^{\gamma}$ 
distribution, although in most bins an agreement is obtained
within the large experimental errors (Fig. \ref{fignojet.etagamma}). 
For $E_T^{\gamma}$ distribution (Fig. \ref{fignojet.ptgamma}) 
the data in two bins are 2 - 3 standard deviations above the predictions,
while in other bins the the data and predictions agree well. 

\begin{figure}[th]
\vskip 9.2cm\relax\noindent\hskip -1cm
       \relax{\includegraphics{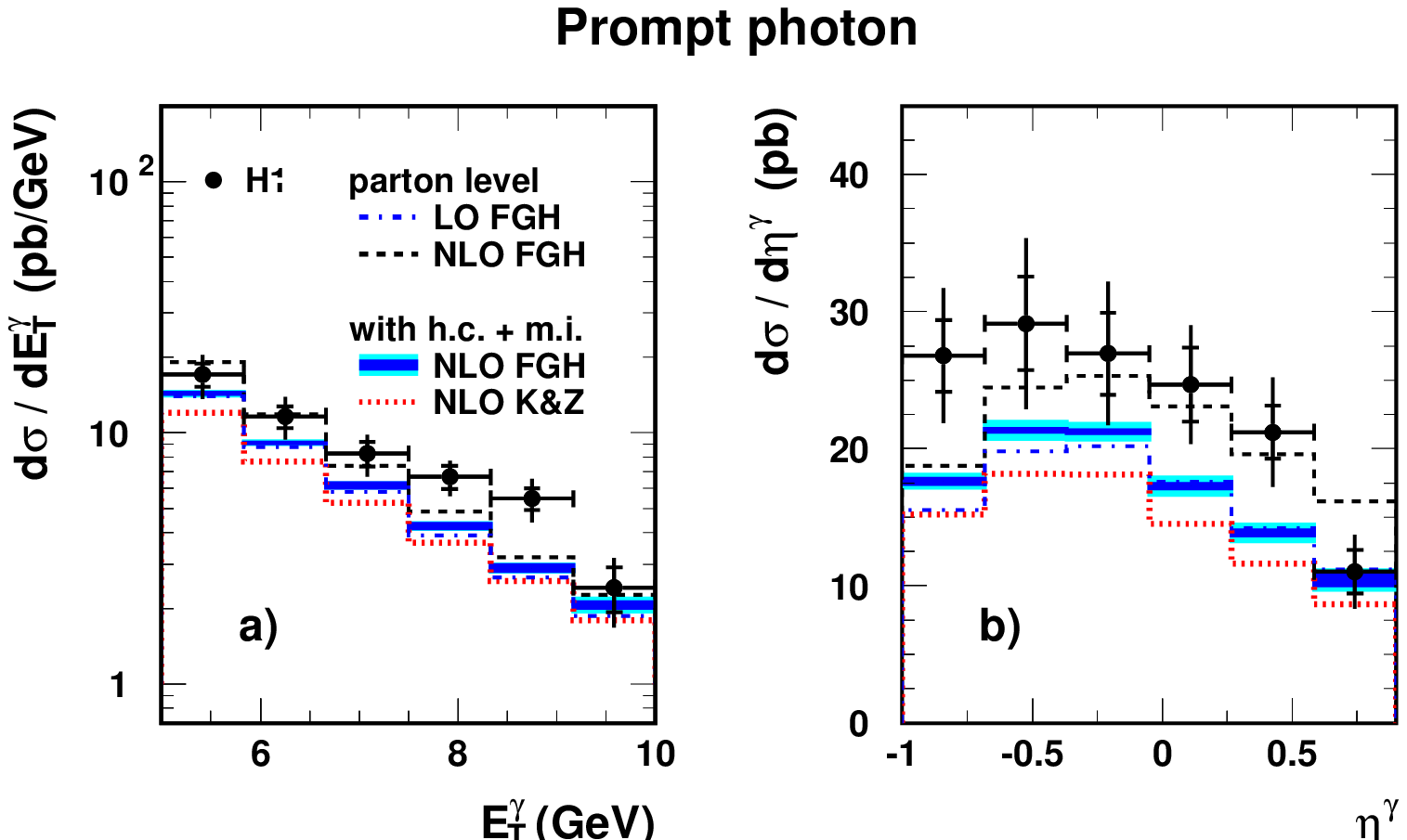}}
\vspace{0cm}
\caption{The cross sections $d\sigma /dE_T^{\gamma}$ (a) and
$d\sigma /d\eta^{\gamma}$ (b) for the $ep\ra e\gamma X$ process
with $-1<\eta^{\gamma}<0.9$, $5<E_T^{\gamma}<10$ GeV and $0.2<y<0.7$.
The H1 data \cite{unknown:2004uv} are compared with K\&Z 
\cite{Zembrzuski:2003nu} (dotted lines) and FGH 
\cite{Fontannaz:2001ek,Fontannaz:2001nq} (solid lines) predictions
obtained using MRST99\hspace{0.2mm}$^{p}$ \cite{Martin:1999ww}, 
AFG$^{\gamma}$ \cite{Aurenche:1994in}, BFG$^{frag}$ \cite{Bourhis:1997yu}
parton densities. Both predictions are corrected for hadronization
and multiple interactions (h.c.+m.i.) effects. The shaded areas show
changes of FGH predictions due to the $\mu$ scale variation between
$0.5\cdot E_T^{\gamma}$  and $2\cdot E_T^{\gamma}$ (inner bands)
and added linearly
uncertainties on h.c.+m.i. corrections (outer bands).
The LO (dashed-dotted) and NLO (dashed) FGH results without
h.c.+m.i. are also shown. The figure is taken 
from the H1 paper \cite{unknown:2004uv}}
\label{fig4abh1}
\end{figure}

Our predictions shown in Figs. \ref{fignojet.ptgamma}, \ref{fignojet.etagamma}
are obtained 
%at the parton-level, i.e. 
for the final photon 
and partons production (Eqs. \ref{223}, \ref{23eq}). According 
to the analysis done by the H1 Collaboration \cite{unknown:2004uv}, 
for a realistic comparison with 
%\PP 
their data, 
the hadronization corrections 
and multiple interactions (h.c.+m.i.) should be taken into account
in the predictions.
Our exact predictions, after h.c.+m.i. corrections implemented by the
H1 Collaboration \cite{unknown:2004uv}, are shown in Fig. \ref{fig4abh1}
%~\footnote{This figure is taken from the H1 paper \cite{unknown:2004uv}.}
(dotted lines) together with the H1 data \cite{unknown:2004uv} and with 
%\PP 
predictions of Fontannaz, Guillet and Heinrich (FGH) 
\cite{Fontannaz:2001ek,Fontannaz:2001nq}. Both, our and FGH, predictions
are obtained with with MRST99\hspace{0.2mm}$^{p}$ 
\cite{Martin:1999ww}, AFG$^{\gamma}$ \cite{Aurenche:1994in}, 
BFG$^{frag}$ \cite{Bourhis:1997yu} parton densities.
To show the 
%\PP 
effect of h.c.+m.i., the FGH predictions are 
presented with (solid lines) and without (dashed lines) these 
corrections~\footnote{The correction factors are typically
0.75-0.90 depending on a bin. In both K\&Z and FGH predictions the
same factors are applied.\label{foot}}.
%\PP It is not clear why this  effect is so large, and to waht extent it is 
%purly experimental effect. 

The FGH and our predictions are below the data typically by 30\% and 40\%,
respectively \cite{unknown:2004uv}.
The discrepancy is seen in almost whole range of $\eta^{\gamma}$
with exception of the last bin, where the QCD calculations and data
are in agreement.
Note, that the prompt photon cross sections measured previously
by the ZEUS Collaboration \cite{Breitweg:1999su} were well described
by predictions at $\eta^{\gamma}>0.1$  (Sec. \ref{results2}).
On the other hand, both the ZEUS \cite{Breitweg:1999su} and
the H1 \cite{unknown:2004uv} data are consistent, as it was shown 
in \cite{unknown:2004uv,Lemrani:thesis}.
It seems, that the above different 
%\PP 
conclusions of comparisons 
between the predictions and the ZEUS data, and between the predictions and
the H1 data are due to large effects of h.c.+m.i.
corrections applied in the H1 analysis
\cite{unknown:2004uv}.

Note, that the dependence on the choice of parton
parametrizations is relatively large: 
results obtained using GRV$^{p,\gamma,frag}$ 
\cite{Gluck:1995uf,Gluck:1992ee,Gluck:1993zx} (not shown)
are 10\% higher on average than these obtained with
MRST99\hspace{0.2mm}$^{p}$ 
\cite{Martin:1999ww}, AFG$^{\gamma}$ \cite{Aurenche:1994in}, 
BFG$^{frag}$ \cite{Bourhis:1997yu} densities, and slightly
decrease the size of discrepancies between the predictions and the data.

~ \\ ~ \\ ~ 

%%%%%%%%%%%%%%%%%%%%%%%%%%%%%%%%%%%%%%%%%%%%%%%%%%%%%%%%%%%%%%%%%%%%%%%
\subsection{Comparison with other QCD predictions (FGH)}\label{Sisol:fgh}
%%%%%%%%%%%%%%%%%%%%%%%%%%%%%%%%%%%%%%%%%%%%%%%%%%%%%%%%%%%%%%%%%%%%%%%

The calculation of Fontannaz, Guillet and Heinrich (FGH) 
\cite{Fontannaz:2001ek,Fontannaz:2001nq}
for the photoproduction of isolated photons at the HERA collider 
includes the $\mathcal{O}$$(\alpha_S)$ corrections to the
processes with the resolved initial photon or/and with the fragmentation
into the final photon, which are not included in our
calculation (Chapter \ref{nlo}).
Both our and FGH calculations take into account 
the $\mathcal{O}$$(\alpha^2\alpha_S^2)$ box diagram (Fig. \ref{figbox})
and all other diagrams shown in Figs. \ref{figborn}-\ref{fig23} and 
\ref{figsingi}-\ref{figbox}.
 
The predictions of FGH presented in their papers
\cite{Fontannaz:2001ek,Fontannaz:2001nq} 
were obtained in the kinematic ranges considered by the 
ZEUS Collaboration \cite{Breitweg:1999su,Chekanov:2001aq}
(see Sec. \ref{results5}) 
and one can not use them for the direct comparison with our
predictions corresponding to the H1 measurements
\cite{unknown:2004uv}.
So, to compare the predictions, we use the FGH and
K\&Z (Zembrzuski and Krawczyk) \cite{Zembrzuski:2003nu}
results 
%\PP 
based on the theoretical predictions and corrected for h.c+m.i.,
as  presented in the
paper of the H1 Collaboration \cite{unknown:2004uv}
\footnote{The applied correction factors are the same for both
K\&Z and FGH predictions, see footnote~$^{\ref{foot}}$. So the relative
differences between both ``corrected'' predictions are the same as the
differences between pure QCD (with no h.c.+m.i.) K\&Z and FGH predictions
\label{foot2}}.
Both predictions are obtained using the same parton parametrizations,
namely MRST99\hspace{0.2mm}$^{p}$ \cite{Martin:1999ww}, 
AFG$^{\gamma}$ \cite{Aurenche:1994in}, BFG$^{frag}$ \cite{Bourhis:1997yu}.

The FGH predictions are 15-20\% larger than ours for the $\eta^{\gamma}$
and $E_T^{\gamma}$ distributions shown in Fig. \ref{fig4abh1}.
and this difference is of a similar order as the difference
between our predictions and the predictions of Gordon (LG) 
\cite{Gordon:1998yt} which was discussed in Sec. \ref{results5}.
It means, that the higher order terms 
included in the LG and FGH calculations and not included in the
K\&Z approach,
give a sizable contribution to the cross section 
in the kinematic range considered 
for the prompt photon photoproduction.
This is not fully supported by our study of the dependence on the
renormalization/factorization scale, which seemed to suggest smaller effect
due to missing higher order terms (Sec. \ref{results3}).

%%%%%%%%%%%%%%%%%%%%%%%%%%%%%%%%%%%%%%%%%%%%%%%%%%%%%%%%%%%%%%%%%%%%%%%%%
\chapter{Isolated photon plus jet production}\label{jet}
%%%%%%%%%%%%%%%%%%%%%%%%%%%%%%%%%%%%%%%%%%%%%%%%%%%%%%%%%%%%%%%%%%%%%%%%%

The partons produced in hard processes (Figs. \ref{figborn}-\ref{fig23},
\ref{figsingi}-\ref{figbox}) are not observed experimentally, 
since they recombine into colorless jets. In previous parts of the work, 
the predictions for the prompt photon production were discussed
and the presented cross sections were integrated over partons/jets
momenta. Now, the isolated photon 
%\PP 
\underline{plus jet} production is studied for limited 
ranges of both photon and jet rapidities and transverse energies.
Such a process with two objects observed in the final state can be
a source of more detailed information concerning the dynamics
of the interaction.
 
%%%%%%%%%%%%%%%%%%%%%%%%%%%%%%%%%%%%%%%%%%%%%%%%%%%%%%%%%%%%%%%%%%%%%%%%%
\section{Jet algorithm}
%%%%%%%%%%%%%%%%%%%%%%%%%%%%%%%%%%%%%%%%%%%%%%%%%%%%%%%%%%%%%%%%%%%%%%%%%

As it was discussed in Chapter~\ref{nlo} and in Sec. \ref{Sapprox}, 
in the calculation 
we deal with $2\ra 2$ and $2\ra 3$ hard processes.
In $2\ra 2$ processes (\ref{223}) the jet originates from the parton $d$
and the momentum of the jet can be identified with the momentum of this parton.

Two partons, $d_1$ and $d_2$, produced in $2\ra 3$ processes (\ref{23eq})
may lead to two separate jets or they may form one jet. The number of 
observed jets and their momenta depend on the jet definition. 
In this work the inclusive $k_T$-jet finding 
algorithm~\cite{Ellis:tq,Catani:1993hr} 
is applied in accordance with the jet definition used in the
H1 Collaboration measurements \cite{unknown:2004uv}.

Since in the presented calculation there are no more than two partons
forming the jet or jets,
the algorithm becomes very simple. If the distance between the partons,
$R_{12}$, defined as
\be
R_{12}=\sqrt{(\eta^{d_1}-\eta^{d_2})^2+(\phi^{d_1}-\phi^{d_2})^2} ~,
\label{Rjet}
\ee
is larger than an arbitrary parameter $R_J$,
then two separate jets arise with transverse energies, rapidities, 
and azimuthal angles of the $d_i$-partons:
\be
E_T^{jet_i}=E_T^{d_i}\makebox[1cm]{,}
\eta^{jet_i}=\eta^{d_i}\makebox[1cm]{,}
\phi^{jet_i}=\phi^{d_i}\makebox[1cm]{,}
i=1, ~2.\nonumber
\ee
For $R_{12} < R_J$ both partons are treated as components of
one jet with
\be
E_T^{jet}=E_T^{d_1}+E_T^{d_2} ,\label{jaE}
\ee
\be
\eta^{jet}=(E_T^{d_1}\eta^{d_1}+E_T^{d_2}\eta^{d_2})/E_T^{jet}\label{jan} ,
\ee
\be
\phi^{jet}=(E_T^{d_1}\phi^{d_1}+E_T^{d_2}\phi^{d_2})/E_T^{jet}\label{jap} .
\ee

The algorithm can be easily applied in numerical calculations using
the phase space slicing (Sec. \ref{pss}). In the first part of the phase 
space, where $w\sim 1$, the kinematics is the same as in $2\ra 2$ processes.
In part 2 (3, 4) one parton moves parallel to the initial electron
(initial proton, final photon) and it, for sufficiently small $\theta_{cut}$
($\theta_{cut}$, $R_{cut}$), does not enter the cone defining the jet,
so the jet consists (on the partonic level) of the second
parton alone. In part 5 one or two jets may arise 
depending on the value of $R_{12}$ (\ref{Rjet}).

% ZEUS:
% 
% The $k_T$-clustering algorithm with $R_J=1$ will be used in this
% dissertation everythere where the cross section for a jet production
% is considered unless explicity stated otherwise. Another choice of the
% jet definition will be applied for comparison with the ZEUS Collaboration
% measurement presented at the Vancouver Conferencion~\cite{}, where the
% cone algorithm~\cite{cone} was used.
% 
% In the cone algorithm the $d_i$-partons form one jet if the distance 
% between 
% each parton and the jet axis defined in Eqs.~\ref{jaE}-\ref{jap} 
% is less than $R_J$:
% \bea
% \sqrt{(\eta^{d_i}-\eta^{jet})^2+(\phi^{d_i}-\phi^{jet})^2} < R_J
% \makebox[1cm]{,} i=1,2.
% \eea

Following experimental analyses \cite{unknown:2004uv},
$R_J=1$ is used in our numerical calculations.

%%%%%%%%%%%%%%%%%%%%%%%%%%%%%%%%%%%%%%%%%%%%%%%%%%%%%%%%%%%%%%%%%%%%%%%%%
\section{Numerical results and discussion}\label{Sjet:res}
%%%%%%%%%%%%%%%%%%%%%%%%%%%%%%%%%%%%%%%%%%%%%%%%%%%%%%%%%%%%%%%%%%%%%%%%%

%\PP 
We perform numerical calculations in kinematic regions as used in 
experimental analysis.
There are two publications of the ZEUS Collaboration presenting results 
of measurements of the isolated photon plus jet photoproduction
at the HERA collider~\cite{Breitweg:1997pa,Chekanov:2001aq}.
In the first paper~\cite{Breitweg:1997pa}
the cross section integrated over some kinematic range 
is given. The aim of the second one~\cite{Chekanov:2001aq} was to study
transverse momentum of partons in the proton, and no results
for cross sections were presented (distributions of events, 
not corrected for the detector
effects, were shown). There is also a conference paper of the ZEUS
Collaboration where data for a differential cross section for the 
isolated photon plus a jet photoproduction are given, however these data
are still preliminary \cite{unknown:uj}.

In the new paper of the H1 Collaboration~\cite{unknown:2004uv}
(see also~\cite{Lemrani:thesis}), the
photoproduction data 
for the isolated final photon ($\epsilon =0.1$ and $R=1$)
with the initial energies $E_e=27.6$ GeV and $E_p=920$ GeV
are presented for various 
differential cross sections of both $ep\ra e\gamma X$ (considered
in Chapter~\ref{isol}) and $ep\ra e\gamma ~jet ~X$ processes.
Herein, the predictions for the kinematic limits as in \cite{unknown:2004uv}
are compared with these recent final data.
The cross sections are integrated over $0.2<y<0.7$, and 
$-1<\eta^{\gamma}<0.9$ and/or $5<E_T^{\gamma}<10$~GeV
with the jet rapidity and jet transverse energy in the
range $-1<\eta^{jet}<2.3$ and 4.5 GeV $<E_T^{jet}$, respectively.
If two jets are found within the above region, that
with higher $E_T^{jet}$ is taken.
Other parameters are specified in Sec. \ref{det}.

%%%%%%%%%%%%%%%%%%%%%%%%%%%%%%%%%%%%%%%%%%%%%%%%%%%%%%%%%%%%%%%%%%%%%%%%%
\subsection{Asymmetric cuts}\label{ressym}
%%%%%%%%%%%%%%%%%%%%%%%%%%%%%%%%%%%%%%%%%%%%%%%%%%%%%%%%%%%%%%%%%%%%%%%%%

As it is discussed in~\cite{Fontannaz:2001nq}, the symmetric 
cuts for the photon and the jet transverse energy,
$E_{T,min}^{jet}=E_{T,min}^{\gamma}$,
lead to unphysical
results in next-to-leading or higher orders of calculations
due to constraints imposed on soft gluons
(see also \cite{Aurenche:1997im,Frixione:1997ks,Catani:1997xc}). 
We have decided to study this effect for the cross section
as a function the photon transverse energy, 
%\PP
% (Fig.~\ref{fig:ptgamma}),
not analyzed in~\cite{Fontannaz:2001nq}.
This 
%method happend to be very useful since it 
allows to understand 
the effect of the symmetric cuts in more details. 
The results of this study was first presented in our paper 
\cite{Zembrzuski:2003nu}.
Next, similar study of the effect of the symmetric cuts
was 
%\PP 
presented in~\cite{Fontannaz:2003yn}.

\begin{figure}[t]
\vskip 24.5cm\relax\noindent\hskip -2cm
       \relax{\includegraphics{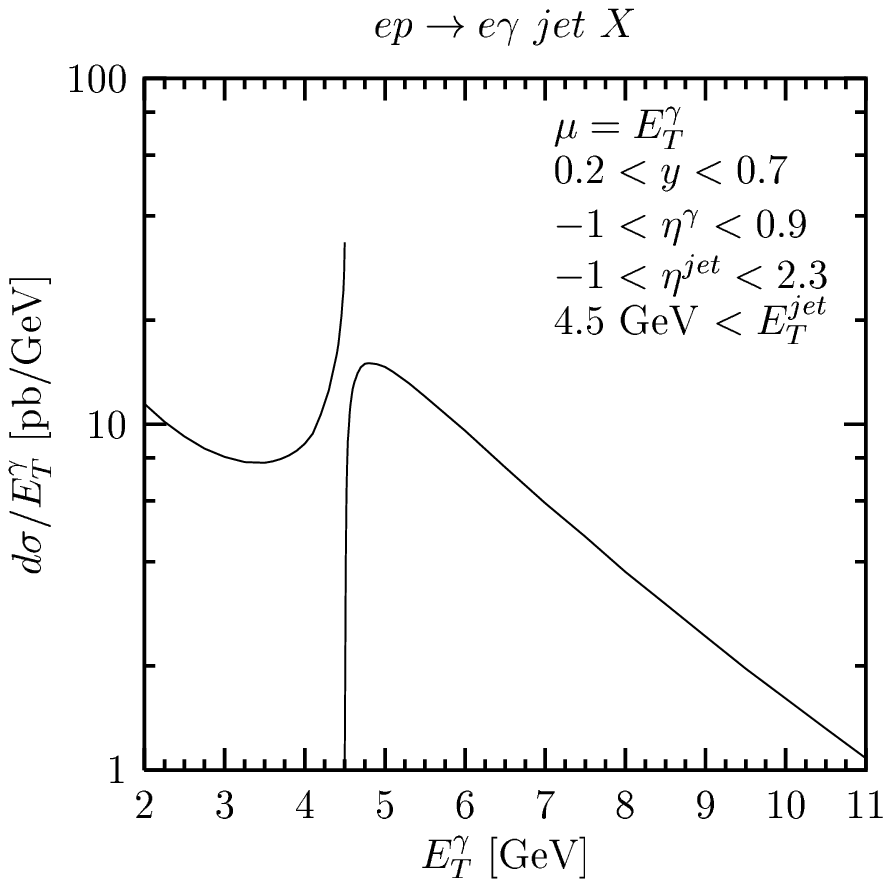}}
\vspace{-15.5cm}
\vskip 24.3cm\relax\noindent\hskip -2cm
       \relax{\includegraphics{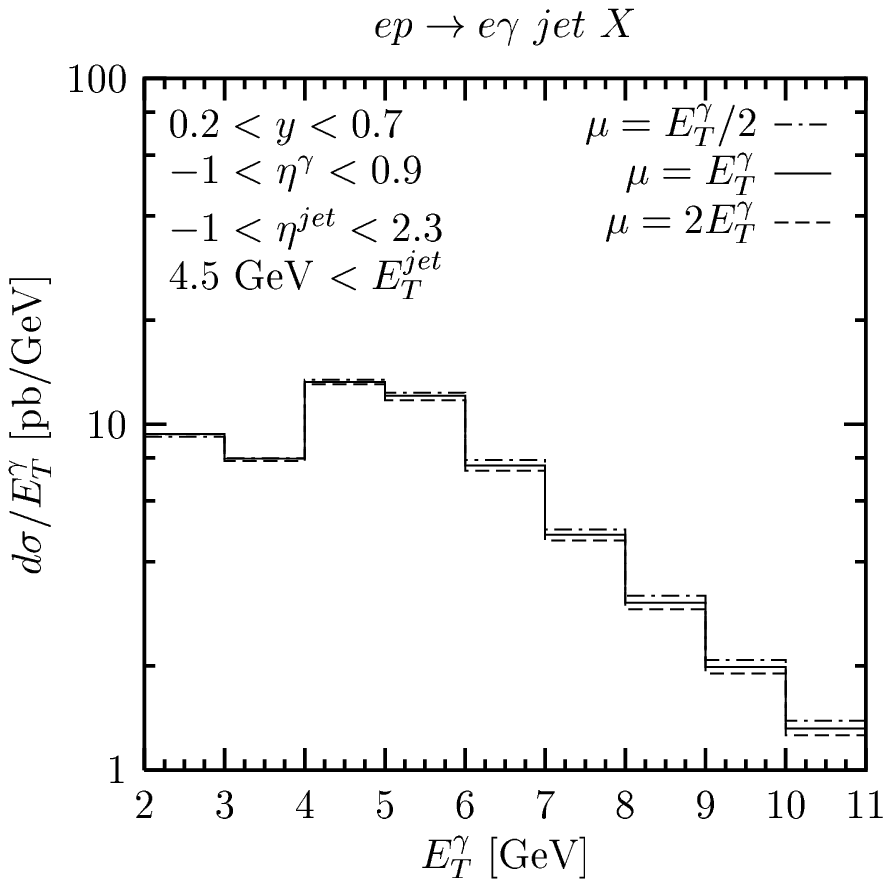}}
\vspace{-16.cm}
\caption{\small\sl The differential cross section $d\sigma /dE_T^{\gamma}$ 
(a) and the differential cross section $d\sigma /dE_T^{\gamma}$ averaged
over $E_T^{\gamma}$-bins (b) for the $ep\ra e\gamma ~jet ~X$ process.
The NLO predictions for $E_T^{\gamma}=\mu$ (a) and for $\mu$ between 
$E_T^{\gamma}/2$ and $2E_T^{\gamma}$ (b) are shown.}
\label{fig:ptgamma}
\end{figure}

%\PP 
Our findings are illustrated in Fig.~\ref{fig:ptgamma}a,b.
We study the dependence of the NLO predictions on the photon
transverse energy 
%is shown in Fig.~\ref{fig:ptgamma}a,
in the $E_T^{\gamma}$-range wider than 
the range considered by the H1 Collaboration.
At $E_T^{\gamma}$ values close to the minimal jet transverse energy,
$E_{T,min}^{jet} = 4.5$ GeV, the NLO differential cross section 
has a discontinuity, see Fig.~\ref{fig:ptgamma}a. 
For $(E_T^{\gamma})_-\ra 4.5$ GeV the cross section has 
a strong maximum while a minimum for $(E_T^{\gamma})_+\ra 4.5$ GeV.
In the minimum the value of the cross section is even negative.
This unphysical fluctuation is due to processes with soft gluons:
large terms corresponding to the soft gluon emission and the virtual
gluon exchange do not cancel properly if the photon transverse
energy is close to the minimal transverse energy of the jet,
because some of this terms corresponds to $E_T^{\gamma}$ slightly below
%\PP and above 
the cut-off $E_{T,min}^{jet}$ 
and the other correspond to $E_T^{\gamma}$
slightly above $E_{T,min}^{jet}$.

However,
if we integrate the cross section over some $E_T^{\gamma}$-bins
then this fluctuation disappears.
% we found that
%to avoid theoretical instabilities, one could consider a cross section
%averaged over some $E_T^{\gamma}$-bins. 
As it is shown in Fig.~\ref{fig:ptgamma}b,
the NLO predictions are well defined if one takes
the cross section integrated
over $E_T^{\gamma}$
from $E_{T,min}^{jet} - \Delta$ to $E_{T,min}^{jet} + \Delta$,
provided that $\Delta$ is sufficiently large in the comparison with
the gap in Fig.~\ref{fig:ptgamma}a, say $\Delta > 0.3$ GeV.
So, the bins of a length 1 GeV presented in Fig.~\ref{fig:ptgamma}b
are large enough to avoid errors corresponding to the fluctuation
around $E_T^{\gamma}=E_{T,min}^{jet}$.

An integration of the differential cross section 
(Fig.~\ref{fig:ptgamma}a) over the photon
transverse energy higher than the minimal jet transverse energy,
$E_T^{\gamma} \ge E_{T,min}^{jet}$ (symmetric cuts), 
leads to underestimated 
predictions in NLO. However numerically this effect is not very important,
being at level of roughly 5\% if the integration is performed
in the range from $E_T^{\gamma} = E_{T,min}^{jet} =4.5$ GeV to
$E_T^{\gamma} = 10$ GeV. 

Note that in the previous measurements
\cite{Breitweg:1997pa,unknown:uj,Chekanov:2001aq}
the symmetric cuts 
%\PP 
$E_T^{\gamma} > 5$ GeV
and  $E_T^{jet} > 5$ GeV were used,  while in the new
H1 analysis \cite{unknown:2004uv} the asymmetric cuts 
%\PP 
$E_T^{\gamma} > 5$ GeV
and  $E_T^{jet} > 4.5$ are taken. This latter choice of cuts
allows to avoid the theoretical errors 
in QCD calculations and is applied in calculations presented in next
sections (Sec. \ref{res}-\ref{fgh}).

It is worth mentioning
that the cross section for $E_T^{\gamma}<E_{T,min}^{jet}$ 
(Fig.~\ref{fig:ptgamma}a) is dominated
by the contribution due to the 
$2\ra 3$ processes in the $\mathcal{O}$$(\alpha_S)$ 
corrections (Figs. \ref{figreal}, \ref{fig23}), since 
the contributions of the $2\ra 2$ processes (Figs. \ref{figborn}, 
\ref{figsingi}-\ref{figbox})
are suppressed by the requirement
$E_T^{jet}>E_{T,min}^{jet}$ and by the isolation.
For $E_T^{\gamma}<E_{T,min}^{jet}/(1+\epsilon)\approx 4$~GeV the $2\ra 2$
processes do not contribute at all. On the other hand, for larger
$E_T^{jet}$, say $E_T^{jet}>5$~GeV, the contribution of 
$\mathcal{O}$$(\alpha_S)$ corrections
is very small, see Tab. \ref{tab1}.

%%%%%%%%%%%%%%%%%%%%%%%%%%%%%%%%%%%%%%%%%%%%%%%%%%%%%%%%%%%%%%%%%%%%%%%%%
\subsection{Theoretical uncertainties}\label{res}
%%%%%%%%%%%%%%%%%%%%%%%%%%%%%%%%%%%%%%%%%%%%%%%%%%%%%%%%%%%%%%%%%%%%%%%%%
\begin{figure}[th]
\vskip 24.5cm\relax\noindent\hskip -2cm
       \relax{\includegraphics{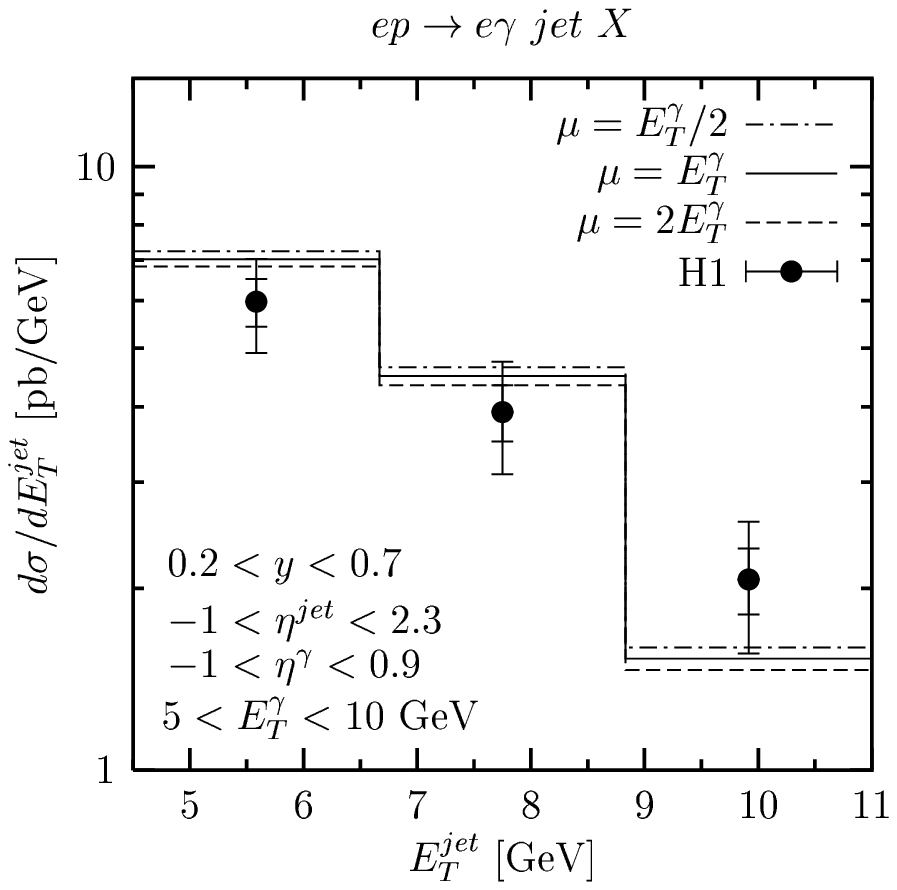}}
\vspace{-16cm}
%\caption{\small\sl The }
%\label{fig:ptjet}
%\end{figure}
%
%\begin{figure}[ht]
\vskip 24.8cm\relax\noindent\hskip -2cm
       \relax{\includegraphics{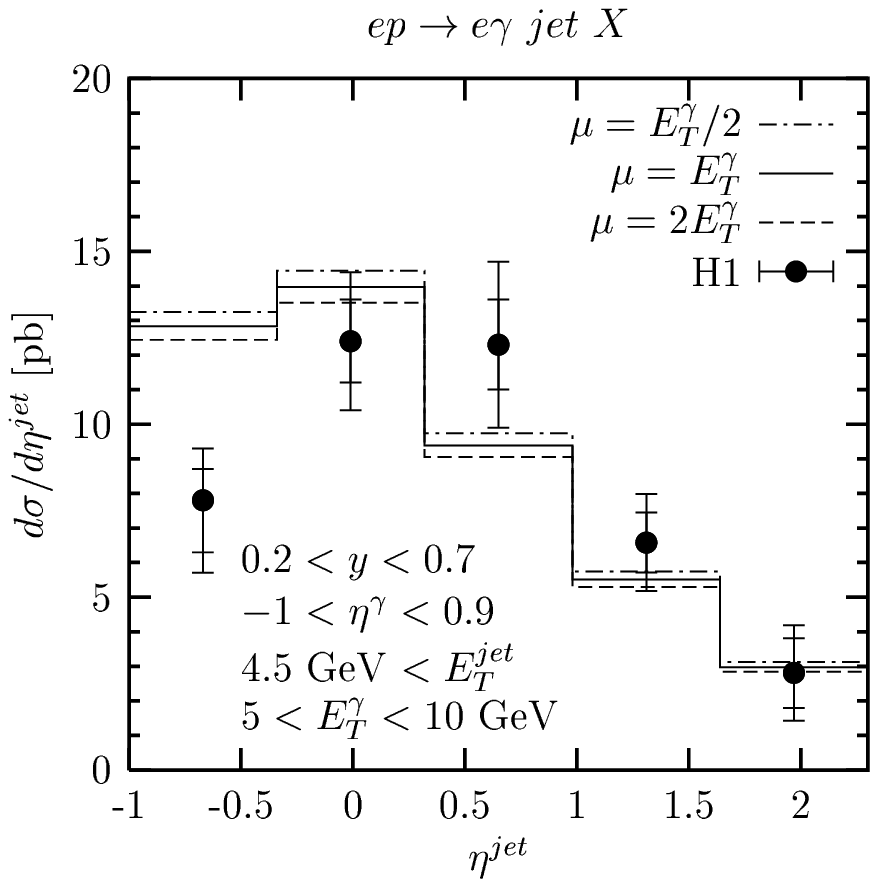}}
\vspace{-16cm} 
\caption{\small\sl The cross section $d\sigma /dE_T^{jet}$ (a)
and $d\sigma /d\eta^{jet}$ (b) for $\mu =0.5\cdot E_T^{\gamma}$ 
(dashed-dotted lines), $\mu = E_T^{\gamma}$ (solid lines) and
$\mu =2\cdot E_T^{\gamma}$ (dashed lines). The H1 Collaboration data
\cite{unknown:2004uv} are shown for comparison.}
\label{fig:etajet}
\end{figure}

\subsubsection{\boldmath $\mu$}

The dependence of the considered cross sections on the
choice of the
re\-normal\-ization/ factor\-ization scale, $\mu$, is not strong:
variations of $\mu$ from $E_T^{\gamma}$ to $E_T^{\gamma}/2$ or $2E_T^{\gamma}$ 
lead to changes of the cross section less than
3\% for $E_T^{\gamma}<E_{T,min}^{jet}$ and up to 5\% 
for $E_T^{\gamma}>E_{T,min}^{jet}$ (Fig. \ref{fig:ptgamma}b).
The predictions for various $\mu$ are also shown
in Fig. \ref{fig:etajet}.
In each presented here bin the dependence on the choice of $\mu$ is less
than $\pm$5\% for both $E_T^{jet}$ (Fig. \ref{fig:etajet}a) and 
$\eta_{jet}$ (Fig. \ref{fig:etajet}b) distributions. 
The total cross section integrated over the kinematic range considered by
the H1 Collaboration \cite{unknown:2004uv} vary by $\pm$ 3.4\%.
Since the effect of the variation of the 
$\mu$ scale is not large, the calculation seems to be stable, and
one can expect that the contributions of higher orders
are not sizable. 

\begin{figure}[th]
\vskip 24.5cm\relax\noindent\hskip -2cm
       \relax{\includegraphics{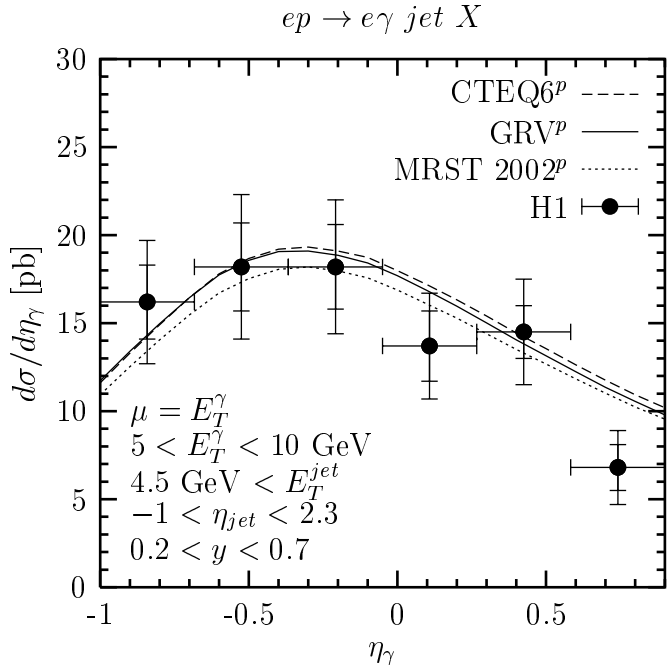}}
\vspace{-16cm} 
\caption{\small\sl The cross section $d\sigma /d\eta^{\gamma}$ for
MRST2002$^p$ (NLO)~\cite{Martin:2002aw} (dotted line), 
GRV$^p$ (NLO)~\cite{Gluck:1995uf} (solid line) and
CTEQ6$^p$ (NLO)~\cite{Pumplin:2002vw} (dashed line) parton densities in the
proton used with GRV$^{\gamma, frag}$ (NLO)~\cite{Gluck:1992ee,Gluck:1993zx}
densities in the photon and fragmentation functions. The H1 Collaboration
data \cite{unknown:2004uv} are also shown.}
\label{fig:etagamma}
\end{figure}

\subsubsection{\boldmath $f_p$}

Next, we have checked the sensitivity to the choice of parametrizations. 
In Fig.~\ref{fig:etagamma} the results obtained using different 
parton densities in the proton are shown. The predictions of 
CTEQ6$^p$ (NLO)~\cite{Pumplin:2002vw} are 6\% higher than
the predictions of MRST2002$^p$ (NLO)~\cite{Martin:2002aw}. The 
GRV$^p$ (NLO)~\cite{Gluck:1995uf} densities 
give results higher than MRST2002$^p$ by 5-7\% at negative $\eta_{\gamma}$,
and 3-5\% at positive $\eta_{\gamma}$. Differencies between CTEQ6$^p$
and GRV$^p$ do not exceed 4\%.

\subsubsection{\boldmath $f_{\gamma}$}

\begin{figure}[t]
\vskip 24.5cm\relax\noindent\hskip -2cm
       \relax{\includegraphics{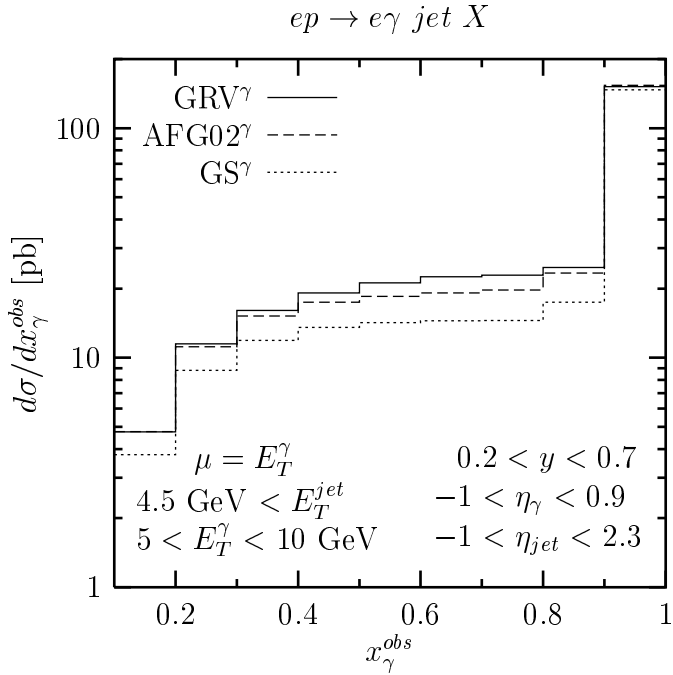}}
\vspace{-16cm} 
\caption{\small\sl The cross section $d\sigma /dx_{\gamma}^{obs}$
for GRV$^{\gamma}$ (NLO)~\cite{Gluck:1992ee} (solid line),
AFG02$^{\gamma}$ (NLO)~\cite{Fontannaz:2002nu} (dashed line) and
GS$^{\gamma}$ \cite{Gordon:1997pm} (dotted line) parton densities in the 
photon used with GRV$^{p, frag}$ (NLO) \cite{Gluck:1995uf,Gluck:1993zx}
densities in the proton and fragmentation functions.}
\label{fig:x}
\end{figure}

The most important variables for testing the structure
of colliding particles are the fractional momenta of partons
in these particles. Below we consider the distribution of 
the fractional momentum in the photon, 
%\PP 
however since the theoretical
variable $x_{\gamma}$ is not a good observable, in experimental analyses 
some estimations of $x_{\gamma}$ are used instead.
%\PP 
We  consider here the observable $x_{\gamma}^{obs}$, which
is defined as \cite{Breitweg:1997pa,h12003}
\footnote{The variable $x_{\gamma}^{obs}$
is equal to the ``theoretical'' one, $x_{\gamma}$, for $2\ra 2$ processes with
the direct final photon. For the processes with a larger number of partons
in the final state and for the processes with the parton-to-photon
fragmentation, the $x_{\gamma}$ and $x_{\gamma}^{obs}$ differ.}:
\be
x_{\gamma}^{obs}=(E_T^{jet}e^{-\eta^{jet}}+E_T^{\gamma}e^{-\eta^{\gamma}})
/2yE_e .
%= [(E^{jet}-p_z^{jet})+(E^{\gamma}-p_z^{\gamma})] /2yE_e .
\ee
In Fig.~\ref{fig:x} the $x_{\gamma}^{obs}$ distributions are shown
for different parton densities in the photon.
The GS$^{\gamma}$ (NLO) parametrization~\cite{Gordon:1997pm} give predictions
lower than GRV$^{\gamma}$ (NLO)~\cite{Gluck:1992ee} by 20-36\%
for $x_{\gamma}^{obs}<0.9$. This large difference is due to
the specific treatment of the charm contribution
in the GS parametrization (see Sec. \ref{results2}).
The results obtained with use of AFG$^{\gamma}$ (NLO)~\cite{Aurenche:1994in} 
and AFG02$^{\gamma}$ (NLO)~\cite{Fontannaz:2002nu} are very similar, and only 
the latter is shown in Fig.~\ref{fig:x}. It gives predictions 
up to 15\% lower than GRV$^{\gamma}$ for $x_{\gamma}^{obs}<0.9$. At 
large-$x_{\gamma}^{obs}$ region, $x_{\gamma}^{obs}>0.9$, the 
cross section is dominated by the contribution of processes
with the direct initial photons, and the differences between
predictions obtained using various parametrizations are small.
For the total cross section integrated over all $x_{\gamma}^{obs}$, 
within the considered range of $y$, $\eta^{\gamma}$, $E_T^{\gamma}$,
$\eta^{jet}$ and $E_T^{jet}$, the difference between results based on
GRV$^{\gamma}$ and AFG02$^{\gamma}$ (G$^{\gamma}$S) is 4\% (16\% ).

Note, that there are new LO \cite{Cornet:2002iy,Cornet:2003ry,Cornet:2004ng}
and NLO \cite{Cornet:2004nb} parametrizations of the real photon 
structure with a special treatment of heavy quark contributions. 
For the first time for the photon they use the ACOT$_{\chi}$ scheme introduced
previously for the proton \cite{Tung:2001mv} and include the newest 
$F_2^{\gamma}$ data, never used in constructing
other parametrizations for the photon. 
We have compared the results obtained using the LO parametrization
of Cornet, Jankowski, Krawczyk and Lorca 
(CJKL$^{\gamma}$)~\cite{Cornet:2002iy} with results obtained using
GRV$^{\gamma}$(LO) \cite{Gluck:1992ee}. 
Despite the fact that parton densities in both
parametrizations differ considerably, they lead to similar results
for the prompt photon production at HERA. Predictions for
CJKL$^{\gamma}$(LO) are about 3\% lower than the predictions for
GRV$^{\gamma}$(LO) (in this comparison the $\mathcal{O}$$(\alpha_S)$ 
corrections 
were not taken into account and the GRV$^p$(LO)~\cite{Gluck:1995uf} 
densities in the proton were used).

\subsubsection{\boldmath $D_{\gamma}$}

We have also compared predictions of DO$^{frag}$ (LO)~\cite{Duke:1982bj}, 
GRV$^{frag}$ (NLO)~\cite{Gluck:1993zx} and BFG$^{frag}$ 
(NLO)~\cite{Bourhis:1997yu} fragmentation functions. 
The isolation requirement reduces the contribution of processes
involving the resolved final photon (Sec. \ref{results1}, Tabs.
\ref{tab1}, \ref{tab2}),
so the dependence on the choice of fragmentation functions is weak, even if
the fragmentation functions differ considerably from one another.
The total cross sections for the isolated photon plus jet production
obtained with DO$^{frag}$ and BFG$^{frag}$ (set I and set II)
are lower than the predictions of GRV$^{frag}$ by 2\% and 4\%, respectively.

\subsubsection{\boldmath $f_p$, $f_{\gamma}$, $D_{\gamma}$}

\begin{figure}[ht]
\vskip 24.5cm\relax\noindent\hskip -2cm
       \relax{\includegraphics{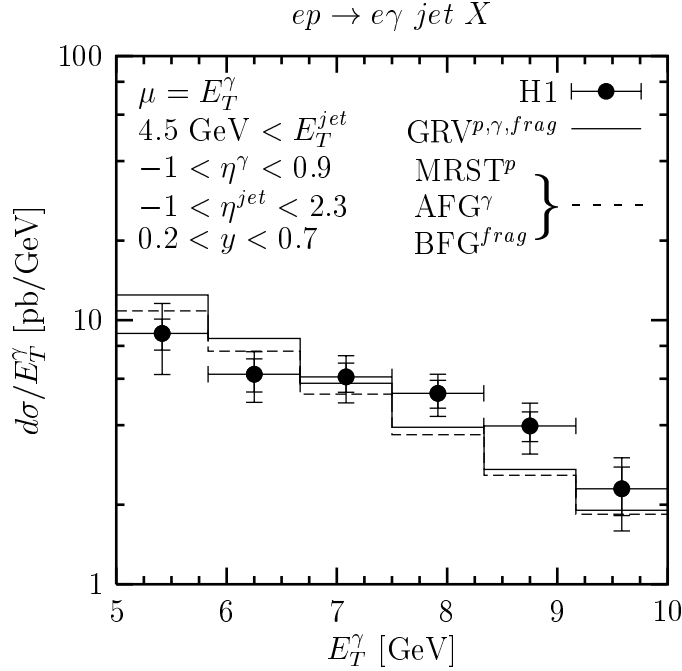}}
\vspace{-16cm} 
\caption{\small\sl The cross section $d\sigma /dE_T^{\gamma}$ averaged over
bins corresponding to the H1 Collaboration data \cite{unknown:2004uv}.
The predictions are obtained using GRV$^{p,\gamma,frag}$ 
\cite{Gluck:1995uf,Gluck:1992ee,Gluck:1993zx} (solid line) and
MRST1999$^p$ \cite{Martin:1999ww}, AFG$^{\gamma}$ \cite{Aurenche:1994in}, 
BFG$^{frag}$ \cite{Bourhis:1997yu} (dashed line) parton densities.}
\label{fig:ptgamma.param}
\end{figure}

The GRV distributions for the proton \cite{Gluck:1995uf}, 
photon \cite{Gluck:1992ee} and fragmentation \cite{Gluck:1993zx}
have been used as a reference in calculations discussed above,
and while performing the comparison each time only one parametrization has 
been changed. The differences observed in the total cross
section (i.e. in the cross section integrated within the considered
kinematic range) are not large (with an exception of the GS densities, which
give predictions considerably lower than the other densities
in the photon, as we discussed before). 
However, the differences can be larger if
one changes simultaneously all the used distributions.
For instance predictions of the MRST1999$^p$ \cite{Martin:1999ww}, 
AFG$^{\gamma}$ \cite{Aurenche:1994in} and BFG$^{frag}$ \cite{Bourhis:1997yu} 
set~\footnote{These parametrizations are used for comparison with
the H1 data and with the FGH predictions in Secs. \ref{Sisol:h1}, 
\ref{Sisol:fgh}, \ref{h1} and \ref{fgh}.}
are lower than the GRV$^{p,\gamma ,frag} $predictions by 10\% on average.
The comparison between both results is presented in Fig. 
\ref{fig:ptgamma.param}. 
The differences are large, up to 13\%, for 
$E_T^{\gamma}
\begin{minipage}[t]{10pt} \raisebox{3pt}{$<$} \makebox[-17pt]{}
\raisebox{-3pt}{$\sim$}\end{minipage}
7.5$ GeV, while for higher $E_T^{\gamma}$ both sets of parton densities
give predictions relatively close to each other and differences are
below 6\%.

\begin{figure}[ht]
\vskip 24.5cm\relax\noindent\hskip -2cm
       \relax{\includegraphics{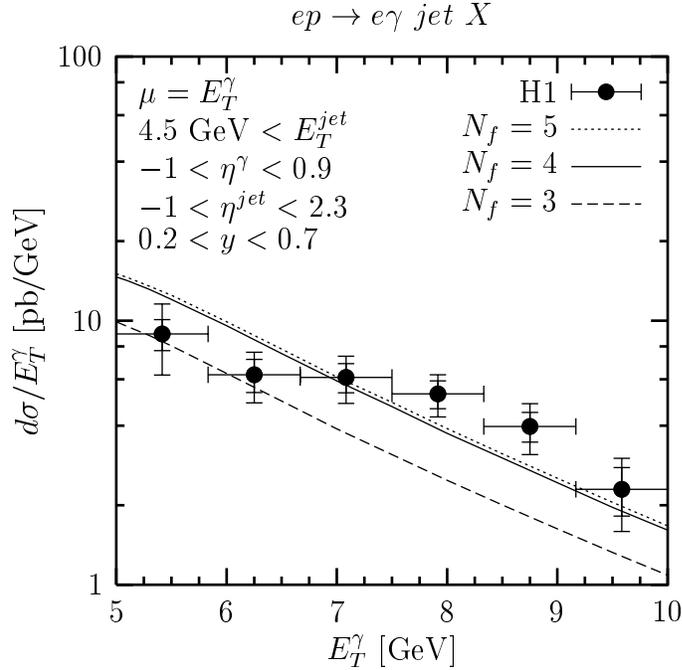}}
\vspace{-16cm} 
\caption{\small\sl The cross section $d\sigma /dE_T^{\gamma}$ obtained with
GRV$^{p,\gamma,frag}$ \cite{Gluck:1995uf,Gluck:1992ee,Gluck:1993zx}
parametrizations for $N_f$= 3 (dashed line), 4 (solid line) and 5 
(dotted line). The H1 Collaboration data \cite{unknown:2004uv} are shown.}
\label{fig:ptgamma.nf}
\end{figure}

\subsubsection{\boldmath $N_f$}

In the calculation the renormalization/factorization
scale $\mu = E_T^{\gamma}$ is used for $E_T^{\gamma}$ between 5 and
10 GeV (with exception of Fig. \ref{fig:ptgamma}, where a wider rage of
$E_T^{\gamma}$ is shown). 
It is a standard assumption 
that the number of active massless flavours at this scale 
is $N_f=4$ \cite{Gordon:1995km}-\cite{Fontannaz:2003yn}.
For a comparison, in Fig.
\ref{fig:ptgamma.nf} the results obtained with $N_f=3$ and $N_f=5$
are also shown. 
The contribution of the bottom quark is very small and the
results for $N_f=4$ and $N_f=5$ are similar.
Differences between these two results are much smaller than the
standard deviations of the H1 data \cite{unknown:2004uv} which
are also presented in Fig. \ref{fig:ptgamma.nf}.
On the other hand,
the contribution of the charm quark is large:
the results obtained using  $N_f=3$ are about 35\% below
the predictions for $N_f=4$. 
Neglecting of the charm mass may lead to a slight overestimation 
of the production rate, especially in the box contribution which is 
particularly sensitive to the change from $N_f=3$ to $N_f=4$.
However we do not expect that an improved treatment of the charm
contribution
would change our results qualitatively, since the energy  
$\mu = E_T^{\gamma}\ge 5$ GeV is several times larger than the charm mass.

%%%%%%%%%%%%%%%%%%%%%%%%%%%%%%%%%%%%%%%%%%%%%%%%%%%%%%%%%%%%%%%%%%%%%%%%%
\subsection{Comparison with the H1 data}\label{h1}
%%%%%%%%%%%%%%%%%%%%%%%%%%%%%%%%%%%%%%%%%%%%%%%%%%%%%%%%%%%%%%%%%%%%%%%%%

The QCD results for the photon plus jet production
shown in Figs.~\ref{fig:etajet}, \ref{fig:etagamma},
\ref{fig:ptgamma.param} and \ref{fig:ptgamma.nf}
are in reasonable agreement with the
data of the H1 Collaboration~\cite{unknown:2004uv} 
although some discrepancies
are present, especially for $\eta_{jet}<-0.3$ (Fig. \ref{fig:etajet})
and for the $E_T^{\gamma}$ distribution (Figs. \ref{fig:ptgamma.param}, 
\ref{fig:ptgamma.nf}).

\begin{figure}[t]
\vskip 9.2cm\relax\noindent\hskip -1cm
       \relax{\includegraphics{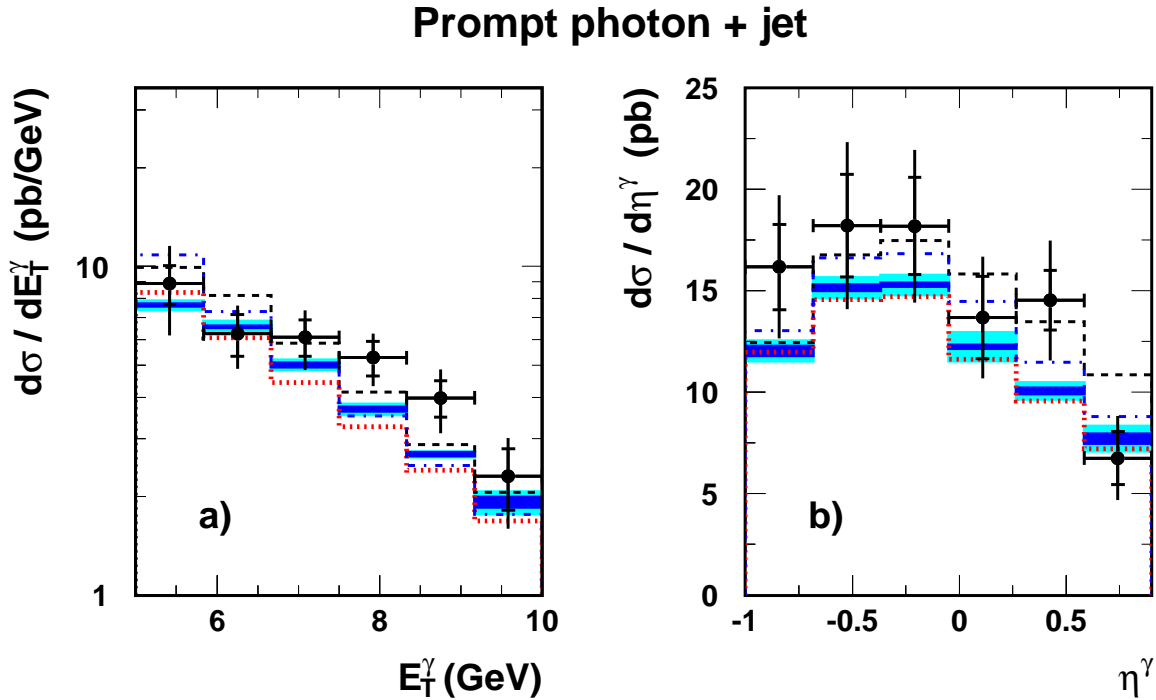}}
\vspace{0cm} 
\caption{\small\sl As in Fig.~\ref{fig4abh1} for the
$ep\ra e\gamma ~jet ~X$ process. The additional cuts for the jet are:
$E_T^{jet}>4.5$ GeV and $-1<\eta^{jet}<2.3$.
The figure is taken from \cite{unknown:2004uv}.}
\label{fig4cdh1}
\end{figure}

\begin{figure}[ht]
\vskip 18.5cm\relax\noindent\hskip -1cm
       \relax{\includegraphics{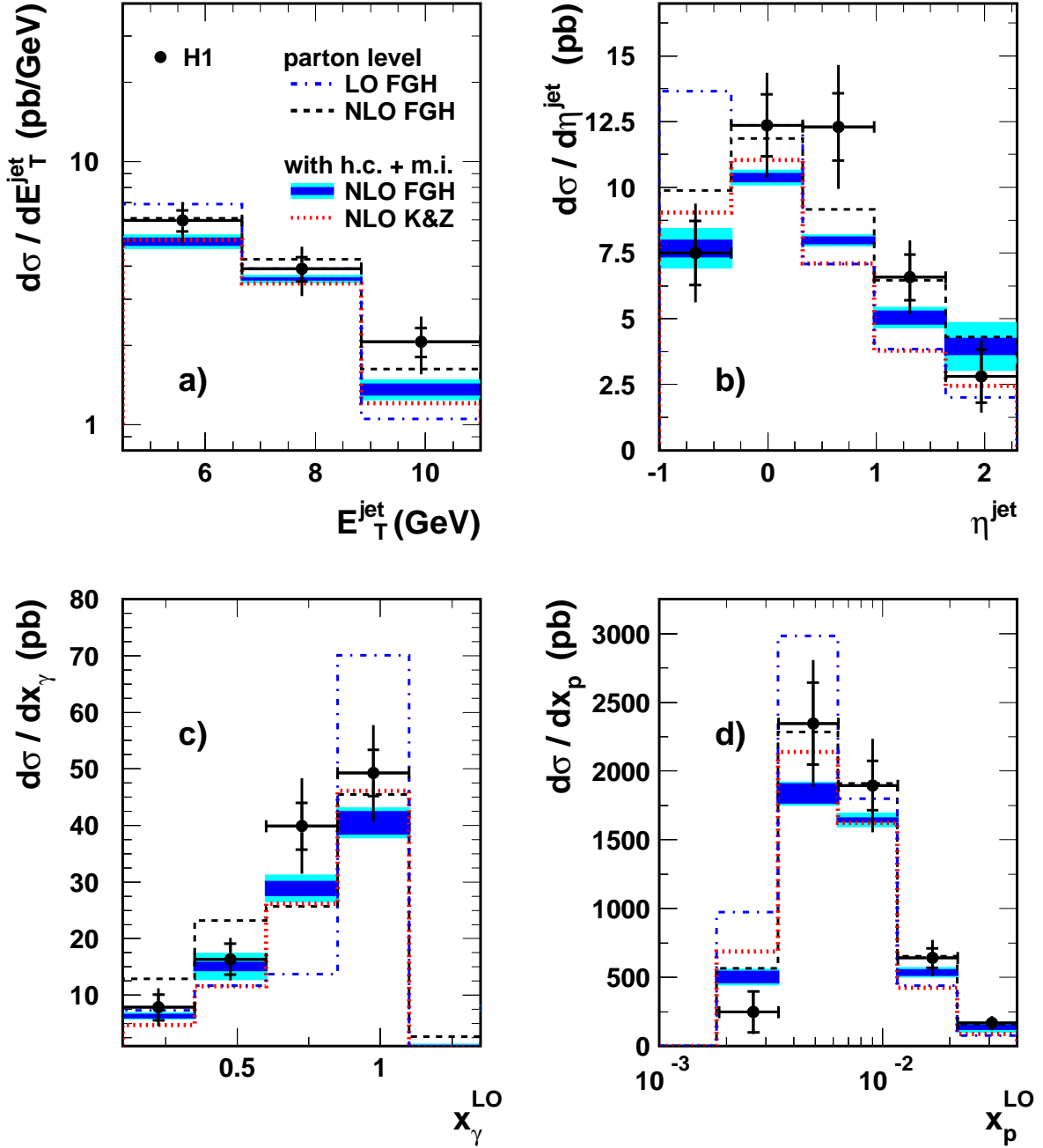}}
\vspace{-0.5cm} 
\caption{\small\sl As in Fig.~\ref{fig4cdh1} for 
$d\sigma /dE_T^{jet}$ (a) $d\sigma /d\eta^{jet}$ (b) 
$d\sigma /dx_{\gamma}^{LO}$ (c) and $d\sigma /dx_p^{LO}$ (d) cross sections.
The figure is taken from \cite{unknown:2004uv}.}
\label{fig5h1}
\end{figure}

However, according to~\cite{unknown:2004uv}, for the realistic 
comparison with the data, the pure perturbative QCD calculations
should be corrected for effects of hadronization and multiple
interactions (h.c.+m.i.), 
%\PP 
as discussed in Sec. \ref{Sisol:h1}. Such a comparison was performed by the
H1 Collaboration \cite{unknown:2004uv}, see
Figs. \ref{fig4cdh1} and \ref{fig5h1}, where
the predictions of Zembrzuski and Krawczyk (K\&Z) \cite{Zembrzuski:2003nu}
(dotted lines) corrected for h.c.+m.i. are presented 
together with the H1 Collaboration data
\cite{unknown:2004uv} and with predictions of Fontannaz, Guillet and Heinrich 
(FGH) \cite{Fontannaz:2001ek,Fontannaz:2001nq}.
To show the size of
the corrections, the FGH NLO results are plotted without (dashed lines)
and with (solid lines) h.c.+m.i~\footnote{See footnote $^{\ref{foot}}$
in Sec. \ref{Sisol:h1}.}.
Both K\&Z and FGH predictions are obtained using the
MRST1999$^p$ \cite{Martin:1999ww}, AFG$^{\gamma}$ 
\cite{Aurenche:1994in} and BFG$^{frag}$ \cite{Bourhis:1997yu} 
parton parametrizations.

The K\&Z and FGH calculations for the $ep\ra e\gamma ~jet ~X$ process
give somewhat better description of the data than in the case of
the $ep\ra e\gamma X$ process (Sec. \ref{pssres}). The agreement is seen
in most bins in Figs. \ref{fig4cdh1} and \ref{fig5h1}, especially when
one takes into account on the one hand the uncertainties 
of h.c.+m.i. corrections and 
on the other hand the theoretical
uncertainties due to the variation of the $\mu$ scale ($\pm 5\%$)
and to the choice of parton densities ($\sim 10\%$) (Sec. \ref{res}).
Nevertheless, the K\&Z and FGH predictions tend to underestimate
the H1 data. In some kinematic ranges the 
data are 1-2 standard deviations above the
K\&Z predictions e.g. for $6.7<E_T^{\gamma}<9.2$ GeV 
(Fig. \ref{fig4cdh1}a) and $0.3<\eta_{jet} < 1.6$ (Fig. \ref{fig5h1}b).

In Figs. \ref{fig5h1}c,d the cross sections
$d\sigma /dx_{\gamma}^{LO}$ and $d\sigma /dx_p^{LO}$ are shown, where
$x_{\gamma}^{LO}$ and $x_p^{LO}$ are defined as follows
\cite{unknown:2004uv}~\footnote{For
$2\ra 2$ processes the experimental variables $x_{\gamma}^{LO}$ and $x_p^{LO}$ 
are equal to corresponding theoretical variables
multiplied by $z$ (defined in Sec. \ref{Snon:x}):  
$x_{\gamma}^{LO}$=$z\cdot x_{\gamma}$,
$x_p^{LO}$=$z\cdot x_p$.}:
\bea
x_{\gamma}^{LO}=E_T^{\gamma}(e^{-\eta^{jet}}+e^{-\eta^{\gamma}})/2yE_e, 
\\
x_{p}^{LO}=E_T^{\gamma}(e^{-\eta^{jet}}+e^{-\eta^{\gamma}})/2E_p.
\eea
The K\&Z predictions are 1-1.5 standard deviations below the data
for $x_{\gamma}^{LO} < 0.85$ (Fig. \ref{fig5h1}c). 
Note, that better agreement is
obtained if GRV$^{p,\gamma,frag}$ 
\cite{Gluck:1995uf,Gluck:1992ee,Gluck:1993zx} parametrizations are
used (not shown in Fig. \ref{fig5h1} which is taken from the H1 paper), 
since they give predictions 18\% higher than
MRST1999$^p$, AFG$^{\gamma}$  
and BFG$^{frag}$ at $x_{\gamma}^{LO} < 0.85 $
(at $x_{\gamma}^{LO} > 0.85 $ the difference is 4\%).

At $x_p^{LO} > 0.012$ ($x_p^{LO} < 0.0034$) the K\&Z predictions are 
2 standard deviations below (above) the data (Fig. \ref{fig5h1}d).
An implementation of GRV$^{p,\gamma,frag}$ 
densities does not improve the description of the data (not shown).

The H1 Collaboration has also presented the cross sections 
$d\sigma /dp_{\perp}$~\cite{unknown:2004uv}, where $p_{\perp}$ 
is the component of the photon momentum perpendicular to the
scattering plane (see~\cite{unknown:2004uv} for the precise definition).
For $p_{\perp}\ne 0$ the $2\ra 2$ processes give no 
contribution,
so the cross section is sensitive to higher order processes only.
For $x_{\gamma}^{LO}>0.85$ the cross section is dominated by
$\mathcal{O}$$(\alpha_S)$ 
corrections to the processes with direct initial and final
photons, which are included in NLO in our calculation, and the data 
are in reasonable agreement with K\&Z predictions. 
On the other hand, for $x_{\gamma}^{LO}<0.85$
the cross section is dominated by $\mathcal{O}$$(\alpha_S)$
corrections to the processes
with resolved photons.
These contributions are not included in the K\&Z calculation,
so our predictions by definition 
can not describe such data.

%%%%%%%%%%%%%%%%%%%%%%%%%%%%%%%%%%%%%%%%%%%%%%%%%%%%%%%%%%%%%%%%%%%%%%%%%
\subsection{Comparison with other QCD predictions (FGH)}\label{fgh}
%%%%%%%%%%%%%%%%%%%%%%%%%%%%%%%%%%%%%%%%%%%%%%%%%%%%%%%%%%%%%%%%%%%%%%%%%

The calculation of Fontannaz, Guillet and Heinrich 
(FGH) \cite{Fontannaz:2001ek,Fontannaz:2001nq}
takes into account the $\mathcal{O}$$(\alpha_S)$ corrections to the 
resolved photon processes, which are not included in the
calculation presented in this work (Chapter \ref{nlo}, Sec. \ref{Sisol:fgh}).
In the considered kinematic range for the photon 
%\PP 
\underline{plus jet} production
the total cross section of FGH is about 4\% higher than our
predictions, so the total contribution of $\mathcal{O}$$(\alpha_S)$ 
corrections to the resolved
photon processes is relatively small.

The K\&Z and FGH results shown in Fig. \ref{fig4cdh1}b
differ by 5\% or less in the whole range of $\eta^{\gamma}$.
The differences are larger for other differential
cross sections (Figs. \ref{fig4cdh1}a, \ref{fig5h1}). 
At $E_T^{\gamma}>6.7$ GeV the FGH predictions are about 13\% higher,
while at $5<E_T^{\gamma}<5.8$~GeV they are 7\% lower than
predictions of K\&Z (Fig. \ref{fig4cdh1}a)
The largest differences are for $1<\eta^{jet}<1.6$ and $1.6<\eta^{jet}<2.3$,
where the FGH predictions are above K\&Z by 33\% and 63\%, respectively.
Despite these divergences, both calculations
lead to a similar description of the H1 data
shown in Figs. \ref{fig4cdh1} and \ref{fig5h1}.

Note that the FGH predictions are closer to the H1 data 
than the K\&Z predictions for $d\sigma /dp_{\perp}$ with 
$x_{\gamma}^{LO}<0.85$~\cite{unknown:2004uv}, since 
we do not include processes which contribute in this region
(Sec. \ref{h1}). 
%%%%%%%%%%%%%%%%%%%%%%%%%%%%%%%%%%%%%%%%%%%%%%%%%%%%%%%%%%%%%%%%%%%%%%%%%%
\chapter{Probing the gluon content of the photon}\label{Sglu}
%%%%%%%%%%%%%%%%%%%%%%%%%%%%%%%%%%%%%%%%%%%%%%%%%%%%%%%%%%%%%%%%%%%%%%%%%%

The photoproduction processes in the electron-proton scattering are
sensitive to the parton densities in the proton as well as in the photon
and provide an opportunity to probe the photon structure at the HERA
collider (for a review of the data see \cite{Krawczyk:2000mf}).
Also the photoproduction of photons with a large transverse momentum
(the Deep Inelastic Compton process) was considered as a possible tool
to test the quark and gluon densities in the photon and proton
\cite{Krawczyk:1990nq}-\cite{Aurenche:1992sb}, 
\cite{Gordon:1994sm}-\cite{Zembrzuski:2003nu}, 
\cite{Krawczyk:1997zv,Krawczyk:1999eq}. 
In particular, it was found that the contribution involving the gluon 
density in the photon is especially large when the final photon is
produced in the forward (proton) direction 
\cite{Krawczyk:1990nq}-\cite{Aurenche:1992sb}, 
\cite{Krawczyk:1997zv,Krawczyk:1999eq}.
Recently new analysis devoted to 
the possibility of measuring the gluon density in the photon and proton
in the $ep\ra e\gamma ~jet ~X$ process at HERA were presented
in \cite{Fontannaz:2003yn,Heinrich:2003vx,Fontannaz:2004qv}.

So far we have considered the production of the photon
with the large transverse momentum (transverse energy) in the
electron-proton scattering where the mediating photons have been
quasi-real, $Q^2\approx 0$, and they spectrum have been given
by the equivalent photon approximation 
\cite{vonWeizsacker:1934sx}-\cite{Frixione:1993yw}
(Sec. \ref{Cnlo:epa}). Now, we extend the
study including the effect of small, but non-zero virtuality,
$Q^2\ne 0$.

The first attempt to describe the Deep Inelastic Compton 
process at HERA using the parton densities in the \underline{virtual} photon 
can be found in \cite{azem,za}, where we tested in LO the validity
of the equivalent photon approximation.
Next, we have examined the usefulness of this process  
to study at the  HERA collider the structure of a 
virtual photon, and in particular its gluonic content 
\cite{Krawczyk:1997zv,Krawczyk:1999eq}.
In this chapter we discuss shortly some of our LO results obtained
in \cite{Krawczyk:1997zv,Krawczyk:1999eq}.

~ \\ ~ \\ ~ \\ ~ \\ ~

%%%%%%%%%%%%%%%%%%%%%%%%%%%%%%%%%%%%%%%%%%%%%%%%%%%%%%%%%%%%%%%%%%%%%%%%%%
\section{Calculation of the cross section}
%%%%%%%%%%%%%%%%%%%%%%%%%%%%%%%%%%%%%%%%%%%%%%%%%%%%%%%%%%%%%%%%%%%%%%%%%%

The invariant differential cross section for the deep inelastic 
electron-proton scattering can be written in 
the following form:
\bea
E^{\prime}_e {d\sigma^{ep\ra eX}\over d^3p_e^{\prime}}= 
\Gamma \Bigl (\sigma^{\gamma^*p\ra X}_T
+\epsilon \sigma^{\gamma^*p\ra X}_L \Bigl),
\label{prob1}
\eea
where $\sigma^{\gamma^*p\ra X}_T$  
($\sigma^{\gamma^*p\ra X}_L$) is the cross section
for the interaction between the proton and the virtual photon 
polarized transversely (longitudinally)
and $p_e^{\prime}$ ($E^{\prime}_e$) stands for the final electron 
momentum (energy).
Coefficients $\Gamma$ and $\epsilon\Gamma$ 
are functions of energies and momenta of the electron in 
initial and final states, see \cite{Halzen:1984mc}. They can
be interpreted as the probability of emitting by the initial 
electron the virtual photon polarized transversely and longitudinally. 
If so, we can use (\ref{prob1}) to obtain the differential cross section
for the $ep\ra e\gamma X$ process taking into account the virtuality
of the exchanged photon:
\bea
E^{\prime}_e  {d\over d^3p_e^{\prime}}
E_{\gamma}{d\sigma^{ep\ra e\gamma X}\over d^3p_{\gamma}}= 
\Gamma \Bigl (E_{\gamma}{d\sigma^{\gamma^*p\ra \gamma X}_T\over 
d^3p_{\gamma}}+\epsilon E_{\gamma}{d\sigma^{\gamma^*p\ra \gamma X}_L\over
d^3p_{\gamma}}\Bigl),
\eea
where $p_{\gamma}$ ($E_{\gamma}$) stands for the final photon
momentum (energy).
Since the cross section for the 
reaction $ep\ra e\gamma X$ is dominated by 
the exchange of photons with small virtuality, 
one  can neglect a contribution due to  
the longitudinal polarization \cite{ula}.
%\footnote{In Ref.\cite{ula} the contribution due to longitudinal photons 
%is discussed for the Born process; 
%this contribution is found to be negligible
%also for $Q^2$ larger than studied by us ).
Assuming that the exchanged photon has only the
transverse polarization we obtain:
\bea
E_{\gamma}{{d\sigma^{ep\ra e\gamma X}}
\over{d^3p_{\gamma }}}
=\int {d^3p^{\prime}_e\over E^{\prime}_e}\,
\Gamma E_{\gamma}{{d\sigma^{\gamma^*p\ra \gamma X}}\over 
{d^3p_{\gamma}}}.
\eea
Our aim is to study the sensitivity of the cross section to the 
gluon distribution in the photon. To achieve this goal we include
the Born process (Fig. \ref{figborn}) and the processes with the 
resolved initial photon (Fig. \ref{figsingi}), and we omit
contributions of other processes:
\bea
E_{\gamma}{d\sigma^{\gamma^*p\ra\gamma X}\over 
d^3p_{\gamma}} =
%\displaystyle
\sum_{q,\bar{q}}\int\limits_0^1dx
f_{q/p}(x, \mu^2)
E_{\gamma}
{d\sigma^{\gamma^*q\ra\gamma q}\over d^3p_{\gamma}}
+ ~ ~ ~ ~ ~ ~ ~ ~ ~ ~ ~ ~ ~ ~ ~ ~ ~ ~ ~ ~ ~ ~ ~ ~ ~ 
\\
\sum_{a=q,\bar{q},g} \int\limits_0^1 dx_{\gamma}
\sum_{b=q,\bar{q},g} \int\limits_0^1 dx 
%\displaystyle
f_{a/{\gamma^*}}(x_{\gamma}, \mu^2,Q^2)
f_{b/p}(x, \mu^2)
E_{\gamma}{d\sigma^{ab\ra\gamma d}\over 
d^3p_{\gamma}} ,
\eea
where $f_{b/p}$ and $f_{a/\gamma^*}$ are the parton densities in the
proton and in the \underline{virtual} photon.
In the partonic cross section $d\sigma^{\gamma^*q\ra\gamma q}$
as well as in the partonic distribution $f_{a/\gamma^*}$
we take the virtuality $Q^2$ of the photon emitted by 
the electron exactly as it follows from the kinematics of the process. 

%%%%%%%%%%%%%%%%%%%%%%%%%%%%%%%%%%%%%%%%%%%%%%%%%%%%%%%%%%%%%%%%%%%%%%%%%%
\section{Numerical results}
%%%%%%%%%%%%%%%%%%%%%%%%%%%%%%%%%%%%%%%%%%%%%%%%%%%%%%%%%%%%%%%%%%%%%%%%%%

In the calculation we use the GRS (LO) \cite{Gluck:1994tv}
and the GRV (LO) \cite{Gluck:1995uf}
parton distributions in the virtual photon and in the proton,
respectively. 
The number of flavours is assumed $N_f = 3$ -
the maximum number of active massless quarks in the GRS
parametrization. We take the QCD parameter $\Lambda_{QCD}=0.2$ GeV
and the hard scale equal to the transverse energy of the final photon,
$\mu =E_T^{\gamma}$.
Calculations are performed for the initial electron and proton
energies in the HERA accelerator: $E_e=30$ GeV$^2$ and $E_p=820$ GeV$^2$
with the virtualities of the mediating photon, $Q^2$ ranging from
$10^{-7}$ GeV$^2$ to 2.5 GeV$^2$. Note that due to a smooth behaviour 
of the GRS parametrization in the limit $Q^2\ra 0$ we were able to perform 
the calculation also for $Q^2$ below $\Lambda^2_{QCD}$.

We study the differential cross section
\bea
E_{\gamma}{d\sigma^{ep\ra e\gamma X}\over {d^3p_{\gamma} dQ^2 dy}},
\label{label.x}
\eea
for the transverse energy of the final photon $E_T^{\gamma}= 5$ GeV and
the energy of the mediating photon $E_{\gamma^*}=y E_e$ with fixed
$y=0.5$.

As it was expected, the cross section decreases for increasing 
values of the virtuality $Q^2$. For example the predictions obtained
for $Q^2=0.25$ GeV$^2$ (2.5 GeV$^2$) are one order (three orders)
of magnitude lower than the predictions for $Q^2=0.03$ GeV$^2$
\cite{Krawczyk:1997zv}.

\begin{figure*}[t]
\vspace{6.5cm}
\vskip 0.cm\relax\noindent\hskip 0cm
       \relax{\includegraphics{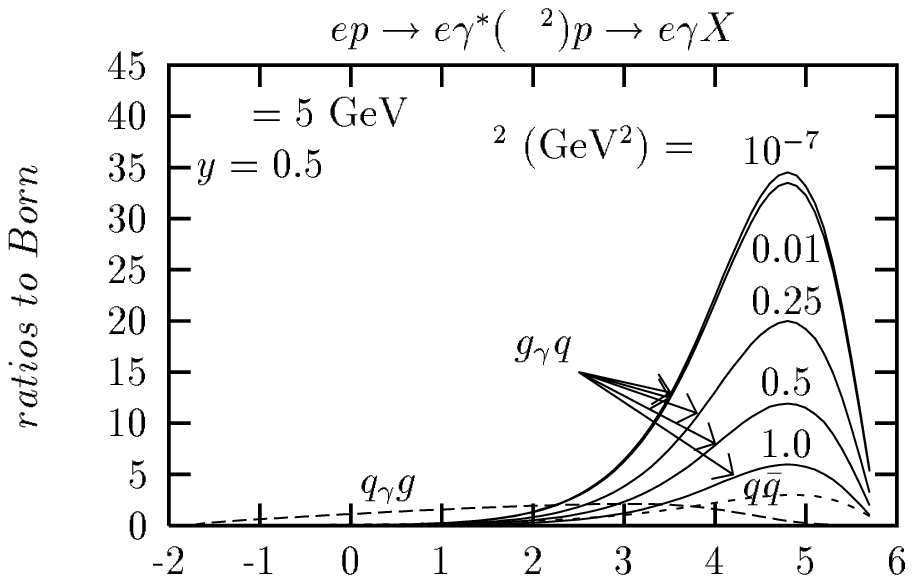}}
\begin{picture}(0,0)
\put(207,0){$\eta^{\gamma}$}
\put(203,168){Q}
\put(189,135){Q}
\put(112,143){$E_T^{\gamma}$}
\end{picture}
\vspace{0cm}
\caption{The results for the $ep \ra e\gamma X$ reaction including
non-zero virtualities, $Q^2$, of the mediating photon obtained 
for $y$=0.5 and $E_T^{\gamma} = 5$ GeV. The curves represent the
cross section (\ref{label.x}) for the
$g_{\gamma^*}q_p\ra \gamma q$ (solid lines),
$q_{\gamma^*}g_p\ra \gamma q$ (long-dashed line)
as well as ${\bar q_{\gamma^*}} q_p \ra \gamma g$ and
$q_{\gamma^*} {\bar q_p}\ra \gamma g$ (short-dashed line)
processes divided by the corresponding Born contribution.
The predictions for $g_{\gamma^*}q_p\ra \gamma q$ process are obtained
with $Q^2=10^{-7}$, 0.01, 0.25, 0.5 and 1 GeV$^2$, while
for the other processes involving resolved photons $Q^2=0.1$ GeV$^2$ was used.}
\label{fig.prob}
\end{figure*}

The processes initiated by the gluonic content of the virtual photon
dominate in the cross section in the forward (proton) direction,
i.e. for large photon rapidities, $\eta_{\gamma}$ 
\cite{Krawczyk:1997zv,Krawczyk:1999eq}. 
This effect is presented in Fig. \ref{fig.prob}, 
where the contributions due to the resolved virtual photon 
divided by the Born contribution are shown as a functions of $\eta_{\gamma}$.
The results for the process $g_{\gamma^*} q_p\ra \gamma q$ are obtained
taking $Q^2$ between $10^{-7}$ GeV$^2$ and 1 GeV$^2$.
For comparison we show the results for $q_{\gamma^*}g_{p}\ra \gamma q$ and 
$q_{\gamma^*}{\bar q_p}(\bar q_{\gamma^*}{q_p})\rightarrow\gamma g$
processes obtained using $Q^2 = 0.1$ GeV$^2$.

The large dominance of the contribution due to the
gluonic content of the virtual photon over the Born contribution 
is seen at large $\eta_{\gamma}$ in the whole considered range of 
the photon virtualities. The largest ratio, up to factor of 35, is 
obtained for the small virtualities. It is important since 
in the limit $Q^2\ra 0$ the value of the cross section is the largest.
Thus the cross section \underline{integrated} over $Q^2$ between 0 and, say,
1 GeV$^2$~~\footnote{This range of $Q^2$ is considered 
for the photoproduction of prompt photons at HERA, see Sec. \ref{Cnlo:epa}.}
is also strongly dominated by the process $g_{\gamma} q_p\ra \gamma q$
in the forward direction.

The process involving the gluonic content of the photon dominates
at large $\eta_{\gamma}$ not only over the Born contribution but also
over the contribution of processes involving quarks from the photon,
see  Fig. \ref{fig.prob} and Ref. \cite{Krawczyk:1997zv}.
This gives an opportunity of measuring the gluon content of the
real and virtual photon at HERA.

%This can have consequences for the possibility of measuring 
%the gluon content of the real and virtual photon at HERA.
%On the other hand, from experimental point of view, such
%measurements can be hard to realize in present experiments
%since the cross section
%at large $\eta_{\gamma}$ is smaller than at central 
%$\eta_{\gamma}$ values (see Figs. \ref{fig.noneg1}, \ref{fig.noneg2}),
%and moreover the backgrounds are large.

%%%%%%%%%%%%%%%%%%%%%%%%%%%%%%%%%%%%%%%%%%%%%%%%%%%%%%%%%%%%%%%%%%%%%%%%%%
\chapter{Summary}\label{Ssum}
%%%%%%%%%%%%%%%%%%%%%%%%%%%%%%%%%%%%%%%%%%%%%%%%%%%%%%%%%%%%%%%%%%%%%%%%%%

We have presented a NLO QCD calculation for the 
photoproduction of the isolated photons and the isolated photons
associated with jets at the DESY HERA $ep$ collider.

Our calculation includes set of diagrams different than
other existing NLO QCD calculations for this process.
This difference arises from different treatment of parton densities
in the photon and parton-to-photon fragmentation functions.
In our approach these parton densities and fragmentation functions
are of order $\mathcal{O}$$(\alpha)$ while in most of other 
calculations they are treated as quantities being of order
$\mathcal{O}$$(\alpha/\alpha_S)$. Our counting originate from the
fact that the ``structure'' of the photon and the fragmentation
into the photon arises from the purely electromagnetic processes
$\gamma \ra q\bar{q}$ and $q\ra \gamma\bar{q}$, respectively.
It ensures a cancellation to a large extend of dependences
on the 
%\PP 
choice of the renormalization/factorization scale $\mu$ in the cross section.

We have presented a method to divide (slice) the three-body
phase space which differs from methods applied in other calculations
for the considered process. Our method allows to obtain 
relatively simple analytical singular-free formulae for the
$\mathcal{O}$$(\alpha_S)$ corrections in each part of the phase space.
%In some parts of the phase space we could simply use the
%same formulae as for the non-isolated photon production.
%We have derived and presented the formulae corresponding to
%$\mathcal{O}$$(\alpha_S)$ in the other parts of the phase space.
Dividing of the phase space is introduced in order to 
implement exactly isolation restrictions and other kinematic cuts.
It was used to obtain predictions for the isolated photon and
isolated photon plus jet production at HERA.

The numerical predictions obtained using the phase space slicing were compared
with the predictions obtained using the small cone approximation.
In this approximation the isolation cuts
in the $\mathcal{O}$$(\alpha_S)$ corrections to the Born process
are implemented in an approximated way
while in other contributions the isolation is implemented exactly.
We confirm the observation of other author, that the small cone
approximation gives very accurate predictions. However we have
found that it is due to the fact that the $\mathcal{O}$$(\alpha_S)$ 
corrections to the Born process are small. These
$\mathcal{O}$$(\alpha_S)$  corrections alone are obtained
within the small cone approximation with very low accuracy.

The theoretical uncertainties of our predictions are well under control.
The predictions vary within 10\% for various used parton parametrizations.
The dependence on the choice of the renormalization/factorization
scale $\mu$ for the isolated photon (isolated photon plus jet) productions
is below $\pm$6\% ($\pm$5\%) in \underline{each} considered kinematic region,
and the average dependence is $\pm$5\% ($\pm$3\%). 
This small sensitivity to the choice of $\mu$ may indicate
that the not included contributions of higher orders
(NNLO or higher) are not sizable.

%\PP 
Other NLO QCD calculations  (including some diagrams which are of
NNLO order from our point of view) give predictions 10-20\%
(4\%) 
higher on the average than ours for the isolated photon 
(isolated photon plus jet) 
production. However there are some kinematic
ranges where the differences are larger, up to 70\% for the
isolated photon production in the range
0.5 $< y <$ 0.9 and $0.7<\eta^{\gamma}< 0.9$ (where our predictions
give the best description of the data). This large difference
is not supported by our study of the sensitivity to the choice of $\mu$,
which seems to suggest smaller effect of higher order contributions.
%\PP It is worth to stress that the corresponding difference for 
%photon + jet final state are 2-5 times smaller (it is equal to 4\%).
To understand 
%\PP 
these differences we should know the dependence of other
predictions on the scale $\mu$ in this particular range,
which however was not provided.

We have studied in details
the influence of the isolation restrictions as well as other 
experimental cuts. In particular, we have shown that the
measurements in the central rapidity range considered in current
experiments are not much sensitive to the parton densities in the
photon in the low-$x_{\gamma}$ region.

Effects of symmetric cuts for the photon and jet transverse energies
were studied. We have found that the NLO differential cross section
$d\sigma /dE_T^{\gamma}$ is a discontinuous function at 
$E_T^{\gamma}=E_{T,min}^{jet}$. This is due to constraints imposed
on the phase space of soft gluons.
We confirm the known result that 
to avoid theoretical uncertainties the asymmetric cuts are preferred.

Our predictions and computer program were used by 
experimental groups for comparison with data.
The QCD predictions for the photon plus jet
production tend to lie below
the data, nevertheless they agree with the data in most of bins.
On the other hand, none of existing predictions describes 
the data satisfactory well for the photon production with no 
cuts for the jet. 
%\PP
It means that the cuts for the observed jet remove from the cross section 
contributions of such kinematic configurations for which the disagreement 
with the data and the differences between predictions are the largest.

%To summarize the results of the comparisons: for the photon plus jet 
%production the NLO QCD predictions give reasonably description of the data,
%and the differences between our and other predictions 
%are usually not large.
%On the other hand, for the photon production the differences between 
%predictions are larger and all the predictions disagree with data
%in some kinematic ranges. 

%Recently large effects of multiple interactions and hadronization 
%corrections have been found in experimental analyses. These 
%effects have been included by the experimental group
%in a comparison between data and predictions (including our predictions).
%These corrections do not improve considerably the description of the 
%data for the prompt photon plus jet production. Moreover, the differences
%between the predictions and the data for the prompt photon
%production (with no jet requirement) are larger if the effects
%of hadronization corrections and multiple interactions is included.
%\PP {\bf kt effect...}

Finally, we 
%\PP 
have pointed out the sensitivity of the 
%\PP 
Deep Inelastic Compton cross section
to the gluon density in the photon also for the \underline{virtual}
photon. We 
%\PP 
presented a formula which allows to include 
the parton densities in the virtual photon in the cross section for 
the $ep\ra e\gamma X$ (or  $ep\ra e\gamma ~jet ~X$) reaction.
We have found that, as it was expected, the processes
initiated by the gluon arising from the photon 
dominate in the cross section in the forward direction, i.e. for
large $\eta_{\gamma}$.
% the possibility 
%of observing the effect of the gluonic content in 
%the real and virtual photon in $ep$ collisions.
\chapter*{Tables}
\setcounter{chapter}{0}
\addcontentsline{toc}{chapter}{Tables}

The tables contain our predictions for the $ep\ra e\gamma X$ and
$ep\ra e\gamma ~jet ~X$ photoproduction processes at the HERA collider.
The initial electron and proton energies, the final photon isolation
parameters ($\epsilon=0.1$, $R=1$) as well as the kinematic ranges of $y$,
$\eta^{\gamma}$, $\eta^{jet}$, $E_T^{\gamma}$ and $E_T^{jet}$
are taken from the H1 \cite{unknown:2004uv} (Tabs. \ref{tab1}, \ref{tab3})
and ZEUS \cite{Breitweg:1999su} (Tab. \ref{tab2}) Collaborations papers
and are described in the captions of corresponding tables.
\\
~\\
The following notation is used for various contributing processes:
\\ 
$\bullet$ $\mathcal{O}$($\alpha_S$) = $\mathcal{O}$($\alpha_S$) 
corrections to the Born process,
\\
$\bullet$ dir,dir = direct initial and final $\gamma$,
\\
$\bullet$ dir,frag = direct initial $\gamma$ and fragmentation into final 
$\gamma$,
\\
$\bullet$ res,dir = resolved initial $\gamma$ and direct final $\gamma$,
\\
$\bullet$ res,frag = resolved initial $\gamma$ and fragmentation into final 
$\gamma$.

\begin{table*}[h]
\begin{center}
\caption{The cross section for the non-isolated (inclusive) photon, 
the isolated photon, 
and the isolated photon + jet production. The photon transverse energy is 
integrated over the range $5\le E_T^{\gamma}\le 10$ GeV. The initial 
electron and proton energies are $E_e=27.6$ GeV and $E_p=920$ 
GeV~\cite{unknown:2004uv}. 
The results for the non-isolated photon are integrated over the whole
range of $y$ and $\eta^{\gamma}$.
The isolated photon cross section is calculated for
%within the limits
$0.2\le y\le 0.7$ and $-1.\le\eta^{\gamma}\le 0.9$. For the 
photon+jet production there are additional cuts, $4.5$ GeV $\le E_T^{jet}$
and $-1.\le\eta^{jet}\le 2.3$.}
\label{tab1}
\vspace{0.5cm}
\begin{tabular}{|c|c|c|c|c|c|c|c|} 
\hline \hline
\raisebox{0pt}[12pt][6pt]{[pb]} & 
\raisebox{0pt}[12pt][6pt]{total} & 
\multicolumn {3}{|c|}{dir,dir}&
\raisebox{0pt}[12pt][6pt]{res,dir}&
\raisebox{0pt}[12pt][6pt]{dir,frag}&
\raisebox{0pt}[12pt][6pt]{res,frag} \\
\cline {3-5}
\raisebox{0pt}[12pt][6pt]{} & 
\raisebox{0pt}[12pt][6pt]{} & 
\raisebox{0pt}[12pt][6pt]{Born} &
\raisebox{0pt}[12pt][6pt]{$\mathcal{O}$($\alpha_S$)} & 
\raisebox{0pt}[12pt][6pt]{box} &
\raisebox{0pt}[12pt][6pt]{}&
\raisebox{0pt}[12pt][6pt]{}&
\raisebox{0pt}[12pt][6pt]{} \\
\hline
\raisebox{0pt}[12pt][6pt]{non-isolated $\gamma$} & 
\raisebox{0pt}[12pt][6pt]{240.6} & 
\raisebox{0pt}[12pt][6pt]{85.8} & 
\raisebox{0pt}[12pt][6pt]{5.0} & 
\raisebox{0pt}[12pt][6pt]{15.2} & 
\raisebox{0pt}[12pt][6pt]{60.0} &
\raisebox{0pt}[12pt][6pt]{27.0} &
\raisebox{0pt}[12pt][6pt]{47.6}\\
\raisebox{0pt}[0pt][6pt]{$E_T^{\gamma}$ cut} & 
\raisebox{0pt}[0pt][6pt]{} & 
\raisebox{0pt}[0pt][6pt]{(35.6\%)} & 
\raisebox{0pt}[0pt][6pt]{(2.1\%)} & 
\raisebox{0pt}[0pt][6pt]{(6.3\%)} & 
\raisebox{0pt}[0pt][6pt]{(25.0\%)} & 
\raisebox{0pt}[0pt][6pt]{(11.2\%)} &
\raisebox{0pt}[0pt][6pt]{(19.8\%)}\\
\hline
\raisebox{0pt}[12pt][6pt]{isolated $\gamma$} & 
\raisebox{0pt}[12pt][6pt]{37.77} & 
\raisebox{0pt}[12pt][6pt]{15.23} & 
\raisebox{0pt}[12pt][6pt]{1.76} & 
\raisebox{0pt}[12pt][6pt]{4.34} & 
\raisebox{0pt}[12pt][6pt]{12.63} &
\raisebox{0pt}[12pt][6pt]{1.49} &
\raisebox{0pt}[12pt][6pt]{2.33}\\
\raisebox{0pt}[0pt][6pt]{$y, E_T^{\gamma}, \eta_{\gamma}$ cuts} & 
\raisebox{0pt}[0pt][6pt]{} & 
\raisebox{0pt}[0pt][6pt]{(40.3\%)} & 
\raisebox{0pt}[0pt][6pt]{(4.7\%)} & 
\raisebox{0pt}[0pt][6pt]{(11.5\%)} & 
\raisebox{0pt}[0pt][6pt]{(33.4\%)} &
\raisebox{0pt}[0pt][6pt]{(3.9\%)} &
\raisebox{0pt}[0pt][6pt]{(6.2\%)}\\
\hline
\raisebox{0pt}[12pt][6pt]{isolated $\gamma$+jet} & 
\raisebox{0pt}[12pt][6pt]{29.45} & 
\raisebox{0pt}[12pt][6pt]{11.60} & 
\raisebox{0pt}[12pt][6pt]{0.19} & 
\raisebox{0pt}[12pt][6pt]{3.41} & 
\raisebox{0pt}[12pt][6pt]{11.45} &
\raisebox{0pt}[12pt][6pt]{1.20} &
\raisebox{0pt}[12pt][6pt]{1.59}\\
\raisebox{0pt}[0pt][6pt]{$y, E_T^{\gamma}, \eta_{\gamma}$ cuts} & 
\raisebox{0pt}[0pt][6pt]{} & 
\raisebox{0pt}[0pt][6pt]{(39.4\%)} & 
\raisebox{0pt}[0pt][6pt]{(0.6\%)} & 
\raisebox{0pt}[0pt][6pt]{(11.6\%)} & 
\raisebox{0pt}[0pt][6pt]{(38.9\%)} &
\raisebox{0pt}[0pt][6pt]{(4.1\%)} &
\raisebox{0pt}[0pt][6pt]{(5.4\%)}\\
\raisebox{0pt}[0pt][6pt]{$E_T^{jet}, \eta_{jet}$ cuts} & 
\raisebox{0pt}[0pt][6pt]{} & 
\raisebox{0pt}[0pt][6pt]{} & 
\raisebox{0pt}[0pt][6pt]{} & 
\raisebox{0pt}[0pt][6pt]{} & 
\raisebox{0pt}[0pt][6pt]{} &
\raisebox{0pt}[0pt][6pt]{} &
\raisebox{0pt}[0pt][6pt]{}\\
\hline\hline
\end{tabular}
\end{center}
\end{table*}
%\vspace{0.4cm}
\begin{table*}[b]
\begin{center}
\caption{The cross section for the non-isolated (inclusive) and isolated 
final photon with $5\le E_T^{\gamma}\le 10$ GeV.
The initial energies are $E_e=27.6$ GeV and $E_p=820$ 
GeV~\cite{Breitweg:1999su}.
The results for the isolated
photon are obtained using the small cone approximation (Sec.~\ref{small})
without and with cuts $0.2\le y\le 0.9$ and $-0.7\le\eta^{\gamma}\le 0.9$.}
\label{tab2}
\vspace{0.3cm}
\begin{tabular}{|c|c|c|c|c|c|c|c|} 
\hline \hline
\raisebox{0pt}[12pt][6pt]{[pb]} & 
\raisebox{0pt}[12pt][6pt]{total} & 
\multicolumn {3}{|c|}{dir,dir}&
\raisebox{0pt}[12pt][6pt]{res,dir}&
\raisebox{0pt}[12pt][6pt]{dir,frag}&
\raisebox{0pt}[12pt][6pt]{res,frag} \\
\cline {3-5}
\raisebox{0pt}[12pt][6pt]{} & 
\raisebox{0pt}[12pt][6pt]{} & 
\raisebox{0pt}[12pt][6pt]{Born} &
\raisebox{0pt}[12pt][6pt]{$\mathcal{O}$($\alpha_S$)} & 
\raisebox{0pt}[12pt][6pt]{box} &
\raisebox{0pt}[12pt][6pt]{}&
\raisebox{0pt}[12pt][6pt]{}&
\raisebox{0pt}[12pt][6pt]{} \\
\hline
\raisebox{0pt}[12pt][6pt]{non-isolated $\gamma$} & 
\raisebox{0pt}[12pt][6pt]{222.0} & 
\raisebox{0pt}[12pt][6pt]{82.0} & 
\raisebox{0pt}[12pt][6pt]{4.8} & 
\raisebox{0pt}[12pt][6pt]{13.9} & 
\raisebox{0pt}[12pt][6pt]{54.8} &
\raisebox{0pt}[12pt][6pt]{24.6} &
\raisebox{0pt}[12pt][6pt]{42.0}\\
\raisebox{0pt}[0pt][6pt]{$E_T^{\gamma}$ cut} & 
\raisebox{0pt}[0pt][6pt]{} & 
\raisebox{0pt}[0pt][6pt]{(36.9\%)} & 
\raisebox{0pt}[0pt][6pt]{(2.2\%)} & 
\raisebox{0pt}[0pt][6pt]{(6.3\%)} & 
\raisebox{0pt}[0pt][6pt]{(24.7\%)} & 
\raisebox{0pt}[0pt][6pt]{(11.1\%)} &
\raisebox{0pt}[0pt][6pt]{(18.9\%)}\\
\hline
\raisebox{0pt}[12pt][6pt]{isolated $\gamma$} & 
\raisebox{0pt}[12pt][6pt]{178.1} & 
\raisebox{0pt}[12pt][6pt]{82.0} & 
\raisebox{0pt}[12pt][6pt]{13.1} & 
\raisebox{0pt}[12pt][6pt]{13.9} & 
\raisebox{0pt}[12pt][6pt]{54.8} &
\raisebox{0pt}[12pt][6pt]{5.1} &
\raisebox{0pt}[12pt][6pt]{9.4}\\
\raisebox{0pt}[0pt][6pt]{$E_T^{\gamma}$ cut} & 
\raisebox{0pt}[0pt][6pt]{} & 
\raisebox{0pt}[0pt][6pt]{(46.0\%)} & 
\raisebox{0pt}[0pt][6pt]{(7.4\%)} & 
\raisebox{0pt}[0pt][6pt]{(7.8\%)} & 
\raisebox{0pt}[0pt][6pt]{(30.8\%)} &
\raisebox{0pt}[0pt][6pt]{(2.9\%)} &
\raisebox{0pt}[0pt][6pt]{(5.3\%)}\\
\hline
\raisebox{0pt}[12pt][6pt]{isolated $\gamma$} & 
\raisebox{0pt}[12pt][6pt]{71.95} & 
\raisebox{0pt}[12pt][6pt]{23.60} & 
\raisebox{0pt}[12pt][6pt]{6.02} & 
\raisebox{0pt}[12pt][6pt]{6.53} & 
\raisebox{0pt}[12pt][6pt]{28.18} &
\raisebox{0pt}[12pt][6pt]{2.34} &
\raisebox{0pt}[12pt][6pt]{5.28}\\
\raisebox{0pt}[0pt][6pt]{$y, E_T^{\gamma}$ cuts} & 
\raisebox{0pt}[0pt][6pt]{} & 
\raisebox{0pt}[0pt][6pt]{(32.8\%)} & 
\raisebox{0pt}[0pt][6pt]{(8.4\%)} & 
\raisebox{0pt}[0pt][6pt]{(9.1\%)} & 
\raisebox{0pt}[0pt][6pt]{(39.2\%)} &
\raisebox{0pt}[0pt][6pt]{(3.3\%)} &
\raisebox{0pt}[0pt][6pt]{(7.3\%)}\\
\hline
\raisebox{0pt}[12pt][6pt]{isolated $\gamma$} & 
\raisebox{0pt}[12pt][6pt]{35.34} & 
\raisebox{0pt}[12pt][6pt]{13.64} & 
\raisebox{0pt}[12pt][6pt]{3.26} & 
\raisebox{0pt}[12pt][6pt]{3.41} & 
\raisebox{0pt}[12pt][6pt]{11.88} &
\raisebox{0pt}[12pt][6pt]{1.20} &
\raisebox{0pt}[12pt][6pt]{1.93}\\
\raisebox{0pt}[0pt][6pt]{$y, E_T^{\gamma}, \eta_{\gamma}$ cuts} & 
\raisebox{0pt}[0pt][6pt]{} & 
\raisebox{0pt}[0pt][6pt]{(38.6\%)} & 
\raisebox{0pt}[0pt][6pt]{(9.2\%)} & 
\raisebox{0pt}[0pt][6pt]{(9.6\%)} & 
\raisebox{0pt}[0pt][6pt]{(33.6\%)} &
\raisebox{0pt}[0pt][6pt]{(3.4\%)} &
\raisebox{0pt}[0pt][6pt]{(5.5\%)}\\
\hline\hline
\end{tabular}
\end{center}
\end{table*}

\begin{table*}[h]
\begin{center}
\caption{The cross section for the isolated final photon
with  $E_e=27.6$ GeV, $E_p=920$~GeV, $0.2\le y\le 0.9$, 
$-0.7\le\eta^{\gamma}\le 0.9$ and 
$5\le E_T^{\gamma}\le 10$ GeV~\cite{unknown:2004uv}. 
The results are obtained with (upper row) and without (lower row) 
the small cone approximation (Sec.~\ref{isol}).}
\label{tab3}
\vspace{0.5cm}
\begin{tabular}{|c|c|c|c|c|c|c|c|} 
\hline \hline
\raisebox{0pt}[12pt][6pt]{[pb]} & 
\raisebox{0pt}[12pt][6pt]{total} & 
\multicolumn {3}{|c|}{dir,dir}&
\raisebox{0pt}[12pt][6pt]{res,dir}&
\raisebox{0pt}[12pt][6pt]{dir,frag}&
\raisebox{0pt}[12pt][6pt]{res,frag} \\
\cline {3-5}
\raisebox{0pt}[12pt][6pt]{} & 
\raisebox{0pt}[12pt][6pt]{} & 
\raisebox{0pt}[12pt][6pt]{Born} &
\raisebox{0pt}[12pt][6pt]{$\mathcal{O}$($\alpha_S$)} & 
\raisebox{0pt}[12pt][6pt]{box} &
\raisebox{0pt}[12pt][6pt]{}&
\raisebox{0pt}[12pt][6pt]{}&
\raisebox{0pt}[12pt][6pt]{} \\
\hline
\raisebox{0pt}[12pt][6pt]{approximated} & 
\raisebox{0pt}[12pt][6pt]{38.93} & 
\raisebox{0pt}[12pt][6pt]{15.23} & 
\raisebox{0pt}[12pt][6pt]{2.94} & 
\raisebox{0pt}[12pt][6pt]{4.34} & 
\raisebox{0pt}[12pt][6pt]{12.63} &
\raisebox{0pt}[12pt][6pt]{1.49} &
\raisebox{0pt}[12pt][6pt]{2.33}\\
\hline
\raisebox{0pt}[12pt][6pt]{exact} & 
\raisebox{0pt}[12pt][6pt]{37.77} & 
\raisebox{0pt}[12pt][6pt]{15.23} & 
\raisebox{0pt}[12pt][6pt]{1.76} & 
\raisebox{0pt}[12pt][6pt]{4.34} & 
\raisebox{0pt}[12pt][6pt]{12.63} &
\raisebox{0pt}[12pt][6pt]{1.49} &
\raisebox{0pt}[12pt][6pt]{2.33}\\
\hline\hline
\end{tabular}
\end{center}
\end{table*}

\appendix

%%%%%%%%%%%%%%%%%%%%%%%%%%%%%%%%%%%%%%%%%%%%%%%%%%%%%%%%%%%%%%%%%%%%%%%%%%
\chapter{Kinematics and notation}\label{AA}
%%%%%%%%%%%%%%%%%%%%%%%%%%%%%%%%%%%%%%%%%%%%%%%%%%%%%%%%%%%%%%%%%%%%%%%%%%
\section{$2\ra 2$ processes}\label{Anot}
%%%%%%%%%%%%%%%%%%%%%%%%%%%%%%%%%%%%%%%%%%%%%%%%%%%%%%%%%%%%%%%%%%%%%%%%%%

We study the production of  photons and jets
in the electron-proton scattering:
\bea
e(p_e) ~p(p_p) \ra 
e(p_e^{\prime}) ~\gamma (p_{\gamma}) ~jet(p_{jet}) ~X(p_X),
\eea
where four-momenta of particles are given in brackets,
and assume that the mediating photon is quasi-real,
\bea
Q^2=-(p_e-p_e^{\prime})^2\approx 0.
\eea
In such processes the emission of the mediating photon from the
electron can be factored out (Sec. \ref{epa}) and we can consider
the photon-proton scattering:
\bea
\gamma (q) ~p(p_p) \ra \gamma (p_{\gamma}) ~jet(p_{jet}) ~X(p_X),
\eea
where $q=y p_p$ is the four-momentum of the mediating photon.
This photon may interact with a parton from the proton
directly or as a resolved one, see Eq. (\ref{223}).
The corresponding $2\ra 2$ partonic processes are:
\bea
a(p_a) ~b(p_b)\ra c(p_c) ~d(p_d) ,
\eea
where
\bea
p_a=x_{\gamma} q =x_{\gamma}yp_e \makebox[1cm]{,}
p_b=x p_p \makebox[1cm]{,}
p_c = p_{\gamma}/z.
\eea
We introduce the standard variables:
\bea
s=(p_a+p_b)^2 \makebox[1cm]{,} t=(p_a-p_c)^2 \makebox[1cm]{,} u=(p_b-p_c)^2,
\label{mand}
\eea
which are used in next Appendices and in formulae for the cross sections.

%%%%%%%%%%%%%%%%%%%%%%%%%%%%%%%%%%%%%%%%%%%%%%%%%%%%%%%%%%%%%%%%%%%%%%%%%%
\section{$2\ra 3$ processes}\label{Anot2}
\begin{figure}[h]
\vskip 6.5cm\relax\noindent\hskip 0cm
       \relax{\includegraphics{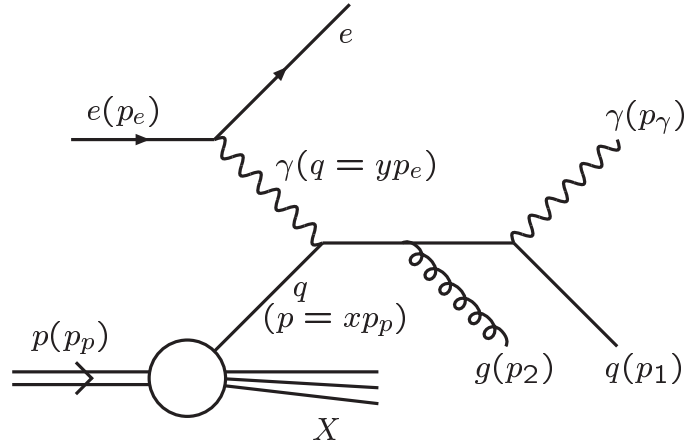}}
\vspace{-0.5cm}
\caption{\small\sl An example of the $2\ra 3$ NLO process contributing
to the $ep\ra e\gamma ~(jet) ~X$ reaction. The four-momenta of particles
are given in brackets.}
\label{23}
\end{figure}

%%%%%%%%%%%%%%%%%%%%%%%%%%%%%%%%%%%%%%%%%%%%%%%%%%%%%%%%%%%%%%%%%%%%%%%%%%
Now, we consider the $2\ra 3$ partonic processes contributing to the
$ep\ra e\gamma ~(jet) ~X$ reaction: $\gamma q\ra \gamma q g$ 
(Fig. \ref{figreal}) and $\gamma g\ra \gamma q \bar{q}$ 
(Fig. \ref{fig23}).
We write
\be
\gamma (q) + q(p) \ra \gamma (p_{\gamma}) + q(p_1) + g(p_2)
\makebox[0.5cm]{} {\rm (see\makebox[0.3cm]{} Fig.~\ref{23}),}
\label{ea7}
\ee
and
\be
\gamma (q) + g(p) \ra \gamma (p_{\gamma}) + q(p_1) + \bar{q}(p_2).
\label{ea8}
\ee
We use the variables $s$, $t$ and $u$ defined in Appendix \ref{Anot},
which for the direct both initial and final photon ($x_{\gamma}=z=1$) 
are given by
\bea
s=(q+p)^2 \makebox[1cm]{,} t=(q-p_{\gamma})^2 \makebox[1cm]{}
u=(p-p_{\gamma})^2.
\label{mand2}
\eea
Finally, we define the scaled variables $v$ and $w$:
\be
v=1+t/s\makebox[2cm]{\rm ,}w=-u/(t+s).
\ee
The variables $v$ and $w$ are in the range from 0 to 1: 
$0\le v\le 1$ and $0\le w\le 1$.
\\ Note, that for massless particles 
\bea
(p_1+p_2)^2=sv(1-w),
\eea 
and in the limit $(p_1+p_2)^2\ra 0$ one obtains $w\ra 1$
($v$ can not be too low, since the final photon has a large
transverse momentum). For the $2\ra 2$ processes one has $w=1$ by definition.
 
%%%%%%%%%%%%%%%%%%%%%%%%%%%%%%%%%%%%%%%%%%%%%%%%%%%%%%%%%%%%%%%%%%%%%%%%%%
\chapter{Cross sections for $2\ra 2$ processes}\label{lox}
%%%%%%%%%%%%%%%%%%%%%%%%%%%%%%%%%%%%%%%%%%%%%%%%%%%%%%%%%%%%%%%%%%%%%%%%%%

In this Appendix we have collected all the tree-level cross sections
for $2\ra 2$ processes which are included in our calculation.
The virtual gluon corrections 
to the Born process are included in Appendix \ref{xsec:} together with the real
gluon corrections.

The cross sections for $2\ra 2$ processes shown in
Figs. \ref{figborn}, \ref{figsingi}-\ref{figbox}
are given by the equation
\bea
E_{\gamma}{d^3\sigma^{ab\ra cd}\over d^3p_{\gamma}} (s,t,u)=
{1\over (4\pi)^2s}\cdot |\overline{M}^{ab\ra cd}|^2(s,t,u)\cdot\delta (s+t+u),
\label{eq.lox}
\eea
where $|\overline{M}^{ab\ra cd}|^2$ are the squared matrix elements:
\bea
|\overline{M}^{\gamma q\ra\gamma q}|^2(s,t,u)=
-2(4\pi)^2\alpha^2e_q^4({u\over s}+{s\over u}),
\eea
\bea
|\overline{M}^{qg\ra\gamma q}|^2(s,t,u)=
-{1\over 3}(4\pi)^2\alpha\alpha_Se_q^2({t\over s}+{s\over t}),
\eea
\bea
|\overline{M}^{q\bar{q}\ra\gamma g}|^2(s,t,u)=
{8\over 9}(4\pi)^2\alpha\alpha_Se_q^2({t\over u}+{u\over t}),
\eea
\bea
|\overline{M}^{gq\ra\gamma q}|^2(s,t,u)=
|\overline{M}^{qg\ra\gamma q}|^2(s,u,t),
\eea
\bea
|\overline{M}^{\gamma q\ra qg}|^2(s,t,u)=
-{8\over 3}(4\pi)^2\alpha\alpha_Se_q^2({t\over s}+{s\over t}),
\eea
\bea
|\overline{M}^{\gamma q\ra gq}|^2(s,t,u)=
|\overline{M}^{\gamma q\ra qg}|^2(s,u,t),
\eea
\bea
|\overline{M}^{\gamma g\ra q\bar{q}}|^2(s,t,u)=(4\pi)^2
\alpha\alpha_Se_q^2({t\over u}+{u\over t}),
\eea
\bea
|\overline{M}^{gq\ra qg}|^2(s,t,u)=(4\pi)^2\alpha_S^2\left[
-{4\over 9}({t\over s}+{s\over t})+{s^2+t^2\over u^2}
\right],
\eea
\bea
|\overline{M}^{gq\ra gq}|^2(s,t,u)=
|\overline{M}^{gq\ra qg}|^2(s,u,t),
\eea
\bea
|\overline{M}^{qg\ra qg}|^2(s,t,u)=
|\overline{M}^{gq\ra qg}|^2(s,u,t),
\eea
\bea
|\overline{M}^{qg\ra gq}|^2(s,t,u)=
|\overline{M}^{gq\ra qg}|^2(s,t,u),
\eea
\bea
|\overline{M}^{qq'\ra qq'}|^2(s,t,u)=(4\pi)^2\alpha_S^2\left[
{4\over 9}{s^2+u^2\over t^2}
\right],
\eea
\bea
|\overline{M}^{qq'\ra q'q}|^2(s,t,u)=
|\overline{M}^{qq'\ra qq'}|^2(s,u,t),
\eea
\bea
|\overline{M}^{qq\ra qq}|^2(s,t,u)=(4\pi)^2\alpha_S^2\left[
{4\over 9}({s^2+u^2\over t^2}+{s^2+t^2\over u^2})
-{8\over 27}{s^2\over tu}
\right],
\eea
\bea
|\overline{M}^{q\bar{q}\ra q\bar{q}}|^2(s,t,u)=(4\pi)^2\alpha_S^2\left[
{4\over 9}({s^2+u^2\over t^2}+{u^2+t^2\over s^2})
-{8\over 27}{u^2\over st}
\right],
\eea
\bea
|\overline{M}^{q\bar{q}\ra\bar{q}q}|^2(s,t,u)=
|\overline{M}^{q\bar{q}\ra q\bar{q}}|^2(s,u,t),
\eea
\bea
|\overline{M}^{q\bar{q}\ra q'\bar{q}'}|^2(s,t,u)=(4\pi)^2\alpha_S^2\left[
{4\over 9}{t^2+u^2\over s^2}
\right],
\eea
\bea
|\overline{M}^{gg\ra q\bar{q}}|^2(s,t,u)=(4\pi)^2\alpha_S^2\left[
{1\over 6}({t\over u}+{u\over t})
-{3\over 8}{t^2+u^2\over s^2}
\right],
\eea
\bea
|\overline{M}^{q\bar{q}\ra gg}|^2(s,t,u)=(4\pi)^2\alpha_S^2\left[
{32\over 27}({t\over u}+{u\over t})
-{8\over 3}{t^2+u^2\over s^2}
\right],
\eea
\bea
|\overline{M}^{gg\ra gg}|^2(s,t,u)={9\over 2}(4\pi)^2\alpha_S^2
(3-{tu\over s^2}-{su\over t^2}-{st\over u^2}).
\label{Mdouble}
\eea
The squared matrix element for the box process (Fig. \ref{figbox})
has a form:
\bea
|\overline{M}^{\gamma g\ra\gamma g}|^2(s,t,u) = 
{1 \over 16} 
\left(\alpha_S\over\alpha\right)^2\left(\sum_{q,\bar{q}}e_q^2\right)^2
\left( |M_{11\ra 11}|^2+|M_{22\ra 22}|^2+\right.
\nonumber
\\
\left. 2|M_{11\ra 22}|^2+2|M_{12\ra 12}|^2+2|M_{12\ra 21}|^2\right),
\label{Combridge}
\eea
where the amplitudes $M_{ij\ra kl}$ $(\sim\alpha^4)$
are given by Eq. (1) in Ref. \cite{Combridge:1980sx}, and
\\ $M_{12\ra 21}(s,t,u)=M_{12\ra 12}(s,u,t)$.

%%%%%%%%%%%%%%%%%%%%%%%%%%%%%%%%%%%%%%%%%%%%%%%%%%%%%%%%%%%%%%%%%%%%%%%%%%
\chapter{Three-body phase space}\label{A3b}
%%%%%%%%%%%%%%%%%%%%%%%%%%%%%%%%%%%%%%%%%%%%%%%%%%%%%%%%%%%%%%%%%%%%%%%%%%
The cross section for a $2\ra 3$ process is given by a general formula:
\bea
d\sigma^{2\ra 3}= {1\over 2s} d(PS)_3 |\overline{M}^{2\ra 3}|^2,
\label{ec1}
\eea
where $(PS)_3$ is the three-body {\sl phase space} in $n$ dimensions:
\bea
(PS)_3 = \int {d^np_{\gamma}\over (2\pi)^{n-1}}
         \int {d^np_1\over (2\pi)^{n-1}}
         \int {d^np_2\over (2\pi)^{n-1}}
\makebox[5cm]{}\nonumber \\  \makebox[3cm]{}\cdot
(2\pi)^n \delta^n(q+p-p_{\gamma}-p_1-p_2)
\delta^+(p_{\gamma}^2)\delta^+(p_1^2)\delta^+(p_2^2).
\label{eq:gen:a}
\eea
The integration of $(PS)_3$ over all four-momenta of two final
partons is discussed e.g. in \cite{Ellis:1979sj,Aurenche:1984hc,jan}.
However, in order to impose the isolation of the final photon
as well as cuts for the jet,
we need to restrict the momenta of the final partons.
To achieve this goal we integrate $(PS)_3$ considering various
configurations of the final particles momenta,
e.g. some collinear configurations, as described in Sec. \ref{pss}.
The integration of $(PS)_3$ including a collinear configuration,
where two final particles move (almost) parallel to each other,
was previously discussed e.g. in \cite{Furman:1981kf}. 
In comparison with \cite{Furman:1981kf} we simplified the integration
over the final energies. Below we derive all formulae for $(PS)_3$
which are applied in Appendix \ref{xsec:} to obtain the cross sections for 
the $\mathcal{O}$$(\alpha_S)$ corrections to the Born process
including e.g. the configurations where a final parton is collinear
with the final photon or with the initial electron or proton.

The integration of $(PS)_3$ as well as the calculations of the
cross sections for the $\mathcal{O}$$(\alpha_S)$ corrections
to the Born process are performed in $n=4-2\varepsilon$ dimensions
using the {\sl dimensional regularization}, see e.g. \cite{Sterman:1994ce}.

First, we integrate (\ref{eq:gen:a}) over $p_2$:
\bea
(PS)_3 = {1\over (2\pi)^{2n-3}}\int d^np_{\gamma}\int d^np_1
\delta^+(p_{\gamma}^2)\delta^+(p_1^2)\delta^+((q+p-p_{\gamma}-p_1)^2).
\label{n1}
\eea
Next, we choose the $z$-axis along the initial proton momentum in any
frame of reference, in which the collision is central, e.g. in the
laboratory frame:
\bea
p=xE_p(1,...0,1),
\eea
\bea
q=yE_e(1,...0,-1)
\eea
(the unspecified components are equal to zero). The collision has the 
rotational symmetry and we can perform an integration over $(n-3)$
azimuthal angles of the photon. Performing this integration as well as
the integration over the photon energy, and making the change of variables
we obtain:
\bea
\int d^np_{\gamma}\delta^+(p_{\gamma}^2)=
{\pi^{1-\varepsilon}\over 2\Gamma (1-\varepsilon )} 
\int dv\int dw[vws(1-v)]^{-\varepsilon}sv.
\label{n2}
\eea
Next, we chose the $z$ and $y$ axes this way that momenta $\vec{p}$,
$\vec{q}$ and $\vec{p_{\gamma}}$ are in the $zy$-plane:
\bea
p=xE_p(1,...0,\sin\alpha,\cos\alpha),
\label{nn1}
\eea
\bea
q=yE_e(1,...0,-\sin\alpha,-\cos\alpha),
\eea
\bea
p_{\gamma}=E_{\gamma}(1,...0,\sin\alpha^{\prime},\cos\alpha^{\prime}),
\label{nn3}
\eea
where $\alpha$ and $\alpha^{\prime}$ are arbitrary, and the unspecified
components in the additional $(n-4)$ dimensions are equal to zero.
We write $p_1$ and $p_2$ in this frame:
\bea
p_1=E(1,...\cos\theta_3\sin\theta_2\sin\theta_1,
\cos\theta_2\sin\theta_1,\cos\theta_1),
\eea
\bea
p_2=p+q-p_{\gamma}-p_1.
\eea
Since the four-momenta $p$, $q$ and $p_{\gamma}$ have only two non-zero
space components, one can perform the integration over $(n-4)$
azimuthal angles of the final parton:
\bea
\int d^np_1\delta^+(p_1^2)=
{\pi^{{1\over 2}-\varepsilon}\over \Gamma ({1\over 2}-\varepsilon )} 
\int dE\,E^{1-2\varepsilon}\int d\theta_1\sin^{1-2\varepsilon}\theta_1
\int d\theta_2\sin^{-2\varepsilon}\theta_2.
\label{n3}
\eea
From (\ref{n1}), (\ref{n2}) and (\ref{n3}) one has
\bea
{d(PS)_3\over dvdw}=
{1\over 4(2\pi)^n}{1\over \Gamma (1-2\varepsilon )} [vws(1-v)]^{-\varepsilon}sv
\cdot
\nonumber
\eea
\bea
\makebox[0.5cm]{} 
\cdot\int d\theta_1\sin^{1-2\varepsilon}\theta_1
\int d\theta_2\sin^{-2\varepsilon}\theta_2
\int dE\,E^{1-2\varepsilon}\delta^+((q+p-p_{\gamma}-p_1)^2).
\label{ps3E}
\eea
Note, that the integration over $E$ is straightforward as the argument
of the $\delta$ function is a linear function of $p_1$:
$(q+p-p_{\gamma}-p_1)^2 = (q+p-p_{\gamma})^2 -2 p_1(q+p-p_{\gamma})$.
The expression (\ref{ps3E}) was obtained with
no approximations and can be used e.g. to calculate the cross section
for $2\ra 3$ processes in Part 5 of the phase space described in 
Sec. \ref{pss}. 

In order to obtain $(PS)_3$ in a form suitable
for considering collinear configurations in Part 2, 3 and 4, we
assume that $\theta_1$ is small:
\bea
\theta_1\approx 0,
\label{col1ap}
\eea
and take in Eqs. (\ref{nn1}-\ref{nn3}) $\alpha=\pi$ or $\alpha=0$
or $\alpha^{\prime}=0$.
With these assumptions we obtain:
\bea
{d(PS)_3\over dvdw}=
\nonumber
\eea
\bea
=\Theta(1-w)
{[svw(1-v)]^{-\varepsilon}\over 4(2\pi)^n \Gamma (1-2\varepsilon )} 
\cdot{E^{2-2\varepsilon}\over 1-w}
\int d\theta_1 \int d\theta_2
\sin^{1-2\varepsilon}\theta_1\sin^{-2\varepsilon}\theta_2,
\label{ps3Ec}
\eea
where
\begin{itemize}
\item for $\alpha=\pi$:
\bea
E=(1-w)yE_e,
\label{ec16}
\eea
\item for $\alpha=0$:
\bea
E={v(1-w)\over 1-vw}xE_p,
\label{ec17}
\eea
\item for $\alpha^{\prime}=0$:
\bea
E={v(1-w)\over 1-v+vw}E_{\gamma}.
\label{ec18}
\eea
\end{itemize}
Of course, the same formulae for $(PS)_3$ 
are valid if we replace $p_1$ with $p_2$:
\bea
p_2=E(1,...\cos\theta_3\sin\theta_2\sin\theta_1,
\cos\theta_2\sin\theta_1,\cos\theta_1),
\eea
\bea
p_1=p+q-p_{\gamma}-p_2.
\eea

%%%%%%%%%%%%%%%%%%%%%%%%%%%%%%%%%%%%%%%%%%%%%%%%%%%%%%%%%%%%%%%%%%%%%%%%%%
\chapter{Corrections of order $\mathcal{O}$$(\alpha_s)$ to the Born process} 
\label{xsec:}
%%%%%%%%%%%%%%%%%%%%%%%%%%%%%%%%%%%%%%%%%%%%%%%%%%%%%%%%%%%%%%%%%%%%%%%%%%

In this Appendix we present formulae for $\mathcal{O}$$(\alpha_S)$ 
corrections (Figs. \ref{figvirt}-\ref{fig23})
to the Born process (Fig. \ref{figborn}).
The general (unintegrated) formulae (Appendix \ref{xsec:gen})
and the formulae for the non-isolated photon production 
(Appendix \ref{xsec:w1}) are taken from 
the literature \cite{Aurenche:1984hc,jan}. 
The cross sections corresponding to
various collinear configurations described in Sec. \ref{pss}
are derived briefly in 
Appendices \ref{xsec:col2}, \ref{xsec:col3} and \ref{xsec:col1}.

%%%%%%%%%%%%%%%%%%%%%%%%%%%%%%%%%%%%%%%%%%%%%%%%%%%%%%%%%%%%%%%%%%%%%%%%%%
\section{General formulae for $2\ra 3$ processes}\label{xsec:gen}
%%%%%%%%%%%%%%%%%%%%%%%%%%%%%%%%%%%%%%%%%%%%%%%%%%%%%%%%%%%%%%%%%%%%%%%%%%

The general formula for the squared matrix element for the 
$\gamma q\ra\gamma qg$ 
%and  $\gamma g\ra\gamma q\bar{q}$
process has the form \cite{Aurenche:1984hc,jan}:
\bea
|\overline{M}^{\gamma q\ra\gamma qg}|^2
=
{2(4\pi)^3\alpha^2\alpha_S\hat{\mu}^{6\varepsilon}c\over a_1a_2a_3b_1b_2b_3}
\,e_q^4\,C_F\,[A_1-\varepsilon (2A_1-A_2+2A_3-8A_4)
\nonumber
\eea
\bea
\makebox[9cm]{}+{\mathcal{O}}(\varepsilon^2)],
\label{eq:xsec:m2}
\eea
%\bea
%|\overline{M}^{\gamma g\ra\gamma q\bar{q}}|^2=
%{2(4\pi)^3\alpha^2\alpha_S\mu^{6\varepsilon}c^{\prime}\over 
%a_1^{\prime}a_2^{\prime}a_3^{\prime}b_1^{\prime}b_2^{\prime}b_3^{\prime}}
%\,e_q^4\,{1\over 2}\,[A_1^{\prime}+\varepsilon (-2A_1^{\prime}+A_2^{\prime}-
%2A_3^{\prime}+8A_4^{\prime})],
%\label{eq:ysec:m2}
%\eea
where $\hat{\mu}$ is an arbitrary mass scale, $C_F=4/3$ 
and the terms $A_i$, $a_i$,
$b_i$, $c$ are given by
\bea\label{eq:xsec:A1}
A_1=a_1b_1(a_1^2+b_1^2)+a_2b_2(a_2^2+b_2^2)+a_3b_3(a_3^2+b_3^2)
\eea
%\bea
%A_1^{\prime}=a_1^{\prime}b_1^{\prime}(a_1^{\prime 2}+b_1^{\prime 2})+
%a_2^{\prime}b_2^{\prime}(a_2^{\prime 2}+b_2^{\prime 2})+a_3^{\prime}
%b_3^{\prime}(a_3^{\prime 2}+b_3^{\prime 2})
%\eea
\bea
A_2=a_1b_1(a_2+b_2)(a_3+b_3)-a_2b_2(a_1+b_1)(a_3+b_3)
\makebox[0cm]{}
\nonumber\\
-a_3b_3(a_1+b_1)(a_2+b_2)
\eea
%\bea
%A_2^{\prime}=a_1^{\prime}b_1^{\prime}(a_2^{\prime}+b_2^{\prime})
%(a_3^{\prime}+b_3^{\prime})-a_2^{\prime}b_2^{\prime}(a_1^{\prime}+
%b_1^{\prime})(a_3^{\prime}+b_3^{\prime})-a_3^{\prime}b_3^{\prime}
%(a_1^{\prime}+b_1^{\prime})(a_2^{\prime}+b_2^{\prime})
%\eea
\bea
A_3=a_1^2b_1^2+a_2^2b_2^2+a_3^2b_3^2
\eea
%\bea
%A_3^{\prime}=a_1^{\prime 2}b_1^{\prime 2}+a_2^{\prime 2}b_2^{\prime 2}+
%a_3^{\prime 2}b_3^{\prime 2}
%\eea
\bea\label{eq:xsec:A4}
A_4=a_1b_1a_2b_2+a_1b_1a_3b_3+a_2b_2a_3b_3
\eea
%\bea
%A_4^{\prime}=a_1^{\prime}b_1^{\prime}a_2^{\prime}b_2^{\prime}+
%a_1^{\prime}b_1^{\prime}a_3^{\prime}b_3^{\prime}+
%a_2^{\prime}b_2^{\prime}a_3^{\prime}b_3^{\prime}
%\eea
and
\bea
a_1=pq={s\over 2},
%\makebox[0.8cm]{}
\label{ed6}
\eea
\bea
a_2=pp_{\gamma}={s\over 2}vw,
%\makebox[0.8cm]{}
\eea
\bea
a_3=pp_2,
\eea
\bea
b_1=p_1q,
%\makebox[0.8cm]{}
\eea
\bea
b_2=p_1p_{\gamma},
%\makebox[0.8cm]{}
\eea
\bea
b_3=p_1p_2={s\over 2}v(1-w),
%\makebox[0.8cm]{}
\eea
\bea
c=pp_1.
\label{ed12}
\eea
The same formula (\ref{eq:xsec:m2}) is valid for the 
squared matrix element of the $\gamma g\ra\gamma q\bar{q}$ process,
$|\overline{M}^{\gamma g\ra\gamma q\bar{q}}|^2$,
but with the factor 1/2 instead of $C_F$, with no
${\mathcal{O}}(\varepsilon^2)$ term, and with the coefficients
$a_1^{\prime}$, $a_2^{\prime}$, $a_3^{\prime}$, $b_1^{\prime}$, 
$b_2^{\prime}$, $b_3^{\prime}$ and $c^{\prime}$ instead of  
$a_1$, $a_2$, $a_3$, $b_1$, $b_2$, $b_3$ and $c$, where the 
coefficients with the prime are given by:
\bea
a_1^{\prime}=-p_2q,
\label{ed13}
\eea
\bea
a_2^{\prime}=-p_2p_{\gamma},
\eea
\bea
a_3^{\prime}=a_3,
\eea
\bea
b_1^{\prime}=b_1,
\eea
\bea
b_2^{\prime}=b_2,
\eea
\bea
b_3^{\prime}=-c,
\eea
\bea
c^{\prime}=-b_3.
\label{ed19}
\eea
The above squared matrix elements are used (with $\varepsilon=0$)
to calculate the cross sections
for the ${\mathcal{O}}(\alpha_S)$ corrections to the Born process
in Part 5 of the phase space (see Sec. \ref{pss}):
\bea
E_{\gamma}{d\sigma^{2\ra 3}_{\alpha_S (5)}\over d^3p_{\gamma}}=
{1\over sv\pi}\cdot
{1\over 2s} {d(PS)_3 \over dvdw} 
|\overline{M}^{2\ra 3}|^2,
\label{xeq:og}
\eea
where the three-body phase space is given by Eq. \ref{ps3E}.

%%%%%%%%%%%%%%%%%%%%%%%%%%%%%%%%%%%%%%%%%%%%%%%%%%%%%%%%%%%%%%%%%%%%%%%%%%
\section{Inclusive photon cross section}\label{xsec:w1}
%%%%%%%%%%%%%%%%%%%%%%%%%%%%%%%%%%%%%%%%%%%%%%%%%%%%%%%%%%%%%%%%%%%%%%%%%%
The cross sections for the $\mathcal O$$(\alpha_S)$ corrections to the
Born process, integrated over all momenta of final partons are given by
\cite{Aurenche:1984hc,jan}:
\bea
E_{\gamma}{d\over d^3p_{\gamma}}
\left[\sigma^{\gamma q\ra\gamma q}_{\alpha_S}
+\sigma^{\gamma q\ra\gamma qg}_{\alpha_S}\right]=\Theta(1-w)
{\alpha^2\alpha_S\over \pi s^2}e_q^4C_F\cdot
\left[ c_1\delta (1-w)+
\raisebox{4mm}{} \raisebox{-4mm}{}\right.
\nonumber
\eea
\bea
\makebox[1cm]{}c_2\left( {1\over 1-w}\right)_+
+c_3\left( {\ln (1-w)\over 1-w}\right)_++
\left( cD_4\delta (1-w)+
\raisebox{4mm}{} \raisebox{-4mm}{}\right.
\nonumber
\eea
\bea
\makebox[1cm]{}\left.\raisebox{4mm}{} \raisebox{-4mm}{}
cW_4\left( {1\over 1-w}\right)_++c_4\right) 
\ln {s\over\mu^2}+c_5\ln v
+c_6\ln (1-vw)+
\nonumber
\eea
\bea
\makebox[1cm]{}c_7\ln (1-v+vw)+c_8\ln (1-v)+c_9\ln w 
+c_{10}\ln (1-w)+c_{11}+
\nonumber
\eea
\bea
\makebox[1cm]{}c_{12}{\ln (1-v+vw)\over 1-w}+c_{13}{\ln w\over 1-w}
\left.
+c_{14}{\ln ((1-vw)/(1-v))\over 1-w} \right],
\label{eq:non:k}
\eea
\bea
E_{\gamma}{d\sigma^{\gamma q\ra\gamma q\bar{q}}_{\alpha_S}\over d^3p_{\gamma}}
=\Theta(1-w) {\alpha^2\alpha_S\over \pi s^2}e_q^4{1\over 2}
%\sum_{q\bar{q}}
\left[ 
c'_4\ln {s\over\mu^2}+c'_5\ln v+c'_6\ln (1-vw)+
\right.
\nonumber
\eea\bea
\makebox[1cm]{}
c'_7\ln (1-v+vw)+c'_8\ln (1-v)+c'_9\ln w +c'_{10}\ln (1-w)+c'_{11}+
\nonumber
\eea\bea
\makebox[1cm]{}\left.
c'_{12}{\ln (1-v+vw)\over 1-w}+c'_{13}{\ln w\over 1-w}
+c'_{14}{\ln ((1-vw)/(1-v))\over 1-w} 
\right],
\label{eq:non:kp}
\eea
where the coefficients $c$ and $c^{\prime}$ are given in Appendix B in the
paper~\cite{Aurenche:1984hc}. We use formulae (\ref{eq:non:k})
and (\ref{eq:non:kp}) to obtain the $\mathcal O$$(\alpha_S)$ corrections
for the non-isolated photon production (Chapters \ref{non}, \ref{small})
as well as to obtain the $\mathcal O$$(\alpha_S)$ corrections
for the isolated photon
in Part 1 of the phase space, as defined in Sec. \ref{pss}
(Chapters \ref{isol}, \ref{jet}).

%%%%%%%%%%%%%%%%%%%%%%%%%%%%%%%%%%%%%%%%%%%%%%%%%%%%%%%%%%%%%%%%%%%%%%%%%
\section{Collinear configuration ($\vec{p}_1 || \vec{p}_e$ or 
$\vec{p}_2 || \vec{p}_e$)}\label{xsec:col2}
%%%%%%%%%%%%%%%%%%%%%%%%%%%%%%%%%%%%%%%%%%%%%%%%%%%%%%%%%%%%%%%%%%%%%%%%%

In this Appendix we consider the configuration corresponding 
to the Part 2 of the phase space defined in Sec. \ref{pss}.
First, we assume that the momentum of the final quark, $\vec{p}_1$, is almost
parallel to the momentum of the initial photon originating from the electron, 
$\vec{q}$ $(=y\,\vec{p}_e)$.
We orient the axes this way that
\bea
q=y E_e(1,...,0,0,1),
\eea
\bea
p=x E_p (1,...,0,0,-1),
\eea
\bea
p_{\gamma}=E_{\gamma}(1,...,0,\sin\alpha^{\prime},\cos\alpha^{\prime}),
\eea
\bea
p_1=E (1,...,\cos\theta_2\sin\theta_1,\cos\theta_1),
\eea
\bea
p_2=q+p-p_{\gamma}-p_1,
\eea
where 
\bea
E_{\gamma}=xE_p(1-v)+yE_evw,
\eea
\bea
\sin\alpha^{\prime} = {\sqrt{s(1-v) vw} \over E_{\gamma}},
\eea
\bea
\cos\alpha^{\prime} = - {x E_p (1-v) - y E_e vw\over E_{\gamma}}.
\eea
\bea
E=(1-w)yE_e,
\eea
\bea
0\le\theta_1\le\theta_{cut}\ll 1.
\label{col2ap}
\eea
From Eqs. (\ref{ec1}, \ref{ps3Ec}, \ref{ec16}, \ref{eq:xsec:m2}-\ref{ed19}) 
we obtain the cross sections  which contain terms
$\sim {1/\varepsilon}$ being singular in 4 dimensions:
\bea
E_{\gamma}{d\sigma^{\gamma q\ra\gamma qg}_{\alpha_S (2)}\over d^3p_{\gamma}}
\makebox[0.1cm]{}| \makebox[-3.4pt]{} \raisebox{-5pt}{$|$}_{singular}
=
{1\over sv\pi}\cdot
{1\over 2s} {d(PS)_3 \over dvdw} 
|\overline{M}^{\gamma q\ra\gamma qg}|^2
\sim {\mathcal{O}}({1\over \varepsilon}) +{\mathcal{O}}(1)+...,
\eea

\bea
E_{\gamma}{d\sigma^{\gamma g\ra\gamma q\bar{q}}_{\alpha_S (2)}
\over d^3p_{\gamma}}
\makebox[0.1cm]{}| \makebox[-3.4pt]{} \raisebox{-5pt}{$|$}_{singular}
=
{1\over sv\pi}\cdot
{1\over 2s} {d(PS)_3 \over dvdw} 
|\overline{M}^{\gamma g\ra\gamma q\bar{q}}|^2
\sim {\mathcal{O}}({1\over \varepsilon}) +{\mathcal{O}}(1)+...,
\eea
We remove these singularities applying the standard factorization 
procedure (see e.g. \cite{Aurenche:1984hc,jan}):
\bea
E_{\gamma}{d\sigma^{\gamma q\ra\gamma qg}_{\alpha_S (2)}\over d^3p_{\gamma}}
=
E_{\gamma}{d\sigma^{\gamma q\ra\gamma qg}_{\alpha_S (2)}\over d^3p_{\gamma}}
\makebox[0.1cm]{}| \makebox[-3.4pt]{} \raisebox{-5pt}{$|$}_{singular}
-\int\limits_0^1 d\xi H_{q\gamma}(\xi,\mu)
E_{\gamma}{d\sigma^{q\bar{q}\ra\gamma g}\over d^3p_{\gamma}}(\xi s,\xi t,u),
\eea
\bea
E_{\gamma}{d\sigma^{\gamma g\ra\gamma q\bar{q}}_{\alpha_S (2)}\over 
d^3p_{\gamma}}=
E_{\gamma}{d\sigma^{\gamma g\ra\gamma q\bar{q}}_{\alpha_S (2)}\over 
d^3p_{\gamma}}
\makebox[0.1cm]{}| \makebox[-3.4pt]{} \raisebox{-5pt}{$|$}_{singular}
-\int\limits_0^1 d\xi H_{q\gamma}(\xi,\mu)
E_{\gamma}{d\sigma^{qg\ra\gamma q}\over d^3p_{\gamma}}(\xi s,\xi t,u),
\eea
with
\bea
H_{q\gamma}(\xi,\mu)=-{1\over\varepsilon}\,{\alpha\over 2\pi}
3e_q^2[\xi^2+(1-\xi)^2]\left({4\pi\hat{\mu}^2\over\mu^2}\right)^{\varepsilon}
{\Gamma (1-\varepsilon)\over\Gamma (1-2\varepsilon)}
+{\mathcal{O}}(\varepsilon),
\eea
\bea
E_{\gamma}{d\sigma^{q\bar{q}\ra\gamma g}\over d^3p_{\gamma}}(s,t,u)=
{2\pi\alpha\alpha_Se_q^2\over 3s}C_F{\hat{\mu}^{2\varepsilon}\over
\Gamma (1-\varepsilon)}\left({4\pi\hat{\mu}^2s\over tu}\right)^{\varepsilon}
\cdot\nonumber
\eea 
\bea
\makebox[6cm]{}\cdot (1-\varepsilon)
\left[(1-\varepsilon)\left({t\over u}+{u\over t}\right) -2\varepsilon \right],
\eea
\bea
E_{\gamma}{d\sigma^{qg\ra\gamma q}\over d^3p_{\gamma}}(s,t,u)=
{2\pi\alpha\alpha_Se_q^2\over 3s}\,{1\over 2}\,{\hat{\mu}^{2\varepsilon}\over
\Gamma (1-\varepsilon)}\left({4\pi\hat{\mu}^2s\over tu}\right)^{\varepsilon}
\cdot\nonumber
\eea 
\bea
\makebox[6cm]{}\cdot (1-\varepsilon)
\left[-(1-\varepsilon)\left({t\over s}+{s\over t}\right) -2\varepsilon \right],
\eea
where $\hat{\mu}$ is an arbitrary mass scale.
Finally, we obtain the singular-free $\mathcal O$$(\alpha_S)$ corrections
to the Born process in the region of the phase space labeled as Part 2
(Sec. \ref{pss}):
\bea
E_{\gamma}{d\sigma^{\gamma q\ra\gamma qg}_{\alpha_S (2)}\over d^3p_{\gamma}}
=
\theta (1-w){\alpha^2\alpha_Se_q^4\over\pi s^2vw}C_F
{v^2+(1-v)^2\over v(1-v)}\cdot
\makebox[3cm]{}
\nonumber\\
\cdot\left[
[w^2+(1-w)^2]\ln{(yE_e)^2\theta_{cut}^2(1-w)^2\over\mu^2}
+1
\right]
\label{xeq:col2}
\eea
\bea
E_{\gamma}{d\sigma^{\gamma g\ra\gamma q\bar{q}}_{\alpha_S (2)}\over 
d^3p_{\gamma}}=
\theta (1-w){\alpha^2\alpha_Se_q^4\over\pi s^2vw}\,{1\over 2}\,
{1+(1-v)^2\over 1-v}\cdot
\makebox[4cm]{}
\nonumber\\
\cdot\left[
[w^2+(1-w)^2]\ln{(yE_e)^2\theta_{cut}^2(1-w)^2\over\mu^2}
+1
\right]
\label{yeq:col2}
\eea
These cross sections (\ref{xeq:col2}, \ref{yeq:col2}) are equivalent
to the results quoted in Sec. \ref{pss} in Eqs. (\ref{co1}, \ref{co2}).

The squared matrix element 
$|\overline{M}^{\gamma q\ra\gamma qg}|^2$ 
calculated for the gluon moving (almost) parallel to the initial
electron, $\vec{p}_2 || \vec{q}=y\vec{p}_e$ (see the notation in Eq. 
(\ref{ea7})), 
contains no collinear singularities and the 
cross section $d\sigma^{\gamma q\ra\gamma qg}_{\alpha_S (2)}$
corresponding to this configuration can be obtained performing
the exact numerical calculations as in Part 5 of the phase
space (see Appendix \ref{xsec:gen}) or it can be neglected as it is a term
of order $\mathcal O$$(\theta_{cut}^2)$.

On the other hand, the configuration corresponding to the final
antiquark collinear to the initial electron, 
$\vec{p}_2 || \vec{q}=y\vec{p}_e$ (see the notation in Eq. (\ref{ea8})), 
leads to the same contribution
in the cross section $d\sigma^{\gamma g\ra\gamma q\bar{q}}_{\alpha_S (2)}$
as the quark collinear to the electron, $\vec{p}_1 || \vec{q}=y\vec{p}_e$,
and this is included in Eq. (\ref{cross5i}) 
in the summation over $2N_f$ flavours.
%$\sum_{q,\bar{q}}$.

%%%%%%%%%%%%%%%%%%%%%%%%%%%%%%%%%%%%%%%%%%%%%%%%%%%%%%%%%%%%%%%%%%%%%%%%%
\section{Collinear configuration ($\vec{p}_1 || \vec{p}_p$ or
$\vec{p}_2 || \vec{p}_p$)}\label{xsec:col3}
%%%%%%%%%%%%%%%%%%%%%%%%%%%%%%%%%%%%%%%%%%%%%%%%%%%%%%%%%%%%%%%%%%%%%%%%%

Herein we derive the cross sections 
$d\sigma^{\gamma q\ra\gamma qg}_{\alpha_S (3)}$ and
$d\sigma^{\gamma g\ra\gamma q\bar{q}}_{\alpha_S (3)}$ corresponding
to the region of the phase space labeled in Sec. \ref{pss} as Part 3.
First, we consider the 
final gluon moving almost parallel to the initial quark
and the final antiquark moving almost parallel to the initial
gluon, $\vec{p_2}||\vec{p}=x\vec{p}_p$ (see the notation in Eqs. 
(\ref{ea7}, \ref{ea8})).
We write the four-momenta in the laboratory frame with the $z$ axis chosen
along the direction of the initial proton momentum:
\bea
q=y E_e(1,...,0,0,-1),
\eea
\bea
p=x E_p (1,...,0,0,1),
\eea
\bea
p_{\gamma}=E_{\gamma}(1,...,0,\sin\alpha^{\prime},\cos\alpha^{\prime}),
\eea
\bea
p_2=E (1,...,\cos\theta_2\sin\theta_1,\cos\theta_1),
\eea
\bea
p_1=q+p-p_{\gamma}-p_2,
\eea
where 
\bea
E_{\gamma}=xE_p(1-v)+yE_evw,
\eea
\bea
\sin\alpha^{\prime} = {\sqrt{s(1-v) vw} \over E_{\gamma}},
\eea
\bea
\cos\alpha^{\prime} = {x E_p (1-v) - y E_e vw\over E_{\gamma}},
\eea
\bea
E={v(1-w)\over 1-vw}xE_p,
\eea
\bea
0\le\theta_1\le\theta_{cut}\ll 1.
\eea
From Eqs. (\ref{ec1}, \ref{ps3Ec}, \ref{ec17}, 
\ref{eq:xsec:m2}-\ref{eq:xsec:A4}, \ref{ed13}-\ref{ed19}) 
we have the cross sections which are singular
in 4 dimensions:
\bea
E_{\gamma}{d\sigma^{\gamma q\ra\gamma qg}_{\alpha_S (3)}\over d^3p_{\gamma}}
\makebox[0.1cm]{}| \makebox[-3.4pt]{} \raisebox{-5pt}{$|$}_{singular}
=
{1\over sv\pi}\cdot
{1\over 2s} {d(PS)_3 \over dvdw} 
|\overline{M}^{\gamma q\ra\gamma qg}|^2
\sim {\mathcal{O}}({1\over \varepsilon}) +{\mathcal{O}}(1)+...,
\eea
\bea
E_{\gamma}{d\sigma^{\gamma g\ra\gamma q\bar{q}}_{\alpha_S (2)}
\over d^3p_{\gamma}}
\makebox[0.1cm]{}| \makebox[-3.4pt]{} \raisebox{-5pt}{$|$}_{singular}
=
{1\over sv\pi}\cdot
{1\over 2s} {d(PS)_3 \over dvdw} 
|\overline{M}^{\gamma g\ra\gamma q\bar{q}}|^2
\sim {\mathcal{O}}({1\over \varepsilon}) +{\mathcal{O}}(1)+...,
\eea
As in Appendix \ref{xsec:col2}, we remove these singularities applying 
the factorization procedure:
\bea
E_{\gamma}{d\sigma^{\gamma q\ra\gamma qg}_{\alpha_S (3)}\over d^3p_{\gamma}}
=
E_{\gamma}{d\sigma^{\gamma q\ra\gamma qg}_{\alpha_S (3)}\over d^3p_{\gamma}}
\makebox[0.1cm]{}| \makebox[-3.4pt]{} \raisebox{-5pt}{$|$}_{singular}
-\int\limits_0^1 d\xi H_{qq}(\xi,\mu)
E_{\gamma}{d\sigma^{\gamma q\ra\gamma q}\over d^3p_{\gamma}}(\xi s,t,\xi u),
\eea
\bea
E_{\gamma}{d\sigma^{\gamma g\ra\gamma q\bar{q}}_{\alpha_S (3)}\over 
d^3p_{\gamma}}=
E_{\gamma}{d\sigma^{\gamma g\ra\gamma q\bar{q}}_{\alpha_S (3)}\over 
d^3p_{\gamma}}
\makebox[0.1cm]{}| \makebox[-3.4pt]{} \raisebox{-5pt}{$|$}_{singular}
-\int\limits_0^1 d\xi H_{qg}(\xi,\mu)
E_{\gamma}{d\sigma^{\gamma q\ra\gamma q}\over d^3p_{\gamma}}(\xi s,t,\xi u),
\eea
with
\bea
H_{qg}(\xi,\mu)={\alpha_S\over 6\alpha e_q^2} H_{q\gamma}(\xi,\mu),
\eea
\bea
H_{qq}(\xi,\mu)=
-{1\over\varepsilon}\,{\alpha_S\over 2\pi}C_F
{1+\xi^2\over 1-\xi}\left({4\pi\hat{\mu}^2\over\mu^2}\right)^{\varepsilon}
{\Gamma (1-\varepsilon)\over\Gamma (1-2\varepsilon)}
+{\mathcal{O}}(\varepsilon)
\eea
and
\bea
E_{\gamma}{d\sigma^{\gamma q\ra\gamma q}\over d^3p_{\gamma}}(s,t,u)=
{2\pi\alpha^2e_q^4\over s}\,{\hat{\mu}^{2\varepsilon}\over
\Gamma (1-\varepsilon)}\left({4\pi\hat{\mu}^2s\over tu}\right)^{\varepsilon}
\cdot\nonumber
\eea 
\bea
\makebox[6cm]{}\cdot (1-\varepsilon)
\left[-(1-\varepsilon)\left({u\over s}+{s\over u}\right) -2\varepsilon \right].
\eea
The final singular-free formulae for the $\mathcal O$$(\alpha_S)$ corrections
to the Born process in Part 3 of the phase space are:
\bea 
E_{\gamma}{d\sigma^{\gamma q\ra\gamma qg}_{\alpha_S (3)}\over d^3p_{\gamma}}
=\Theta (1-w){\alpha^2\alpha_Se_q^4\over \pi s^2(1-v)}C_F
\left(vw+{1\over vw}\right)\cdot
%\makebox[4cm]{}
\nonumber\eea\bea\makebox[4.5cm]{}
\cdot\left[
{1+\hat{x}^2\over 1-\hat{x}}\ln{(xE_p)^2\theta_{cut}^2(1-\hat{x})^2\over\mu^2}
+1-\hat{x}
\right],
\label{xeq:col3}
\eea
\bea
E_{\gamma}{d\sigma^{\gamma g\ra\gamma q\bar{q}}_{\alpha_S (3)}\over 
d^3p_{\gamma}}
=\Theta (1-w) {\alpha^2\alpha_Se_q^4\over \pi s^2(1-v)}\,{1\over 2}\,
\left(vw+{1\over vw}\right)\cdot
%\makebox[4cm]{}
\nonumber\eea\bea\makebox[4.5cm]{}
\cdot\left[
(\hat{x}^2+ (1-\hat{x})^2)\ln{(xE_p)^2\theta_{cut}^2(1-\hat{x})^2\over\mu^2}
+1\right],
\label{yeq:col3}
\eea
where
\bea
\hat{x}={1-v\over 1-vw}.
\eea
These formulae are equivalent to the formulae (\ref{co3}, \ref{co4})
in Sec. \ref{isol:a}.

The squared matrix element 
$|\overline{M}^{\gamma q\ra\gamma qg}|^2$ 
calculated for the final quark moving parallel to the initial
quark, $\vec{p}_1 || \vec{p}=x\vec{p}_p$ (see Eq. (\ref{ea7})), 
contains no collinear singularities and the 
cross section $d\sigma^{\gamma q\ra\gamma qg}_{\alpha_S (3)}$
corresponding to this configuration can be either calculated
numerically or neglected.

The configurations corresponding to the final
quark or antiquark collinear to the initial electron, 
$\vec{p}_1 || \vec{p}=x\vec{p}_p$ or
$\vec{p}_2 || \vec{p}=x\vec{p}_p$ (see Eq. (\ref{ea8})), 
give the same contributions
to the cross section $d\sigma^{\gamma g\ra\gamma q\bar{q}}_{\alpha_S (3)}$
and this is included in Eq. (\ref{cross5i}) 
in the summation over $2N_f$ flavours.
%$\sum_{q,\bar{q}}$.

%%%%%%%%%%%%%%%%%%%%%%%%%%%%%%%%%%%%%%%%%%%%%%%%%%%%%%%%%%%%%%%%%%%%%%%%%
\section{Collinear configuration ($\vec{p}_1 || \vec{p}_{\gamma}$)}
\label{xsec:col1}
%%%%%%%%%%%%%%%%%%%%%%%%%%%%%%%%%%%%%%%%%%%%%%%%%%%%%%%%%%%%%%%%%%%%%%%%%

Now, we choose the $z$ axis in the direction of the final photon
momentum:
\bea
q=y E_e(1,...,0,-\sin\alpha,-\cos\alpha),
\eea
\bea
p=x E_p (1,...,0,\sin\alpha,\cos\alpha),
\eea
\bea
p_{\gamma}=E_{\gamma}(1,...,0,0,1),
\eea
\bea
p_1=E (1,...,\cos\theta_2\sin\theta_1,\cos\theta_1),
\eea
\bea
p_2=q+p-p_{\gamma}-p_1,
\eea
with 
\bea
E_{\gamma}=xE_p(1-v)+yE_evw,
\eea
\bea
\sin\alpha = {\sqrt{s(1-v) vw} \over E_{\gamma}},
\eea
\bea
\cos\alpha = {x E_p (1-v) - y E_e vw\over E_{\gamma}},
\eea
\bea
E={v(1-w)\over 1-v+vw}E_{\gamma}\,\,\,\equiv-{s+t+u\over t+u}E_{\gamma}.
\eea
We assume that the angle between the final quark momentum and the final
photon momentum is small:
\bea
0\le \theta_1 \le \delta_{cut}\ll 1,
\eea
where $\delta_{cut}$ is a cut-off parameter.
The partonic cross sections derived from Eqs. 
(\ref{ec1}, \ref{ps3Ec}, \ref{ec18}, \ref{eq:xsec:m2}-\ref{ed12}, 
\ref{xeq:og}) contain the $1/\varepsilon$ singularity:
\bea
E_{\gamma}{d\sigma^{\gamma q\ra\gamma qg}_{\alpha_S (4)}\over d^3p_{\gamma}}
\makebox[0.1cm]{}| \makebox[-3.4pt]{} \raisebox{-5pt}{$|$}_{singular}
=
{1\over sv\pi}\cdot
{1\over 2s} {d(PS)_3 \over dvdw} 
|\overline{M}^{\gamma q\ra\gamma qg}|^2
\sim {\mathcal{O}}({1\over \varepsilon}) +{\mathcal{O}}(1)+...,
\eea
\bea
E_{\gamma}{d\sigma^{\gamma g\ra\gamma q\bar{q}}_{\alpha_S (4)}\over 
d^3p_{\gamma}}
\makebox[0.1cm]{}| \makebox[-3.4pt]{} \raisebox{-5pt}{$|$}_{singular}
=
{1\over sv\pi}\cdot
{1\over 2s} {d(PS)_3 \over dvdw} 
|\overline{M}^{\gamma g\ra\gamma q\bar{q}}|^2
\sim {\mathcal{O}}({1\over \varepsilon}) +{\mathcal{O}}(1)+...
\eea
The singularities are removed in the factorization procedure:
\bea
E_{\gamma}{d\sigma^{\gamma q\ra\gamma qg}_{\alpha_S (4)}\over d^3p_{\gamma}}
=
\nonumber\eea\bea\makebox[1cm]{}=
E_{\gamma}{d\sigma^{\gamma q\ra\gamma qg}_{\alpha_S (4)}\over d^3p_{\gamma}}
\makebox[0.1cm]{}| \makebox[-3.4pt]{} \raisebox{-5pt}{$|$}_{singular}
-\int\limits_0^1 d\xi H_{\gamma q}(\xi,\mu)
E_{\gamma}{d\sigma^{\gamma q\ra qg}\over d^3p_{\gamma}}(s,t/\xi,u/\xi),
\eea
\bea
E_{\gamma}{d\sigma^{\gamma g\ra\gamma q\bar{q}}_{\alpha_S (4)}\over 
d^3p_{\gamma}}=
\nonumber\eea\bea\makebox[1cm]{}=
E_{\gamma}{d\sigma^{\gamma g\ra\gamma q\bar{q}}_{\alpha_S (4)}\over 
d^3p_{\gamma}}
\makebox[0.1cm]{}| \makebox[-3.4pt]{} \raisebox{-5pt}{$|$}_{singular}
-\int\limits_0^1 d\xi H_{\gamma q}(\xi,\mu)
E_{\gamma}{d\sigma^{\gamma g\ra q\bar{q}}\over d^3p_{\gamma}}(s,t/\xi,u/\xi),
\eea
with
\bea
H_{\gamma q}(\xi,\mu)=
-{1\over\varepsilon}\,{\alpha\over 2\pi}e_q^2
\,{1+(1-\xi)^2\over \xi}\left({4\pi\hat{\mu}^2\over\mu^2}\right)^{\varepsilon}
{\Gamma (1-\varepsilon)\over\Gamma (1-2\varepsilon)}
+{\mathcal{O}}(\varepsilon),
\eea
\bea
E_{\gamma}{d\sigma^{\gamma q\ra qg}\over d^3p_{\gamma}}(s,t,u)=
8 E_{\gamma}{d\sigma^{qg\ra \gamma q}\over d^3p_{\gamma}}(s,t,u),
\eea
\bea
E_{\gamma}{d\sigma^{\gamma g\ra q\bar{q}}\over d^3p_{\gamma}}(s,t,u)=
{9\over 8} E_{\gamma}{d\sigma^{q\bar{q} \ra\gamma g}\over d^3p_{\gamma}}(s,t,u),
\eea
where the partonic cross sections $d\sigma^{\gamma q\ra qg}$
and $d\sigma^{\gamma q\ra qg}$ are given in Appendix \ref{xsec:col3}.
Finally, the $\mathcal O$$(\alpha_S)$ corrections
to the Born process in Part 4 of the phase space are
\bea
E_{\gamma}{d\sigma^{\gamma q\ra\gamma qg}_{\alpha_S (4)}\over d^3p_{\gamma}}
=\Theta (1-w) \cdot
\nonumber\eea\bea\makebox[1cm]{}
\cdot{\alpha^2\alpha_Se_q^4\over \pi s^2\hat{z}}C_F
{1+(1-\hat{v})^2\over 1-\hat{v}}
\left[
{1+(1-\hat{z})^2\over\hat{z}}\ln{E_{\gamma}^2\delta_{cut}^2
(1-\hat{z})^2\over\mu^2}
+\hat{z}
\right],
\label{xeq:col1}
\eea
\bea
E_{\gamma}{d\sigma^{\gamma g\ra\gamma q\bar{q}}_{\alpha_S (4)}
\over d^3p_{\gamma}}
=\Theta(1-w)\cdot
\nonumber\eea\bea\makebox[1cm]{}\cdot
{\alpha^2\alpha_Se_q^4\over \pi s^2\hat{z}}\,{1\over 2}\,
{\hat{v}^2+(1-\hat{v})^2\over\hat{v}(1-\hat{v})}
\left[
{1+(1-\hat{z})^2\over\hat{z}}\ln{E_{\gamma}^2\delta_{cut}^2
(1-\hat{z})^2\over\mu^2}
+\hat{z}
\right],
\label{yeq:col1}
\eea
where
\bea
\hat{z}=1-v+vw,
\eea
\bea
\hat{v}={vw\over 1-v+vw}.
\eea
Since $\delta_{cut}$ is small, we have
\bea
E_{\gamma}\delta_{cut}\approx E_T^{\gamma}R_{cut},
\label{k1}
\eea
where $R_{cut}$ is the cut-off radius in the rapidity and azimuthal
angle space, see Sec. \ref{pss}. From Eqs. (\ref{xeq:col1}), (\ref{yeq:col1})
and (\ref{k1}) one obtains Eqs. (\ref{co5}, \ref{co6}) corresponding
to Part 4 of the phase space as well as Eqs. (\ref{small1}, \ref{small3})
for the subtraction terms in the small cone approximation.

%%%%%%%%%%%%%%%%%%%%%%%%%%%%%%%%%%%%%%%%%%%%%%%%%%%%%%%%%%%%%%%%%%%%%%%%%
\section{Collinear configuration ($\vec{p}_2 || \vec{p}_{\gamma}$)}
\label{xsec:col4}
%%%%%%%%%%%%%%%%%%%%%%%%%%%%%%%%%%%%%%%%%%%%%%%%%%%%%%%%%%%%%%%%%%%%%%%%%

Finally, we assume that the momentum $\vec{p}_2$ is almost parallel
with the momentum of the final photon, $\vec{p}_{\gamma}$:
\bea
q=y E_e(1,...,0,-\sin\alpha,-\cos\alpha),
\eea
\bea
p=x E_p (1,...,0,\sin\alpha,\cos\alpha),
\eea
\bea
p_{\gamma}=E_{\gamma}(1,...,0,0,1),
\eea
\bea
p_2=E (1,...,\cos\theta_2\sin\theta_1,\cos\theta_1),
\eea
\bea
p_1=q+p-p_{\gamma}-p_2.
\eea
The calculation of the cross section for the process 
$\gamma g\ra \gamma q\bar{q}$ is the same as in the previous Appendix
\ref{xsec:col1}, and the formula for 
$d\sigma^{\gamma g\ra\gamma q\bar{q}}_{\alpha_S (4)}$ is given by
Eq. (\ref{yeq:col1}). This is included in the cross section
for the $\gamma p$ collision in the summation over quarks 
and antiquarks (2$N_f$ flavours), see Eq. \ref{cross5i}.

The calculation for the process $\gamma q\ra \gamma qg$ with 
the gluon momentum ($\vec{p}_2$) parallel to $\vec{p}_{\gamma}$
is different than the corresponding calculation in 
Appendix \ref{xsec:col1} for the quark momentum ($\vec{p}_1$)
parallel to $\vec{p}_{\gamma}$, since now we deal with the 
cross section which contains no singularities.

The cross section $d\sigma^{\gamma q\ra\gamma qg}_{\alpha_S (4)}$
is equal to
\bea
E_{\gamma}{d\sigma^{\gamma q\ra\gamma qg}_{\alpha_S (4)}\over d^3p_{\gamma}}
={1\over sv\pi}\cdot
{1\over 2s} {d(PS)_3 \over dvdw} 
|\overline{M}^{\gamma q\ra\gamma qg}|^2 ,
\label{las}
\eea
where $(PS)_3$ and $|\overline{M}^{\gamma q\ra\gamma qg}|^2$
are given in Eqs. (\ref{ps3Ec}) and (\ref{eq:xsec:m2}), respectively.
This cross section contains no singularities 
and is proportional to $\theta_{cut}^2$.
In the Part 4 of the phase space we assume that $w<w_{cut}$.
This cut-off restricts the phase space of the final gluon
this way that the gluon can not be too soft, so the cross section
(\ref{las}) contains no large terms due to the emission of soft gluons
and is small $\sim$$\mathcal O$$(\theta_{cut}^2)$.
It can be either calculated numerically 
or neglected if $\theta_{cut}$ is sufficiently small.

Similar configuration (gluon moving parallel to the final photon) was 
considered in Chapter \ref{small}, when the subtraction cross section
in the small cone approximation was discussed. In this case
we had no cut-off for $w$, and large terms due to the soft gluon
emission might contribute to the cross section, so the 
subtraction term $d\sigma_{sub}^{\gamma q\rightarrow\gamma g + q}$
(\ref{small2}) could not be neglected.
We have obtained the formula (\ref{small2}) for the subtraction
term from Eqs. (\ref{ps3Ec}, \ref{eq:xsec:m2}, \ref{las}) 
keeping all terms of order $\mathcal O$$(\theta_{cut}^2)$
and neglecting higher powers of $\theta_{cut}$.

%%%%%%%%%%%%%%%%%%%%%%%%%%%%%%%%%%%%%%%%%%%%%%%%%%%%%%%%%%%%%%%%%%%%%%%%%
%%%%%%%%%%%%%%%%%%%%%%%%%%%%%%%%%%%%%%%%%%%%%%%%%%%%%%%%%%%%%%%%%%%%%%%%%

% %%%%%%%%%%%%%%%%%%  CONE JET  %%%%%%%%%%%%%%%%%%%%%%%%%%%%%%%%%%%%%%%%%%%
%
% \bibitem{cone}
% J. Huth et al., Proc. of the 1990 DPF Summer Study on High Energy Physics,
% Snowmass, Colorado, edited by E.L. Berger (World Scientific, Singapore, 1992)
% p. 134.
%
% {\bf (
% \bibitem{Huth:1990mi}
% J.~E.~Huth {\it et al.},
% %``Toward A Standardization Of Jet Definitions,''
% FERMILAB-CONF-90-249-E
% %\href{http://www.slac.stanford.edu/spires/find/hep/www?r=fermilab-conf-90-249-e}{SPIRES entry}
% {\it Presented at Summer Study on High Energy Physics, Reaearch Directions for the Decade, Snowmass, CO, Jun 25 - Jul 13, 1990}
% ) }
%
% %\cite{Arnison:1983rn}
% %\bibitem{Arnison:1983rn}
% G.~Arnison {\it et al.}  [UA1 Collaboration],
% %``Observation Of Jets In High Transverse Energy Events At The Cern Proton - Anti-Proton Collider,''
% Phys.\ Lett.\ B {\bf 123} (1983) 115.
% %%CITATION = PHLTA,B123,115
%
% %\cite{Derrick:1995bg}
% %\bibitem{Derrick:1995bg}
% M.~Derrick {\it et al.}  [ZEUS Collaboration],
% %``Dijet cross-sections in photoproduction at HERA,''
% Phys.\ Lett.\ B {\bf 348} (1995) 665
% [arXiv:hep-ex/9502008].
% %%CITATION = HEP-EX 9502008

\end{document}